\documentclass[10pt, final, twocolumn, twoside, romanappendices]{IEEEtran}

\usepackage[utf8]{inputenc} 
\usepackage[T1]{fontenc}
\usepackage{url}              
\usepackage[colorlinks,linkcolor=blue,citecolor=blue]{hyperref}

\usepackage[cmex10]{amsmath}  
\usepackage{mleftright}       
\mleftright                   

\usepackage{graphicx}
\usepackage{epsfig}
\usepackage{stfloats}
\ifCLASSOPTIONcompsoc
\usepackage[caption=false,font=normalsize,labelfon
t=sf,textfont=sf]{subfig}
\else
\usepackage[caption=false,font=footnotesize]{subfig}
\fi

\usepackage{booktabs}

\usepackage[table]{xcolor}
\usepackage{amssymb}
\usepackage{epsfig,verbatim}
\usepackage{mathtools}
\usepackage{bbm}
\usepackage{bm}
\usepackage{enumitem}
\usepackage{mathrsfs}
\usepackage{mathtools, cuted}
\usepackage{esint}
\usepackage{amsfonts}
\usepackage{amsmath}
\usepackage{empheq}

\newtheorem{definition}{{{Definition}}}

\newtheorem{lemma}{{{Lemma}}}

\newtheorem{corollary}{{{Corollary}}}

\newtheorem{remark}{{{Remark}}}

\newtheorem{proposition}{{{Proposition}}}
\newcommand{\propositionlabel}[1]{\label{thm:#1}}
\newcommand{\propositionref}[1]{\ref{thm:#1}}

\newtheorem{assumption}{{{Assumption}}}
\newcommand{\assumptionlabel}[1]{\label{thm:#1}}
\newcommand{\assumptionref}[1]{\ref{thm:#1}}

\newtheorem{discussion}{{{Discussion}}}

\allowdisplaybreaks[4]

\usepackage{bbding}

\long\def\comment#1{}

\newcommand{\Vc}{{\cal V}}

\def\imagunit{\mathsf{j}}
\def\rq{\mathsf{r}}
\def\tq{\mathsf{t}}

\usepackage{orcidlink} 

\begin{document}

\title{Near-Field Positioning and Attitude Sensing Based on  Electromagnetic Propagation Modeling}
\author{Ang Chen$^{\orcidlink{0000-0001-9711-0078}}$,~Li Chen$^{\orcidlink{0000-0002-1754-0607}}$,~\IEEEmembership{Senior~Member, IEEE},~Yunfei Chen$^{\orcidlink{0000-0001-8083-1805}}$,~\IEEEmembership{Senior~Member, IEEE},\\Nan Zhao$^{\orcidlink{0000-0002-6497-7799}}$,~\IEEEmembership{Senior~Member, IEEE},~Changsheng You$^{\orcidlink{0000-0003-3245-9361}}$,~\IEEEmembership{Member, IEEE}

\thanks{Manuscript received 27 October 2023; revised 14 March 2024 and
30 April 2024; accepted 8 May 2024. This work of Li Chen was supported in part by the Industrial Technology Basic Project of MIIT (No. TC220A04M) and Anhui Provincial Natural Science Foundation (No. 2308085J24). This work of Yunfei Chen was supported in part by the King Abdullah University of Science and Technology Research Funding (KRF) under Award ORA-2021-CRG10-4696 and EPSRC TITAN (EP/Y037243/1, EP/X04047X/1). \emph{(Corresponding author: Li Chen.)}}

\thanks{
Ang Chen and Li Chen are with the CAS Key Laboratory of Wireless Optical Communication, University of Science and Technology of China (USTC), Hefei 230027, China (e-mail: chenang1122@mail.ustc.edu.cn; chenli87@ustc.edu.cn).
  
 Yunfei Chen is with the Department of Engineering, University of Durham, Durham DH1 3LE, U.K. (e-mail: yunfei.chen@durham.ac.uk).

Nan Zhao is with the School of Information and Communication Engineering, Dalian University of Technology, Dalian 116024, China (e-mail: zhaonan@dlut.edu.cn).
  
 Changsheng You is with the Department of Electronic and Electrical Engineering, Southern University of Science and Technology (SUSTech), Shenzhen 518055, China (e-mail: youcs@sustech.edu.cn).

  }

  }

\makeatletter
\def\ps@IEEEtitlepagestyle{
  \def\@oddfoot{\mycopyrightnotice}
  \def\@evenfoot{}
}
\def\mycopyrightnotice{
  {\scriptsize
  \begin{minipage}{\textwidth}
  \centering
  
\copyright~2024 IEEE. Personal use is permitted, but republication/redistribution requires IEEE permission.
See https://www.ieee.org/publications/rights/index.html for more information.
  \end{minipage}
  }
}

\markboth{IEEE Journal on Selected Areas in Communications}{Chen \MakeLowercase{\textit{et al.}}: Near-Field Positioning and Attitude Sensing Based on Electromagnetic Propagation Modeling}

\maketitle

\begin{abstract}
Positioning and sensing over wireless networks are imperative for many emerging applications. {However, since traditional wireless channel models over-simplify the user equipment (UE) as a point target, they cannot be used for sensing the attitude of the UE, which is typically described by the spatial orientation.} In this paper, a comprehensive electromagnetic propagation modeling (EPM) based on electromagnetic theory is developed to precisely model the near-field channel. For the noise-free case, the EPM model establishes the non-linear functional dependence of observed signals on both the position and attitude of the UE. To address the difficulty in the non-linear coupling, we first propose to divide the distance domain into three regions, separated by the defined
 \textit{Phase ambiguity distance} and \textit{Spacing constraint distance}. Then, for each region, we obtain the closed-form solutions for joint position and attitude estimation with low complexity. Next, to investigate the impact of random noise on the joint estimation performance, the Ziv-Zakai bound (ZZB) is derived to yield useful insights. The expected Cramér-Rao bound (ECRB) is further provided to obtain the simplified closed-form
expressions for the performance lower bounds. Our numerical results demonstrate that the derived ZZB can provide accurate predictions of the performance of estimators in all signal-to-noise ratio (SNR) regimes. More importantly, we achieve the millimeter-level accuracy in position estimation and attain the 0.1-level accuracy in attitude estimation. 
\end{abstract}

\section{Introduction}
\label{sec:intro}
Positioning and sensing over wireless networks play pivotal roles in a variety of emerging applications, including 
location-based services \cite{9729782}, object detection and tracking \cite{9309315}, Internet-of-Things (IoT) \cite{8454389}, smart environments \cite{7397880}, and the integrated sensing and communication (ISAC) systems \cite{Cong2023NearfieldIS,10147248}. Leveraging wireless networks to enhance the accuracy of positioning and sensing is challenging yet profoundly meaningful. Particularly, the accuracy of positioning and sensing is fundamentally contingent on the structure of the observed signals, which is inherently circumscribed by the underlying wireless channel models. Thus, channel modeling is very important.

\begin{figure}[!t]
	\centering
	\subfloat[{UPW and  SWM}]{
	\includegraphics[scale=0.68]{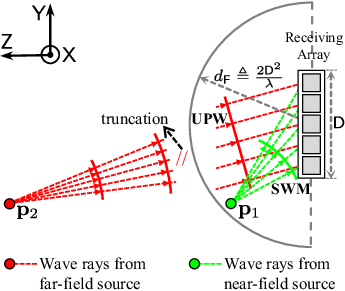}
		\label{fig:jsac11}
	}
	\subfloat[{EPM in this paper}]{
		\includegraphics[scale=0.68]{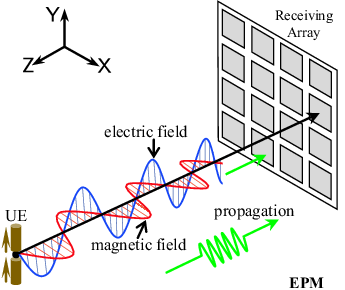}
		\label{fig:jsac12}
	}
 \caption{Illustration of the far-field UPW, near-field SWM, and near-field EPM. Fig. \ref{fig:jsac11} shows the far-field UPW and near-field SWM in the 2D view, {where the approximation process of transition from a spherical wavefront to a planar wavefront is truncated due to space limitations}. Fig. \ref{fig:jsac12} illustrates the near-field EPM based on 3D electromagnetic propagation theory, where the UE is considered as a dipole as an example.}
\end{figure}

The conventional channel modeling method usually adopts the far-field uniform plane wavefront (UPW) assumption \cite{lu2021communicating}, as shown by point $\mathbf{p}_{2}$ in Fig. \ref{fig:jsac11}. This is because the small arrays of today's wireless networks make the UE usually reside
in the far-field region\footnote{
 In the far-field region, the transceiver distance is larger than the \textit{Fraunhofer distance} $d_{\mathsf{F}}\triangleq2\mathsf{D}^{2}/\lambda$ \cite{sherman1962properties}, where $\mathsf{D}$ is the maximum geometric dimension of the receiving array and $\lambda$ is the wavelength.} of the receiving array. The far-field UPW model only allows the extraction of \textit{direction} information since the signals impinging on all the receiving elements are assumed to have the same angle of arrival (AoA). {To obtain the {position} parameter of the UE, one feasible approach is by deploying multiple spatially distributed arrays with known positions for the estimation of AoAs \cite{5762798}, i.e., the triangulation method. Alternatively, the clock synchronization between the UE and the receiver can be employed to measure the time of arrivals (ToAs) \cite{1618619}, i.e., the trilateration method. In addition, the carrier phase-based method has recently been studied as an ultra-accurate positioning scheme \cite{10015771,9601204}.}

Enhancing the accuracy of positioning is achievable through the utilization of large-scale receiving arrays or surfaces \cite{hu2018beyond2,bjornson2019massive,tang2020wireless}. In fact, the 5G standard  was envisioned to operate in the millimeter-wave (mmWave) bands \cite{8316581}, while the ongoing beyond 5G (B5G) research is already focusing on the so-called sub-terahertz (THz) bands \cite{9269931,akyildiz2018combating,rappaport2019wireless}, i.e., in the range of
100--300 GHz. The short wavelength corresponding to high-frequency signals makes it practically possible to realize antenna arrays featuring a significantly large number of small elements. Adopting large-scale arrays in high-frequency bands will shift the electromagnetic diffraction field from the far-field region to the near-field region\footnote{In this paper, the term "near-field" refers to the "radiative near-field", where the transceiver distance is smaller than $d_{\mathsf{F}}$, but larger than the \textit{Fresnel distance} $d_{\mathsf{f}}=0.5\sqrt{\mathsf{D}^{3}/\lambda}$ \cite{selvan2017fraunhofer}.}, 
making the far-field UPW assumption no longer hold. When the UE resides within the near-field region of the receiving array, as exemplified by point $\mathbf{p}_{1}$ in Fig. \ref{fig:jsac11}, the spherical wavefront modeling (SWM) is widely adopted to effectively exploit the spherical wavefront properties for highly accurate position estimation. The SWM-based array manifold encompasses a wealth of information pertaining to the position of the UE, which allows to directly extract both \textit{direction} and \textit{distance} information from the receiving array. Consequently, references \cite{9335528,guerra2021near,9625826,Guidi,zhang2018spherical,yin2017scatterer} studied the SWM-based signal processing algorithm for high-precision near-field positioning. Besides, in \cite{hu2018beyond2,delmas2016crb,alegria2019cramer,6362262}, theoretical performance bounds for the SWM-based near-field positioning were derived.

However, neither the far-field UPW nor the near-field SWM is able to capture the {attitude} parameter of the UE, since both models overlook the geometric shape of the UE. In fact, the attitude parameter
depends on the geometric shape of the UE, which should be modeled as an extended target. For basic electronic devices like dipoles and wire antennas, the attitude parameter can be defined
as the \textit{spatial orientation}, which represents the rotation angle of the UE. 
\begin{table}[!t]
\center
{\caption{Parameters that can be obtained in different channel models.}
\label{tab:parameters}
\begin{tabular}{|c|c|}
\hline \rowcolor{blue!25}
{\textbf{Channel Models}}& {\textbf{Parameters}}\\
\hline
{Far-field} UPW& AoA\\
\hline
Near-field SWM & {position}\\
\hline
Near-field EPM &{position} and {attitude} \\
\hline
\end{tabular}}
\end{table}

To sense the attitude over wireless networks, it is necessary to establish a precise relationship between the wireless channel and the {attitude}. This can be achieved by using the electromagnetic propagation theory since the UE can be regarded as an extended excitation source responsible for creating the wireless equivalent channel. Fig. \ref{fig:jsac12} portrays a refined approach to the near-field channel modeling stemming from the fundamental electromagnetic (Maxwell's) equations, referred to as electromagnetic propagation modeling (EPM). The EPM, in contrast to the traditional UPW and SWM, comprehensively considers the physical characteristics of the UE, i.e., attitude, shape, and size. These physical characteristics directly influence the {source current distribution} within the UE, which generates electric and magnetic fields propagating in three-dimensional (3D) space, causing the observed {electric field} at the receiving array. The ratio of the observed electric field to the initial electric field defines the equivalent channel, which in turn is determined
by the source current distribution of the UE and thus essentially characterizes the functional dependence on the {position} and {attitude} parameters. Consequently, the EPM enables the joint sensing of {position} and {attitude}. In Table \ref{tab:parameters}, the types of parameters that can be
obtained under different channel modeling are summarized.

EPM is considered to be the most accurate electromagnetic theory-based model for investigating signals or channels in the near-field region by far. In \cite{friedlander2019localization}, B. Friedlander first recognized that both the far-field UPW and the near-field SWM ignore the characteristics of the source, and more accurate near-field channel modeling should be established based on electromagnetic theory. A. A. D'Amico \emph{et al.} in \cite{de2021cramer} and \cite{d2022cramertsp} investigated the EPM based on the \textit{radiation vector} \cite[Ch. 15]{orfanidis} and 
obtained the Cramér-Rao Bound (CRB) for the source positioning with high accuracy. In \cite{angchen}, A. Chen \emph{et al.} developed a generic near-field positioning system with the arbitrary position of UE and three different types of observed electric fields. However, the previously mentioned works on EPM are predominantly constrained to near-field positioning. To the best of the authors'
knowledge, the research on the joint {position} and {attitude} estimation utilizing EPM is still lacking in the existing literature. The challenge is that the EPM is a complex non-linear model that encompasses both position and attitude parameters, making it hard to accurately estimate these parameters from noise-free signals. Further, the corresponding performance bounds of joint {position} and {attitude} estimation have never been studied for the noisy case, which makes the fundamental limits of the joint estimation system unclear.

To fill in this gap, in this paper, we take a pioneering step to investigate the joint position and attitude estimation based on the
EPM. To this end, we first develop a comprehensive EPM, which allows us to establish a set of non-linear equations containing both the {position} and {attitude} parameters\footnote{We consider the UE to be a dipole whose position is defined as the spatial coordinate while the attitude is modeled as the spatial orientation.}. Then, we derive the closed-form solutions for this set of non-linear equations. Further, to investigate the impact of random noise on joint estimation performance and the fundamental limits of estimation performance, we derive the Ziv-Zakai bound (ZZB) and the expected CRB (ECRB) for joint position and attitude estimation. Our main contributions are summarized as follows.

\begin{itemize}
	\item 
 \textbf{{Closed-form solutions for noise-free joint estimation.}} In the noise-free case, the joint estimation problem can be transformed into solving a set of non-linear equations to accurately obtain the \textit{unknown} quantities of position and attitude as functions of the \textit{known} observed signals. The above set of non-linear equations is hard to solve directly due to the complex exponential coupling. To tackle this challenge, we first propose to divide the distance domain into three regions, separated by the defined \textit{Phase ambiguity distance} $d_{\mathsf{PA}}$ and  \textit{Spacing constraint distance} $d_{\mathsf{SC}}$. Then, for each region, we reconstruct the above equations by decoupling and provide closed-form solutions. 
\item \textbf{Globally tight performance bound for joint
estimation.} We derive the ZZB for joint position and attitude estimation. Specifically, the ZZB is a \textit{globally tight} bound in all SNR regimes, which incorporates \textit{prior} information about
the unknown parameters and is not limited to unbiased estimates. Our derived ZZBs are the explicit functions of SNR, receiving array size, carrier frequency, \textit{prior} distributions of the unknown parameters, and the \textit{ambiguity function} (AF) related to the \textit{near-field electromagnetic channel}. As such, the ZZBs afford new insights into the impact of the above system parameters on joint estimation performance. Further, to obtain simplified closed-form expressions for the performance lower bounds and facilitate the
comparison with the derived ZZB, we derive the ECRB, which is a \textit{locally tight} bound characterizing a fundamental limit in the high SNR regime.
\end{itemize}

\textit{Organization:} Section \ref{sec:problem} describes the generic joint position and attitude estimation system model, develops a comprehensive  EPM, and derives the \textit{near-field electromagnetic channel} based on the EPM. In Section \ref{sec:method}, the \textit{Phase ambiguity distance} $d_{\mathsf{PA}}$ and  \textit{Spacing constraint distance} $d_{\mathsf{SC}}$ are proposed to finely divide the distance domain into three regions and the closed-form solutions to the position and attitude parameters for each region are provided. In Section \ref{sec:method1}, the closed-form expressions of ZZB and ECRB are derived to evaluate the joint estimation performance. Numerical results are presented in Section \ref{sec:simulation}, and the conclusions are provided in Section \ref{sec:con}.

\textit{Notation:} Vectors and matrices are denoted in bold lowercase and uppercase, respectively, e.g., $\mathbf{a}$ and $\mathbf{A}$. We use $[\mathbf{A}]_{ij}$ to
denote the $(i,j)$-th entry of $\mathbf{A}$ and $\mathbf{a}_{i}$ to denote the $i$-th entry of $\mathbf{a}$. The superscripts $(\cdot)^{\dagger}$, $(\cdot)^{-1}$, and $(\cdot)^{\mathsf{T}}$ represent the matrix hermitian-transpose, inverse, and transpose, respectively. $(\cdot)^{*}$ and $\Re\{\cdot\}$ designate the complex conjugate and the real part of the input operations. The operator $\|\cdot\|$ means to compute $\mathcal{L}_{2}$-norm of the input and $|\cdot|$ stands for the modulo operator. $\otimes$ denotes the cross product. The notations $\mathbb{C}$ and $\mathbb{R}$ represent the sets of complex numbers and of real numbers, respectively. The notation $\imagunit$ denotes the imaginary unit, and $\mathbf{I}$ indicates the identity matrix. $\mathbb{E}_{\kappa}\left\{\cdot\right\}$ and $\mathbb{V}_{\kappa}\left\{\cdot\right\}$ denote the statistical expectation and variance with respect to $\kappa$, respectively. $\mathcal{F}(x)|_{x=\kappa}$ denotes the value of function $\mathcal{F}(x)$ under the condition $x=\kappa$.

\section{System Model}\label{sec:problem}
\subsection{System Description}\label{sec:sysdescription}
\begin{figure}[!t]
\centering
\includegraphics[scale=0.59]{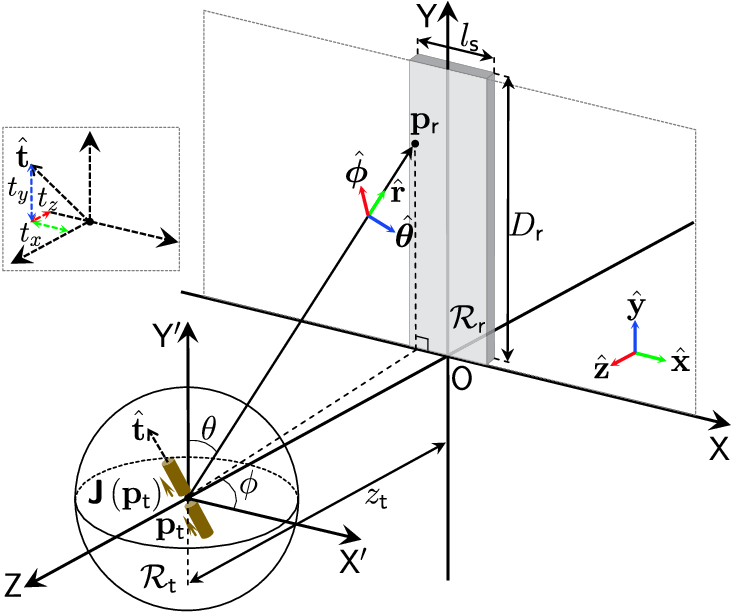}
\caption{Illustration of the joint {position} and {attitude} estimation system.}
\label{system2}
\end{figure}

Consider the system in Fig. \ref{system2}, in which we aim to jointly estimate the {position} and {attitude} of UE based on the {observed signals}. Different electromagnetic components have various geometric shapes, resulting in diverse current distributions and {attitude} descriptions. We consider that the UE is a \textit{Hertzian dipole}\footnote{The \textit{Hertzian dipole}, also known as the \textit{elementary}, \textit{ideal}, or \textit{short} \textit{dipole}, can be regarded as the fundamental component of complex electromagnetic devices such as V-dipole, wire, rectangular loop, biquad, and bowtie antennas.} of length $l_{\mathsf{dip}}$. Thus, the {position} and {attitude} parameters of the UE can be characterized by the spatial coordinates of its mass point and its spatial orientation, respectively.

\textit{Geometric model of the transmitter (UE):} We first establish the $\mathsf{OXYZ}$ cartesian coordinate system. The UE is located at an arbitrary point $\mathbf{p}_{\mathsf{t}}=\left(0,0,z_{\tq}\right)^{\mathsf{T}}$\footnote{This is the {"CPL"} case (the UE is {located} at $\left(0,0,z_{\tq}\right)$), which is frequently discussed in classic near-field positioning works, such as \cite{hu2018beyond2}, \cite{guerra2021near}, \cite{de2021cramer}, \cite{d2022cramertsp}.} on the $\mathsf{Z}$-axis inside the source region $\mathcal{R}_{\mathsf{t}}$ and is pointed in an arbitrary direction $\hat{\mathbf{t}}=t_{x}\hat{\mathbf{x}}+t_{y}\hat{\mathbf{y}}+t_{z}\hat{\mathbf{z}}$, where $\hat{\mathbf{x}}$, $\hat{\mathbf{y}}$, and $\hat{\mathbf{z}}$ are unit vectors along the $\mathsf{X}$-, $\mathsf{Y}$-, and $\mathsf{Z}$-dimension in the $\mathsf{OXYZ}$ system. We assume that the UE is {rotated} in the $\mathsf{YOZ}$ plane (i.e., $t_{x}=0$) and the {attitude} is estimated frequently enough that the deflection angle of the dipole does not change significantly compared to the reference state, whose mathematical description is $t_{y}\in(0,1]$ and $t_{z}\in[0,1)$. Thus, we have $t_{y}=\sqrt{1-t_{z}^2}$ based on the normalization constraint $t_{x}^2+t_{y}^2+t_{z}^2=1$. 
Then, we provide the {position} and {attitude} parameters to be estimated in \textit{Definition} \ref{definition2} and their \textit{prior} distributions in \textit{Assumption} \assumptionref{assum:p}.

\begin{definition}[Parameters to be estimated]\label{definition2} 
   {The {position} and {attitude} parameters of the UE can be written as the unknown vector $\bm{\xi}\triangleq\left({z_{\tq}},t_{z}\right)^{\mathsf{T}}\in \mathbb{R}^{2}$. }
\end{definition}
\begin{assumption}[Prior distributions of $\bm{\xi}$]\assumptionlabel{assum:p}
{$z_{\tq}$ is  uniformly distributed in $[\mathsf{H}_{1}, \mathsf{H}_{2}]$ and $t_{z}$ is  uniformly distributed in $[0, 1)$, i.e., $z_{\tq} \sim \mathcal{U}[\mathsf{H}_{1}, \mathsf{H}_{2}]$ and
$t_{z} \sim \mathcal{U}[0, 1)$, where ${\mathsf{H}_{1}}$ and ${\mathsf{H}_{2}}$ are arbitrary positive constants with ${\mathsf{H}_{1}}<{\mathsf{H}_{2}}$ and $\mathsf{H}_{\tq}\triangleq \mathsf{H}_{2}-\mathsf{H}_{1}$.}
\end{assumption}

\textit{Geometric model of the receiver:} The {long strip observation region} is $\mathcal{R}_{\mathsf{r}}=\left \{(x_{\mathsf{r}},y_{\mathsf{r}},0):|x_{\mathsf{r}}|\leq l_{\mathsf{s}}/2,0 \leq y_{\mathsf{r}}\leq D_{\mathsf{r}}\right \}$ equipped with any observation point $\mathbf{p}_{\mathsf{r}}=(x_{\mathsf{r}},y_{\mathsf{r}},0)^{\mathsf{T}}$, where $l_{\mathsf{s}}$ is the length of the short side, and $D_{\rq}$ is the length of the long side. Then, the maximum geometric dimension of $\mathcal{R}_{\mathsf{r}}$ can be expressed as $\mathsf{D}=\sqrt{l_{\mathsf{s}}^2+D_{\rq}^2}\approx D_{\rq}$. In addition, we consider that  $\mathcal{R}_{\mathsf{r}}$ is filled with small rectangular antenna elements of length and width $l_{\mathsf{s}}$, placed along the linear grid. The spacing between adjacent grid centers is denoted as $d_{\mathsf{sp}}$ with $l_{\mathsf{s}}\leq d_{\mathsf{sp}}$. We define $\varpi\triangleq\frac{l_{\mathsf{s}}}{d_{\mathsf{sp}}}\leq 1$ as the
array occupation ratio (AOR), which clearly signifies the fraction of the total area of the {observation region} that is occupied by the rectangular array elements. We consider the extreme case where the discrete array becomes a continuous holographic surface \cite{huang2020holographic,9724113,9696209} that completely covers $\mathcal{R}_{\rq}$, i.e., $\varpi= 1$. Therefore, the centers of the  array elements are the set of points
of $\mathcal{R}_{\rq}$ given by $\left\{\left(0,y_{\rq},0\right):y_{\rq}=\left(n_{\mathsf{y}}-\frac{1}{2}\right)l_{\mathsf{s}}\triangleq
y_{\mathsf{r};{n_{\mathsf{y}}}}\right\}$ with $1\leq n_{\mathsf{y}}\leq \mathsf{N}$ and $\mathsf{N}=\lfloor {D_{\mathsf{r}}}/{l_{\mathsf{s}}}\rfloor$, and the region of the $n_{\mathsf{y}}$-th element is 
$\mathcal{R}_{\mathsf{r};n_{\mathsf{y}}}=\left\{\left(x_{\rq},y_{\rq},0\right):x_{\rq}\in \left[-\frac{l_{\mathsf{s}}}{2},\frac{l_{\mathsf{s}}}{2}\right],y_{\rq}\in \left[y_{\mathsf{r};{n_{\mathsf{y}}}}-\frac{l_{\mathsf{s}}}{2},y_{\mathsf{r};{n_{\mathsf{y}}}}+\frac{l_{\mathsf{s}}}{2}\right]\right\}$. We also establish the $\mathsf{PX^{\prime}Y^{\prime}Z}$ system  with $\mathbf{p}_{\tq}$ as its origin, which has a pure translational relationship with $\mathsf{OXYZ}$ system, and establish a spherical coordinate system $\left(r,\theta,\phi\right)$ of point $\mathbf{p}_{\mathsf{t}}$ with respect to $\mathsf{PX^{\prime}Y^{\prime}Z}$ system. $\hat{\bm{\theta}}$ and $\hat{\bm{\phi}}$ are unit vectors along the $\theta$ and $\phi$ coordinate curves. ${\hat{\mathbf{r}}}$ is a unit vector denoting the direction of the spatial vector $\mathbf{r}\triangleq\mathbf{p}_{\mathsf{r}}-\mathbf{p}_{\mathsf{t}}$, i.e., ${\hat{\mathbf{r}}}={{\mathbf{r}}}/ {\|{\mathbf{r}}\|}$. 

\textit{Description of electromagnetic propagation:} Based on the electromagnetic theory, the electric
current distribution ${\bm{\mathsf{J}}}(\mathbf{p}_{\tq},t)$ of the UE generates the electric field vector ${\mathbf{e}}\left(\mathbf{r},t\right) \in \mathbb{C}^{3}$ at  $\mathbf{p}_{\mathsf{r}}$, identified through the vector 
 $\mathbf{r}$. We consider time-harmonic fields and introduce the phasor fields: ${\bm{\mathsf{J}}}(\mathbf{p}_{\tq}, t)=\Re\left\{{\bm{\mathsf{J}}}\left(\mathbf{p}_{\tq}\right)\mathrm{e}^{\imagunit \omega t}\right\}$ and ${\mathbf{e}}\left(\mathbf{r},t\right)=\Re\left\{{\mathbf{e}}\left(\mathbf{r}\right)\mathrm{e}^{\imagunit \omega t}\right\}$,
where $\omega$ is the angular frequency in  radians/second. {In this case, Maxwell's equations are considerably simplified and can be written in terms of the current
and field phasors, i.e., ${\bm{\mathsf{J}}}\left(\mathbf{p}_{\tq}\right)$ and ${\mathbf{e}}\left(\mathbf{r}\right)$.} We assume that the electromagnetic field
propagates in a homogeneous and isotropic medium with neither scatterers nor reflectors. In other words, there is only a line-of-sight link from $\mathcal{R}_{\mathsf{t}}$ to  $\mathcal{R}_{\mathsf{r}}$\footnote{An interesting extension of our model is to consider the multi-path, where the electromagnetic reflection and scattering properties of obstacles or clusters need to be further investigated. This is left for future work.}. 

\subsection{Electric Field Model Based on Comprehensive EPM}
According to the fundamental electromagnetic (Maxwell's) equations, the current  distribution ${\bm{\mathsf{J}}}\left(\mathbf{p}_{\mathsf{v}}\right)$ and vector electric field ${\mathbf{e}}\left(\mathbf{p}_{\mathsf{v}}\right)$ at any arbitrary point $\mathbf{p}_{\mathsf{v}}\in\mathbb{R}^{3}$ in space satisfy the following inhomogeneous Helmholtz wave equation \cite{poon2005degrees}:
\begin{equation}
-\nabla_{\mathbf{p}_{\mathsf{v}}} \times \nabla_{\mathbf{p}_{\mathsf{v}}} \times {\mathbf{e}}\left(\mathbf{p}_{\mathsf{v}}\right)+k^2 {\mathbf{e}}\left(\mathbf{p}_{\mathsf{v}}\right)=\imagunit k  \eta{\bm{\mathsf{J}}}\left(\mathbf{p}_{\mathsf{v}}\right),\label{eq:Helmholtz}
\end{equation}
where $\nabla_{\mathbf{p}_{\mathsf{v}}} \times$ is
the curl operation with respect to $\mathbf{p}_{\mathsf{v}}$, $k=2\pi/\lambda$ is the wave number, and $\eta$ is the intrinsic impedance of  spatial medium, which is $376.73~\text{ohm}$ in free space.

To explicitly express the relationship between the
current distribution ${\bm{\mathsf{J}}}\left(\mathbf{p}_{\tq}\right)$ of the excitation
source (UE) and the electric field
${\mathbf{e}}\left(\mathbf{r}\right)$ at the receiver, \textit{Green’s} method is utilized to obtain the inverse map of \eqref{eq:Helmholtz}:
\begin{equation}
{\mathbf{e}}\left({\mathbf{r}}\right)=\iiint_{\mathcal{R}_{\mathsf{t}}}{{\mathbf{G}}_{t}}\left(\mathbf{r}\right){\bm{\mathsf{J}}}\left({\mathbf{p}}_{\mathsf{t}}\right)d\mathbf{p}_{\mathsf{t}},\label{eq:eff}
\end{equation}
where ${\mathbf{G}}_{t}({\mathbf{r}}) \in \mathbb{C}^{3\times 3}$ is referred to as the \textit{tensor Green function} in electromagnetic theory and can be expressed as
\begin{equation}
\frac{{\mathbf{G}_{t}}({\mathbf{r}})}{{G_{s}}(r)}=\left(1+\frac{\imagunit}{kr}-\frac{1}{k^{2}{r}^{2}}\right)\mathbf{I}-\left(1+\frac{\imagunit 3}{kr}-\frac{3}{k^{2}{r}^{2}}\right)\hat{\mathbf{r}}{\hat{\mathbf{r}}}^{\dagger},
\label{eq:green} 
\end{equation}
where $r=\|{\mathbf{r}}\|$ and ${G_{s}}(r)$ is the \textit{scalar Green function}, i.e.,
\begin{equation}
    {G_{s}}(r)=\imagunit\frac{ \eta}{2 \lambda r} \mathrm{e}^{\imagunit  k r}.
\end{equation}
It is evident from \eqref{eq:green} that when $r \geq \lambda$\footnote{When $r=\lambda$, $\left|1+\frac{\imagunit}{k r}-\frac{1}{k^{2}{r^{2}}}\right|^{2}\approx 0.975$, $\left|1+\frac{\imagunit 3}{k r}-\frac{3}{k^{2}{r^{2}}}\right|^{2}\approx 1.082$.}, the second and third terms in the two parentheses in \eqref{eq:green} can be neglected \cite{angchen}, and hence 
\begin{equation}
{{\mathbf{G}}_{t}}({\mathbf{r}}) \simeq {G_{s}}(r)\left(\mathbf{I}-{\hat{\mathbf{r}}{\hat{\mathbf{r}}^{\dagger}}}\right).
\label{eq:Gr}
\end{equation}
Since $r \geq \lambda$ always holds when the UE resides in the near-field region of the receiving array (\textit{observation region} $\mathcal{R}_{\rq}$),  \eqref{eq:Gr} is adopted in the subsequent analysis. Substituting \eqref{eq:Gr} into \eqref{eq:eff} and following the definition of vector cross product, we have
\begin{equation}
{\mathbf{e}}\left({\mathbf{r}}\right)={G_{s}}(r) \left({\hat{\mathbf{r}}} \otimes \iiint_{\mathcal{R}_{\mathsf{t}}}{\bm{\mathsf{J}}}\left({\mathbf{p}}_{\mathsf{t}}\right)d\mathbf{p}_{\mathsf{t}}\right)\otimes {\hat{\mathbf{r}}}.
        \label{eq:E1}
    \end{equation}
Rewrite $\iiint_{\mathcal{R}_{\mathsf{t}}}{\bm{\mathsf{J}}}\left({\mathbf{p}}_{\mathsf{t}}\right)d\mathbf{p}_{\mathsf{t}}$ as ${\bm{\mathsf{J}}}_{\mathcal{R}_{\mathsf{t}}}={\mathsf{J}}_{{r}}{\hat{\mathbf{r}}}+{\mathsf{J}}_{\theta}\hat{\bm{\theta}}+{\mathsf{J}}_{\phi}\hat{\bm{\phi}}$, where ${\mathsf{J}}_{r}$, ${\mathsf{J}_\theta}$, and ${\mathsf{J}_\phi}$ are three components of the \textit{source current integral vector} ${\bm{\mathsf{J}}}_{\mathcal{R}_{\mathsf{t}}}$ along ${\hat{\mathbf{r}}}$, $\hat{\bm{\theta}}$, and $\hat{\bm{\phi}}$ directions. Since ${\hat{\mathbf{r}}}\otimes {\hat{\mathbf{r}}}=\mathbf{0}$, $({\hat{\mathbf{r}}}\otimes \hat{\bm{\theta}})\otimes {\hat{\mathbf{r}}}=\hat{\bm{\theta}}$, and $({\hat{\mathbf{r}}}\otimes \hat{\bm{\phi}})\otimes {\hat{\mathbf{r}}}=\hat{\bm{\phi}}$, we have
\begin{equation}
{\mathbf{e}}\left({\mathbf{r}}\right)={G_{s}}\left(r\right)\left({\mathsf{J}}_{\theta}\hat{\bm{\theta}}+{\mathsf{J}}_{\phi}\hat{\bm{\phi}}\right)\triangleq {G_{s}}\left(r\right){\bm{\mathsf{J}}}_{\mathcal{R}_{\mathsf{t}}}^{\perp}.\label{eq:12}
\end{equation}
Eq. \eqref{eq:12} is the comprehensive EPM, which is the product of ${G_{s}}(r)$ and the transverse component ${\bm{\mathsf{J}}}_{\mathcal{R}_{\mathsf{t}}}^{\perp}$. The reason why ${\mathsf{J}}_{{r}}$ cannot generate the electric field is that the matrix $\mathbf{I}-{\hat{\mathbf{r}}{\hat{\mathbf{r}}^{\dagger}}}$ in \eqref{eq:Gr} restricts the oscillation direction of
the radiated field to be perpendicular to the propagation direction vector $\hat{\mathbf{r}}$. Besides, some interesting insights are provided in \textit{Remarks} \ref{reEPM1} and \ref{reEPM2}.
\begin{remark}[EPM contains SWM] \label{reEPM1}{${G_{s}}({r})$  represents a scalar spherical wave, which 
accounts for the distance $r$ between $\mathbf{p}_{\tq}$
and $\mathbf{p}_{\rq}$. Note that the vast majority of previous near-field signal models are based on the SWM and its received scalar field can be written as $e_{\mathsf{swm}}\left({\mathbf{r}}\right)=\varepsilon {G_{s}}(r)$ \cite{guerra2021near,Guidi,delmas2016crb}, where $\varepsilon$ is a channel power scaling parameter. Thus, EPM includes SWM as a special case when treating ${\bm{\mathsf{J}}}_{\mathcal{R}_{\mathsf{t}}}^{\perp}$ as a scalar constant $\varepsilon$.}
\end{remark}
\begin{remark}[EPM provides attitude information] \label{reEPM2}{${\bm{\mathsf{J}}}_{\mathcal{R}_{\mathsf{t}}}^{\perp}$ essentially characterizes the vector nature of ${\mathbf{e}}\left({\mathbf{r}}\right)$ and its {functional dependence} on the {current distribution} inside $\mathcal{R}_{\mathsf{t}}$. Furthermore, the current distribution is functionally related to the {position} and {attitude} parameters of the UE. Thus, EPM provides not only {position} information but also {attitude} information, which is the essence of 
jointly estimating the {position} and {attitude}.}
\end{remark}

As described in Section \ref{sec:sysdescription}, we consider the UE to be a dipole with the {position} parameter $z_{\mathsf{t}}$ and {attitude} parameter $t_{z}$. Hence,  ${\bm{\mathsf{J}}}\left({\mathbf{p}}_{\mathsf{t}}\right)$ in \eqref{eq:E1} can be written as
\begin{equation}
{\bm{\mathsf{J}}}\left({\mathbf{p}}_{\mathsf{t}}\right)=I_{\mathsf{dip}}l_{\mathsf{dip}}\delta\left(\mathbf{p}_{\mathsf{t}}\right)\hat{\mathbf{t}}, \label{eq:dipole}
\end{equation}
where $I_{\mathsf{dip}}$ is the uniform current level in the dipole, and $\delta(\cdot)$ is the Dirac's
delta function. Then, $\bm{{\mathsf{J}}}_{\mathcal{R}_{\mathsf{t}}}$ can be written as
\begin{equation}
\bm{\mathsf{J}}_{\mathcal{R}_{\mathsf{t}}}=I_{\mathsf{dip}}l_{\mathsf{dip}}\hat{\mathbf{t}}\iiint_{\mathcal{R}_{\mathsf{t}}}\delta\left(\mathbf{p}_{\mathsf{t}}\right)d\mathbf{p}_{\mathsf{t}}=I_{\mathsf{dip}}l_{\mathsf{dip}}\hat{\mathbf{t}}.\label{eq:dipJR}
\end{equation}
Substituting \eqref{eq:dipJR} into \eqref{eq:E1}, it follows that
\begin{equation}
\frac{{\mathbf{e}}\left({\mathbf{r}}\right)}{{G_{s}}\left(r\right) I_{\mathsf{dip}}l_{\mathsf{dip}}}=\left({\hat{\mathbf{r}}} \otimes \hat{\mathbf{t}}\right)\otimes {\hat{\mathbf{r}}}=\underbrace{\hat{\mathbf{t}}-\left(\hat{\mathbf{r}}\cdot\hat{\mathbf{t}}\right)\hat{\mathbf{r}}}_{\text{radiation vector}},\label{eq:dipeR}
\end{equation}
where the radiation vector is determined by the unit vectors $\hat{\mathbf{t}}$ and $\hat{\mathbf{r}}$. It can be seen that the direction of the vector electric
field ${\mathbf{e}}\left({\mathbf{r}}\right)$ is determined by the radiation vector $\hat{\mathbf{t}}-\left(\hat{\mathbf{r}}\cdot\hat{\mathbf{t}}\right)\hat{\mathbf{r}}$ and is parallel to the plane perpendicular to the propagation direction.
Consider the basis vector transformation relationship between the spherical coordinate and the cartesian coordinate: $\hat{\mathbf{r}} =\sin{\theta}\cos{\phi}\hat{\mathbf{x}}+\cos{\theta}\hat{\mathbf{y}}-\sin{\theta}\sin{\phi}\hat{\mathbf{z}}$.
Substituting the above transformation into \eqref{eq:dipeR} and considering three components of the electric field vector along the $\hat{\mathbf{x}}$, $\hat{\mathbf{y}}$, and $\hat{\mathbf{z}}$ directions in the $\mathsf{OXYZ}$ system, we have
\begin{align}
&\frac{{e}_{x}\left({\mathbf{r}}\right)}{\mathcal{E}_{\mathsf{in}}}=\imagunit \frac{\sin{\theta}\cos{\phi}\left(-
\cos{\theta}t_{y}+\sin{\theta}\sin{\phi}t_{z}\right)}{r}\mathrm{e}^{\imagunit k r}
,\label{eq:exftheta}\\
&\frac{{e}_{y}\left({\mathbf{r}}\right)}{{\mathcal{E}}_{\mathsf{in}}}=\imagunit\frac{\sin{\theta}\left(\sin{\theta}t_{y}+\cos{\theta}\sin{\phi}t_{z}\right)}{r}\mathrm{e}^{\imagunit k r},\\
&\frac{{e}_{z}\left({\mathbf{r}}\right)}{\mathcal{E}_{\mathsf{in}}}=\imagunit \frac{\sin{\theta}\cos{\theta}\sin{\phi}t_{y}+\left(1-\sin^{2}{\theta}\sin^{2}{\phi}\right)t_{z}}{r\mathrm{e}^{-\imagunit k r}},\label{eq:ezftheta}
    \end{align}
where ${e}_{x}\left({\mathbf{r}}\right)\triangleq {\mathbf{e}}\left({\mathbf{r}}\right)\cdot \hat{\mathbf{x}}$, ${e}_{y}\left({\mathbf{r}}\right)\triangleq {\mathbf{e}}\left({\mathbf{r}}\right)\cdot \hat{\mathbf{y}}$, ${e}_{z}\left({\mathbf{r}}\right)\triangleq {\mathbf{e}}\left({\mathbf{r}}\right)\cdot \hat{\mathbf{z}}$, and $\mathcal{E}_{\mathsf{in}}=\frac{\eta I_{\mathsf{dip}}l_{\mathsf{dip}}}{2 \lambda}$ is the initial electric intensity measured in volts (${\mathsf{{V}_{olt}}}$). Note that the functional dependence of $e_{x}\left(\mathbf{r} \right)$, $e_{y}\left(\mathbf{r} \right)$, and $e_{z}\left(\mathbf{r} \right)$ on $z_{\tq}$ is hidden in $\left(r,\theta,\phi \right)$. Indeed, we have $\cos \theta=\frac{y_{\mathsf{r}}}{r}$, $\tan \phi=\frac{z_{\mathsf{t}}}{{x_{\mathsf{r}}}}$, and $r=\sqrt{x_{\mathsf{r}}^{2}+y_{\mathsf{r}}^{2}+z_{\mathsf{t}}^{2}}$. Plugging above conversions into \eqref{eq:exftheta}--\eqref{eq:ezftheta} yields the explicit expressions of $e_{x}\left(\mathbf{r} \right)$, $e_{y}\left(\mathbf{r} \right)$, and $e_{z}\left(\mathbf{r} \right)$ about $z_{\mathsf{t}}$ and  $t_{z}$:
\begin{align}
&{e}_{x}\left({\mathbf{r}}\right)=\imagunit \mathcal{E}_{\mathsf{in}}\frac{-x_{\rq}y_{\rq}\sqrt{1-t_{z}^2}+x_{\rq}z_{\tq}t_{z}}{r^{3}}\mathrm{e}^{\imagunit kr},\label{eq:ex2D}
\\
&{e}_{y}\left({\mathbf{r}}\right)=\imagunit \mathcal{E}_{\mathsf{in}}\frac{\left(x_{\rq}^{2}+z_{\tq}^{2}\right)\sqrt{1-t_{z}^2}+y_{\rq}z_{\tq}t_{z}}{r^{3}}\mathrm{e}^{\imagunit kr},\label{eq:ey2D}\\
&{e}_{z}\left({\mathbf{r}}\right)=\imagunit \mathcal{E}_{\mathsf{in}}\frac{y_{\rq}z_{\tq}\sqrt{1-t_{z}^2}+\left(x_{\rq}^{2}+y_{\rq}^{2}\right)t_{z}}{r^{3}}\mathrm{e}^{\imagunit kr}.\label{eq:ez2D}
\end{align}
Further, we consider the scalar electric field defined from the power point
of view. Specifically, we exploit the scalar electric field that
is a component of the \textit{Poynting vector} perpendicular to the \textit{observation region} $\mathcal{R}_{\mathsf{r}}$ as follows\footnote{In \cite{angchen}, we compared the performance of using the vector field like the form of \eqref{eq:ex2D}--\eqref{eq:ez2D} and the scalar field like the form of \eqref{eq:esrrt} for near-field positioning, and the results show that the performance gap of these two electric field types can be negligible. Using the scalar field can avoid the complex decoupling of the three components of the vector field at the receiver.}
\begin{align}
\notag e_{\mathsf{s}}\left(\mathbf{r}\right)&\triangleq \|\mathbf{e}\left(\mathbf{r} \right)\|\underbrace{\sqrt{- {\hat{\mathbf{r}}}\cdot{\hat{\mathbf{z}}}}}_{\text{projection factor}} \mathrm{e}^{\imagunit kr}\\
&=\mathcal{E}_{\mathsf{in}}\underbrace{\frac{\mathrm{e}^{\imagunit kr}}{{r}^{5/2}}\sqrt{z_{\tq}x_{\rq}^{2}+z_{\tq}\left(y_{\rq}t_{z}+z_{\tq}\sqrt{1-t_{z}^2}\right)^2}}_{\triangleq{h}\left(\bm{\xi};x_{\rq},y_{\rq}\right)}\label{eq:esrrt},
\end{align}
where ${{h}}\left(\bm{\xi};x_{\rq},y_{\rq}\right)$ is the \textit{near-field electromagnetic (NF-EM) channel} between the dipole (UE) with the parameter vector $\bm{\xi}$ and the observation point $\mathbf{p}_{\mathsf{r}}$. The projection factor arises from the fact that the \textit{effective signal} is the component of the signal perpendicular to the receiving surface\footnote{In other words, the \textit{effective area} of the receiving surface is the projected area of its physical region perpendicular to the propagation direction.}. In addition, on the $\mathsf{Y}$-axis, i.e., $x_{\rq}=0$, the NF-EM channel is simplified as
    \begin{equation}
{{h}}_{\mathsf{y}}\left(\bm{\xi};y_{\rq}\right)=\frac{\mathrm{e}^{\imagunit k r_{\mathsf{ry}}}}{r_{\mathsf{ry}}^{5/2}}\sqrt{z_{\tq}}\left(y_{\rq}t_{z}+z_{\tq}\sqrt{1-t_{z}^2}\right),\label{eq:hy}
      \end{equation}
where $r_{\mathsf{ry}}\triangleq r|_{x_{\rq}= 0}=\sqrt{y_{\rq}^{2}+z_{\mathsf{t}}^{2}}$. 

\begin{figure}
\centering
\includegraphics[scale=0.44]{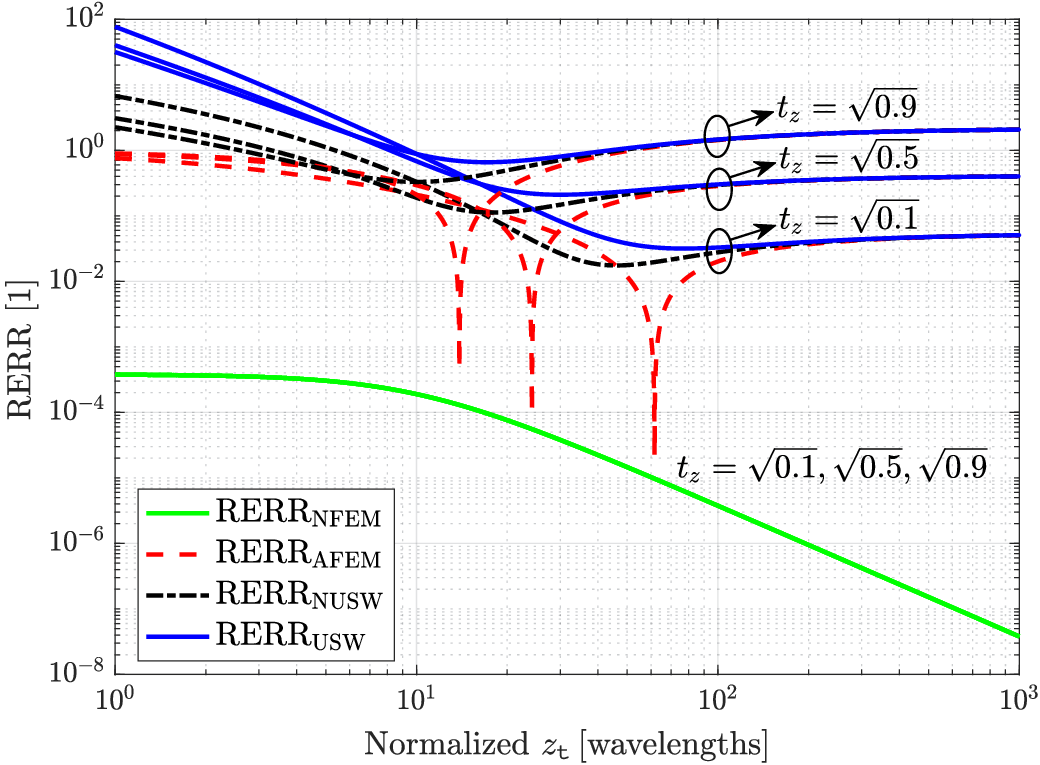}
\caption{{RERRs versus the normalized $z_{\tq}$. Without loss of generality, we set that $\lambda=0.01~\textrm{m}$, $x_{\rq}=0~\textrm{m}$, $y_{\rq}=10\lambda$, and $t_z=\sqrt{0.1}$, $\sqrt{0.5}$, or $\sqrt{0.9}$.}}
\label{fig:RERR}
\end{figure}

{\begin{discussion}[General case]\label{definition21} 
   It is worth mentioning that our system can be easily extended to the more general case, where the UE is located at an arbitrary point in front of $\mathcal{R}_{\mathsf{r}}$ and it has a general orientation in the 3D space. Correspondingly, the general position and attitude parameters can be written as 
   \begin{equation}
    \bm{\xi}_{\mathsf{ar}}\triangleq \left(\mathbf{p}_{\mathsf{ar}}^{\mathsf{T}},{\hat{\mathbf{t}}^{\mathsf{T}}}\right)^{\mathsf{T}}=\left({x_{\tq},y_{\tq},z_{\tq}},{t_x,t_y,t_z}\right)^{\mathsf{T}},
   \end{equation}
   where $\mathbf{p}_{\mathsf{ar}}$ is the general position coordinate of the UE in the $\mathsf{OXYZ}$ system. Then, the NF-EM channel between the UE and $\mathbf{p}_{\mathsf{r}}$ can be derived as
   \begin{align}
\notag h_{\mathsf{ar}}\left(\bm{\xi}_{\mathsf{ar}};\mathbf{p}_{\mathsf{r}}\right)=&\frac{\mathrm{e}^{\imagunit k \|\mathbf{p}_{\mathsf{r}}-\mathbf{p}_{\mathsf{ar}}\|}}{{\|\mathbf{p}_{\mathsf{r}}-\mathbf{p}_{\mathsf{ar}}\|^{5/2}}}\sqrt{z_{\tq}}\left[\left(t_y x_{\rq,\tq}-t_x y_{\rq,\tq}\right)^2+\right.\\
&\left.\left(t_z x_{\rq,\tq}+t_x z_{\tq}\right)^2+\left(t_z y_{\rq,\tq}+t_y z_{\tq}\right)^2\right]^{{1}/{2}},\label{eq:dis1}
\end{align}   
where $x_{\mathsf{r},\mathsf{t}}\triangleq x_{\mathsf{r}}-x_{\mathsf{t}}$ and $y_{\mathsf{r},\mathsf{t}}\triangleq y_{\mathsf{r}}-y_{\mathsf{t}}$. For readability, the proof of \eqref{eq:dis1} is deferred to Appendix~\ref{app:eq:dis1}. Eq.~\eqref{eq:dis1} establishes a functional relationship between the channel model and general position and attitude parameters, {making it possible to estimate these parameters}. Compared with \eqref{eq:esrrt} and \eqref{eq:hy}, the number of the unknown parameters in \eqref{eq:dis1} increases from {two} to {six}. This does not affect the {derivation methods} of closed-form solutions in Section \ref{sec:method} and performance bounds in Section \ref{sec:method1} but will introduce a few tedious yet straightforward algebraic computations into the implementation process. 
\end{discussion}}\begin{discussion}[Explicit electromagnetic simulation platform channel]\label{disminor1} {Common electromagnetic simulation platforms, such as MATLAB Antenna Toolbox and High Frequency Simulator Structure (HFSS), utilize specific numerical calculation methods to solve Maxwell's equations to obtain the electromagnetic characteristics of the antenna. Based on the complete \textit{tensor Green function} given in \eqref{eq:green}, we can derive the electromagnetic channel applicable to the \textit{full-field region} including reactive near-field, radiative near-field, and far-field region, which can be considered as the channel adopted by the electromagnetic simulation platform\footnote{{\cite{minor111} and \cite{minor112} have verified that the relative error between the channel based on the complete Green function and the channel used by the electromagnetic
simulation platform is less than $10^{-3}$.}} \cite{minor111,minor112,minor113}. In particular, the explicit {electromagnetic simulation platform (EM-SIMP) channel} between the UE and $\mathbf{p}_{\mathsf{r}}$ can be expressed as $h_{\mathrm{SIM}}\left(\bm{\xi};x_{\rq},y_{\rq}\right)=\mathcal{F}_{\mathsf{SF}} {h}\left(\bm{\xi};x_{\rq},y_{\rq}\right)$, where $\mathcal{F}_{\mathsf{SF}}\triangleq\sqrt{1+\frac{3}{k^2r^2}+\frac{9}{k^4r^4}}$ is the dimensionless scaling factor. The calculation process of $h_{\mathrm{SIM}}\left(\bm{\xi};x_{\rq},y_{\rq}\right)$ is given in Appendix~\ref{app:eq:disminor1}.}
\end{discussion}
\begin{discussion}[Simplifications of ${{h}}\left(\bm{\xi};x_{\rq},y_{\rq}\right)$ in \eqref{eq:esrrt}]\label{disminor2} {To give an intuitive comparison, we also briefly discuss a series of simplifications of the NF-EM channel ${{h}}\left(\bm{\xi};x_{\rq},y_{\rq}\right)$ as follows.
\begin{itemize}[leftmargin=*]
    \item Attitude-fixed electromagnetic (AF-EM) channel: The works \cite{de2021cramer,angchen,9184098,bjornson2021primer,10366272}  assume that the attitude is fixed along the positive $\mathsf{Y}$-direction, i.e., $t_{y}=1$. Correspondingly, the NF-EM channel ${{h}}\left(\bm{\xi};x_{\rq},y_{\rq}\right)$ degenerates into the AF-EM channel given by ${{h}}_{\mathrm{AFEM}}\left(\mathbf{p}_{\tq};x_{\rq},y_{\rq}\right)=\frac{\sqrt{z_{\tq}}\sqrt{x_{\rq}^{2}+z_{\tq}^2}}{{r}^{5/2}}\mathrm{e}^{\imagunit kr}$.
    \item Non-uniform SWM (NU-SW) channel: Ignoring the attitude mismatch and projection loss, i.e., $\frac{\sqrt{x_{\rq}^2+z_{\tq}^2}}{r}=1$ and ${\sqrt{\frac{z_{\tq}}{r}}}=1$, the AF-EM channel degenerates into the NU-SW channel \cite{9335528,guerra2021near,delmas2016crb} given by $h_{\mathrm{NUSW}}\left(\mathbf{p}_{\tq};x_{\rq},y_{\rq}\right)=\frac{1}{r}\mathrm{e}^{\imagunit k r}$.
\item Uniform SWM (U-SW) channel \cite{10002944}: If $z_{\tq} \sim \mathcal{O}(D_{\rq}^{\mathcal{P}})$ with $\mathcal{P}>1$, $r$ in the amplitude term can be replaced by $z_{\tq}$. The propagation distances corresponding to different receiving elements are identical, i.e., “uniform” power across receiving elements. Thus, the NU-SW channel can be simplified into the U-SW channel given by $h_{\mathrm{USW}}\left(\mathbf{p}_{\tq};x_{\rq},y_{\rq}\right)=\frac{1}{z_{\tq}}\mathrm{e}^{\imagunit k r}$.
\end{itemize}}
\end{discussion}
\begin{discussion}[Accuracy of ${{h}}\left(\bm{\xi};x_{\rq},y_{\rq}\right)$]\label{disminor4} {We calculate the relative error (RERR) between the EM-SIMP channel and the NF-EM channel as $\mathrm{RERR}_{\mathrm{NFEM}}=\frac{\left|h_{\mathrm{SIM}}\left(\bm{\xi};x_{\rq},y_{\rq}\right)-{{h}}\left(\bm{\xi};x_{\rq},y_{\rq}\right)\right|}{\left|h_{\mathrm{SIM}}\left(\bm{\xi};x_{\rq},y_{\rq}\right)\right|}=1-\frac{1}{\mathcal{F}_{\mathsf{SF}}}$, which is a function of the wavelength $\lambda$ and the distance $r$ between $\mathbf{p}_{\tq}$ and $\mathbf{p}_{\rq}$. Also, we use the EM-SIMP channel as the baseline and calculate the RERRs between the EM-SIMP channel and three near-field channels in \textit{Discussion} \ref{disminor2} as $\mathrm{RERR}_{\mathsf{type}}=\frac{\left|h_{\mathrm{SIM}}\left(\bm{\xi};x_{\rq},y_{\rq}\right)-{{h}}_{\mathsf{type}}\left(\mathbf{p}_{\tq};x_{\rq},y_{\rq}\right)\right|}{\left|h_{\mathrm{SIM}}\left(\bm{\xi};x_{\rq},y_{\rq}\right)\right|}$, where the subscript $_\mathsf{type}$ stands for $_\mathrm{AFEM}$, $_\mathrm{NUSW}$, or $_\mathrm{USW}$. Fig. \ref{fig:RERR} illustrates the RERRs versus the normalized $z_{\tq}$ ($z_{\tq}$ is normalized by the wavelength $\lambda$) when $\lambda=0.01~\textrm{m}$, $x_{\rq}=0~\textrm{m}$, $y_{\rq}=10\lambda$, and $t_z=\sqrt{0.1}$, $\sqrt{0.5}$, or $\sqrt{0.9}$. It can be observed that when $\mathbf{p}_{\rq}$ is very close to the UE, i.e., $\lambda \leq z_{\tq} \leq 10\lambda$, $\mathrm{RERR}_{\mathrm{NFEM}}$ can reach about $10^{-4}$, and it further decreases as the transceiver distance increases, which indicates that the NF-EM channel is accurate enough. Besides, the other three RERRs are relatively large at all distances and cannot decrease infinitely like $\mathrm{RERR}_{\mathrm{NFEM}}$ as the transceiver distance increases. This is because the AF-EM channel cannot accurately characterize the attitude of the UE, and the NU-SW and U-SW channels directly ignore the attitude. In fact, even in the far-field region, the attitude information of the UE cannot be overlooked.
}
\end{discussion}

The NF-EM channel in \eqref{eq:esrrt} and \eqref{eq:hy} provides a complex-valued non-linear function of $\bm{\xi}$ described in \textit{Definition} \ref{definition2} and \textit{Assumption} \assumptionref{assum:p}, whose phase part contains $z_{\tq}$ and amplitude part contains both $z_{\tq}$ and $t_z$. This allows us to jointly estimate the {position} and {attitude} parameters of the UE.

\section{Closed-Form Solutions of Joint Estimation}\label{sec:method}
The problem of joint {position} and {attitude} estimation can essentially be transformed into obtaining $z_{\tq}$ and $t_z$ based on the known quantities. For the noise-free case, we use the observed voltages at any two\footnote{The use of two elements is only suitable for the noise-free case since two elements are sufficient to solve the set of equations. For the noisy case, as discussed in Section \ref{sec:method1}, we use all elements to average the noise.} receiving elements to construct a set of non-linear equations, and then solve this set of equations to obtain the closed-form solutions of $z_{\tq}$ and $t_z$. 

In particular, the noise-free voltage (measured in ${\mathsf{{V}_{olt}}}$) at the $n_{\mathsf{y}}$-th element can be obtained by integrating over $\mathcal{R}_{\mathsf{r};n_{\mathsf{y}}}$ the scalar electric field given by \eqref{eq:esrrt}:
\begin{equation}
{\mathrm{v}}_{n_{\mathsf{y}}}=\frac{1}{l_{\mathsf{s}}}\iint_{\mathcal{R}_{\mathsf{r};n_{\mathsf{y}}}}e_{\mathsf{s}}\left(\mathbf{r}\right)d{x_{\mathsf{r}}}d{y_{\mathsf{r}}}\overset{(\mathsf{a})}{\approx}\mathcal{E}_{\mathsf{in}}l_{\mathsf{s}}{{h}}_{\mathsf{y}}\left(\bm{\xi};y_{\mathsf{r};{n_{\mathsf{y}}}}\right),\label{eq:esmxny}
\end{equation}
where $(\mathsf{a})$ follows due to the fact that the size of each element is on the wavelength scale, thus the variation of the scalar electric field across different points on the same element can be ignored. ${{h}}_{\mathsf{y}}\left(\bm{\xi};y_{\mathsf{r};{n_{\mathsf{y}}}}\right)$ is the discrete sample point of ${{h}}_{\mathsf{y}}\left(\bm{\xi};y_{\mathsf{r}}\right)$ in \eqref{eq:hy} at $y_{\rq}=y_{\mathsf{r};{n_{\mathsf{y}}}}$, namely, ${{h}}_{\mathsf{y}}\left(\bm{\xi};y_{\mathsf{r};{n_{\mathsf{y}}}}\right)\triangleq{{h}}_{\mathsf{y}}\left(\bm{\xi};y_{\mathsf{r}}\right)|_{y_{\rq}=y_{\mathsf{r};{n_{\mathsf{y}}}}}$. 

From \eqref{eq:esmxny}, the joint estimation in the noise-free case can be transformed into solving the following non-linear equations:
\begin{subequations}
 \begin{empheq}[left=\empheqlbrace]{align} 
 &{{\mathrm{v}}_{\alpha}}={\mathcal{E}_{\mathsf{in}}l_{\mathsf{s}}}\frac{\sqrt{z_{\tq}}\left(y_{\mathsf{r};{\alpha}}t_{z}+z_{\tq}\sqrt{1-t_{z}^2}\right)}{r_{\alpha}^{5/2}}{\mathrm{e}^{\imagunit k r_{\alpha}}},\label{eq:valp1} \\
&{{\mathrm{v}}_{\beta}}={\mathcal{E}_{\mathsf{in}}l_{\mathsf{s}}}\frac{\sqrt{z_{\tq}}\left(y_{\mathsf{r};{\beta}}t_{z}+z_{\tq}\sqrt{1-t_{z}^2}\right)}{r_{\beta}^{5/2}}{\mathrm{e}^{\imagunit k r_{\beta}}},\label{eq:vbeta1}
 \end{empheq}
 \end{subequations}
where ${\mathrm{v}}_{\alpha}$ and ${\mathrm{v}}_{\beta}$ are \textit{known} voltages obtained by two elements centered at $y_{\mathsf{r};{\alpha}}$ and $y_{\mathsf{r};{\beta}}$ ($1\leq\alpha<\beta \leq\mathsf{N}$), $r_{\alpha}\triangleq r_{\mathsf{ry}}|_{y_{\rq}=y_{\rq;\alpha}}$, $r_{\beta}\triangleq r_{\mathsf{ry}}|_{y_{\rq}=y_{\rq;\beta}}$, while the \textit{unknown} variables are $z_{\tq}$ and $t_z$. 

It is difficult to directly solve the above non-linear equations due to the complex exponential coupling. In order to tackle this challenge, we propose to reconstruct the above equations by using proper decoupling methods. Specifically, we decouple the \textit{known} complex values ${\mathrm{v}}_{\alpha}$ and ${\mathrm{v}}_{\beta}$ into \textit{known} amplitudes and phase angles, because the combination of the real part and imaginary part of a complex number has a {one-to-one} mapping relationship with the combination of its amplitude and phase. However, this decoupling method will cause a \textit{phase ambiguity problem}, which will be discussed in detail in Section \ref{sec:PA11}.

\subsection{Phase Ambiguity Problem} \label{sec:PA11}

The \textit{phase ambiguity problem} is that the obtained \textit{known} phase angle can be different from the actual one. Specifically, the \textit{known} phase angle range is $[0,2\pi)$ because of the periodicity of the phase while the actual phase angle range is $[0,\infty)$. The actual phase can establish a mapping relationship with the position of UE, i.e., $\frac{2\pi}{\lambda}\sqrt{y_{\rq}^{2}+z_{\mathsf{t}}^{2}}=\Theta_{\mathsf{actual}}\in [0,\infty)$, where $\Theta_{\mathsf{actual}}$ is the actual phase.
The existence of \textit{phase ambiguity} makes this relationship no longer hold, i.e., $\frac{2\pi}{\lambda}\sqrt{y_{\rq}^{2}+z_{\mathsf{t}}^{2}}\neq \Theta_{\mathsf{known}} \in [0,2\pi)$, where $\Theta_{\mathsf{known}}$ is the obtained \textit{known} phase. Next, we discuss different cases of \textit{phase ambiguity}.
\subsubsection{Case I \texorpdfstring{($r_{\alpha}<\lambda$ and $r_{\beta}<\lambda$)}{zt}} In \textit{Case I}, the distance from the UE to any receiving element is within a wavelength, so the \textit{known} phase angle is the same as the actual and there is no \textit{phase ambiguity}. Denote ${\mathrm{v}}_{\alpha}\triangleq \Psi_{\alpha}{\mathrm{e}^{\imagunit \Theta_{\alpha}}}$ and ${\mathrm{v}}_{\beta}\triangleq \Psi_{\beta}{\mathrm{e}^{\imagunit \Theta_{\beta}}}$, where $\Psi_{\alpha}$ and $\Psi_{\beta}$ are \textit{known} amplitudes, $\Theta_{\alpha}$ and $\Theta_{\beta}$ are \textit{known} phase angles. Then, \eqref{eq:valp1} and  \eqref{eq:vbeta1} are decoupled as
\begin{subequations}
 \begin{empheq}[left=\empheqlbrace]{align}
&\Theta_{\kappa}=k\sqrt{y_{\mathsf{r};{\kappa}}^2+z_{\tq}^2}, \label{eq:Thetaa}\\
&\Psi_{\kappa}=\mathcal{E}_{\mathsf{in}}l_{\mathsf{s}}\frac{\sqrt{z_{\tq}}\left(y_{\mathsf{r};{\kappa}}t_{z}+z_{\tq}\sqrt{1-t_{z}^2}\right)}{r_{\kappa}^{5/2}}\label{eq:Psib},
 \end{empheq}
 \end{subequations}
where $\kappa\triangleq\alpha~\text{or}~\beta$. \textit{Case I} is usually difficult to occur because the wavelength of wireless systems is often small.
\subsubsection{Case \texorpdfstring{II ($r_{\alpha}\geq\lambda$ and $r_{\beta}\geq\lambda$)}{zt}}  In \textit{Case II}, the distance from the UE to any receiving element is no smaller than the wavelength, thus the \textit{known} phase angle goes through \textit{unknown} integer period and is different from the actual. Further, \eqref{eq:Thetaa} can be written as
\begin{subequations}
 \begin{empheq}[left=\empheqlbrace]{align}
&\Theta_{\alpha}+2\pi\mathsf{N}_{\alpha}=kr_{\alpha},\label{eq:Na} \\
&\Theta_{\beta}+2\pi\mathsf{N}_{\beta}=kr_{\beta},\label{eq:Nb}
 \end{empheq}
 \end{subequations}
where $\Theta_{\alpha}$ and $\Theta_{\beta}$ are \textit{known} phase angles, $\mathsf{N}_{\alpha}$ and $\mathsf{N}_{\beta}$ are the corresponding \textit{unknown} integer periods. Notice that these \textit{unknown} integer periods are tricky because they are introduced as a series of new variables. The  \textit{unknown} integer period varies depending on the position of UE and the receiving elements involved. To deal with \textit{unknown} integer periods, we propose an efficient method that divides the distance domain of UE to explore the relationship of \textit{unknown} integer periods at different receiving elements. The results are provided in \textit{Proposition} \propositionref{prop:PA}.
\begin{proposition}[Phase ambiguity distance $d_{\mathsf{PA}}$ and Spacing constraint distance $d_{\mathsf{SC}}$]\propositionlabel{prop:PA} When $z_{\tq}$ is not less than the \textit{Phase ambiguity distance} $d_{\mathsf{PA}}$, the difference between the \textit{unknown} integer periods at {any} two receiving elements is $0$ or $1$, i.e., 
\begin{subequations}
 \begin{empheq}[left=\empheqlbrace]{align}
&\mathsf{N}_{\beta}-\mathsf{N}_{\alpha}=0,~~\text{if}~\Theta_{\beta}>\Theta_{\alpha},\label{eq:NNN}\\
&\mathsf{N}_{\beta}-\mathsf{N}_{\alpha}=1,~~\text{if}~\Theta_{\beta}\leq\Theta_{\alpha}\label{eq:NNNN}.
 \end{empheq}
 \end{subequations}
where we define $d_{\mathsf{PA}}$ as one quarter of the \textit{Fraunhofer distance} $d_{\mathsf{F}}$, i.e.,
\begin{equation}
    d_{\mathsf{PA}}\triangleq \frac{1}{4}d_{\mathsf{F}}=\frac{D_{\rq}^{2}}{2\lambda}\label{eq:dPA}
\end{equation}
with $D_{\rq}\geq 4.8\lambda$. Thus, we can eliminate $\mathsf{N}_{\alpha}$ and $\mathsf{N}_{\beta}$ in \eqref{eq:Na} and \eqref{eq:Nb} by the \textit{difference method}. When $z_{\tq}< d_{\mathsf{PA}}$, we cannot arbitrarily select two receiving elements to eliminate \textit{phase ambiguity}. The distance between two receiving elements needs to be reduced. In particular, we utilize two elements centered at $y_{\rq;1}=\frac{l_{\mathsf{s}}}{2}$ and $y_{\rq;2}=\frac{3l_{\mathsf{s}}}{2}$. Then, the difference between the \textit{unknown} integer periods at these two receiving elements is $0$ or $1$ when $z_{\tq}\geq d_{\mathsf{SC}}$, i.e.,
\begin{subequations}
 \begin{empheq}[left=\empheqlbrace]{align}
&\mathsf{N}_{2}-\mathsf{N}_{1}=0,~~\text{if}~\Theta_{2}>\Theta_{1},\label{eq:AA}\\
&\mathsf{N}_{2}-\mathsf{N}_{1}=1,~~\text{if}~\Theta_{2}\leq\Theta_{1}\label{eq:AAA}.
 \end{empheq}
 \end{subequations}
where $d_{\mathsf{SC}}$ is defined as the \textit{Spacing constraint distance}:
\begin{equation}
    d_{\mathsf{SC}}\triangleq\max\left\{\frac{l_{\mathsf{s}}^{2}}{\lambda},3.6l_{\mathsf{s}}\right\}.\label{eq:dSC}
\end{equation}
\end{proposition}
\begin{IEEEproof}\begin{figure}[!t]
\centering
\includegraphics[scale=0.59]{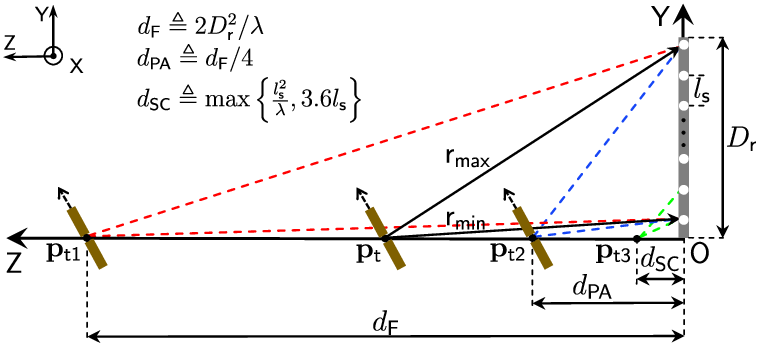}
\caption{Illustration of the UE distance domain division, where $\mathbf{p}_{\mathsf{t1}}$ is located at the near-field boundary defined by the \textit{Fraunhofer distance} $d_{\mathsf{F}}$, $\mathbf{p}_{\mathsf{t2}}$ is located at the boundary defined by the \textit{Phase ambiguity distance} $d_{\mathsf{PA}}$, and $\mathbf{p}_{\mathsf{t3}}$ is located at the boundary defined by the \textit{Spacing constraint distance} $d_{\mathsf{SC}}$.}
\label{system3}
\end{figure}
As shown in Fig. \ref{system3}, the minimum and maximum distances from the UE to the receiving element are
\begin{equation}
\mathsf{r}_{\mathsf{min}}=\sqrt{z_{\tq}^2+l_{\mathsf{s}}^2/4}\overset{(\mathsf{b})}{\approx} z_{\tq},
\end{equation}
and 
\begin{equation}
\mathsf{r}_{\mathsf{max}}\overset{(\mathsf{c})}{\approx}\sqrt{z_{\tq}^2+D_{\rq}^2}\overset{(\mathsf{d})}{\approx}z_{\tq}+\frac{D_{\rq}^2}{2z_{\tq}},
\end{equation}
respectively, where $(\mathsf{b})$ follows by the fact that $l_{\mathsf{s}}$ is small, $(\mathsf{c})$ follows by ignoring the distance between the center of the last element and the upper boundary of the {observation region}, and we utilized the Taylor approximation $\sqrt{1+x}\approx1+\frac{x}{2}$ for $x=\frac{D_{\rq}^2}{z_{\tq}^2}$ in $(\mathsf{d})$. Then, the maximum phase difference between different elements can be written as 
\begin{equation}
    \frac{2\pi}{\lambda}\left(\mathsf{r}_{\mathsf{max}}-\mathsf{r}_{\mathsf{min}}\right)\approx \frac{\pi}{\lambda }\frac{D_{\rq}^2}{z_{\tq}}.\label{eq:3232}
\end{equation}
{Making \eqref{eq:3232} not greater than $2\pi$ leads to $z_{\tq}\geq\frac{D_{\rq}^2}{2\lambda}$.} Besides, the Taylor approximation error in $(\mathsf{d})$ is smaller than $3.5\times 10^{-3}$ if $z_{\tq}\geq 2.4D_{\rq}$, which sets an additional condition $\frac{D_{\rq}^2}{2\lambda}\geq 2.4D_{\rq}$, i.e., $D_{\rq}\geq 4.8\lambda$. Notice that this condition always holds for electrically large arrays, hence we define the \textit{Phase ambiguity distance} $d_{\mathsf{PA}}$ as \eqref{eq:dPA}. Similarly,  the difference between the distances from UE to $\left(0,y_{\rq;1},0\right)$ and $\left(0,y_{\rq;2},0\right)$ is written as
    \begin{equation}
    \sqrt{z_{\tq}^2+y_{\rq;2}^2}-\sqrt{z_{\tq}^2+y_{\rq;1}^2}\overset{(\mathsf{e})}\approx \frac{y_{\rq;2}^2-y_{\rq;1}^2}{2z_{\tq}}=\frac{l_{\mathsf{s}}^2}{z_{\tq}},\label{eq:29}
\end{equation}
where $(\mathsf{e})$ used $\sqrt{1+x}\approx1+\frac{x}{2}$ for $x=\frac{y_{\mathsf{r};{1}}^2}{z_{\tq}^2}$ and $x=\frac{y_{\mathsf{r};{2}}^2}{z_{\tq}^2}$. By making \eqref{eq:29} not greater than $\lambda$, we have $z_{\tq}\geq \frac{l_{\mathsf{s}}^2}{\lambda}$. Moreover, the Taylor approximation error in $(\mathsf{e})$ is smaller than $3.5\times 10^{-3}$ if $z_{\tq}\geq 3.6l_{\mathsf{s}}$. Thus, we define $d_{\mathsf{SC}}$ as \eqref{eq:dSC}. 
\end{IEEEproof}

Note that $z_{\tq}\geq d_{\mathsf{SC}}$ usually holds even for near-field regions since $l_\mathsf{s}$ is on the small wavelength scale. Consequently, based on \textit{Proposition} \propositionref{prop:PA}, we utilize \eqref{eq:NNN} and \eqref{eq:NNNN} to eliminate \textit{phase
ambiguity} if $z_{\tq}\geq d_{\mathsf{PA}}$, and utilize  \eqref{eq:AA} and \eqref{eq:AAA} to eliminate \textit{phase
ambiguity} if $d_{\mathsf{SC}}\leq z_{\tq}<d_{\mathsf{PA}}$.

\subsection{Closed-Form Solutions of \texorpdfstring{$z_{\tq}$ and $t_z$}{zt}} \label{sec:PAclose}
We will derive the closed-form solutions of $z_{\tq}$ and $t_z$ for  \textit{Case I} and \textit{Case II}, respectively.

\subsubsection{Case I \texorpdfstring{($r_{\alpha}<\lambda$ and $r_{\beta}<\lambda$)}{zt}} In \textit{Case I}, only one observation element is needed to obtain $z_{\tq}$. From the spherical wave phase given by \eqref{eq:Thetaa}, we can solve $z_{\tq}$ (denoted as $\mathsf{z}_{\textit{I}}$):
\begin{equation}
\mathsf{z}_{\textit{I}}=\sqrt{\frac{\Theta_{\kappa}^2}{k^2}-y_{\mathsf{r};{\kappa}}^2}.\label{eq:ztsslove}
\end{equation}
{Then, we begin to derive the expression of the parameter $t_z$.} Taking the difference between the equations in \eqref{eq:Psib}, we have 
\begin{equation}
    \left(y_{\mathsf{r};{\alpha}}-y_{\mathsf{r};{\beta}}\right)t_{z}=\frac{\Psi_{\alpha}{r_{\alpha}^{5/2}}-\Psi_{\beta}{r_{\beta}^{5/2}}}{\mathcal{E}_{\mathsf{in}}l_{\mathsf{s}}\sqrt{{z}_{\tq}}}.\label{eq26}
\end{equation}
Substituting \eqref{eq:ztsslove} into \eqref{eq26}, we can solve $t_z$ (denoted as $\mathsf{t}_{\textit{I}}$):
\begin{equation}
\mathsf{t}_{\textit{I}}=\frac{\Psi_{\alpha}{\left(y_{\mathsf{r};{\alpha}}^2+\mathsf{z}_{\textit{I}}^2\right)^{5/4}}-\Psi_{\beta}{\left(y_{\mathsf{r};{\beta}}^2+\mathsf{z}_{\textit{I}}^2\right)^{5/4}}}{\mathcal{E}_{\mathsf{in}}l_{\mathsf{s}}\sqrt{\mathsf{z}_{\textit{I}}}\left(y_{\mathsf{r};{\alpha}}-y_{\mathsf{r};{\beta}}\right)}.\label{eq:tsolve1}
\end{equation}

\subsubsection{Case II \texorpdfstring{($r_{\alpha}\geq\lambda$ and $r_{\beta}\geq\lambda$)}{zt}} In \textit{Case II}, we rewrite \eqref{eq:Na} and \eqref{eq:Nb} as
\begin{subequations}
 \begin{empheq}[left=\empheqlbrace]{align}
\Theta_{\alpha}+2\pi\mathsf{N}_{\alpha}&\overset{(\mathsf{f})}{\approx} k\left(z_{\tq}+\frac{y_{\mathsf{r};{\alpha}}^2}{2z_{\tq}}\right), \label{eq:OO}\\
\Theta_{\beta}+2\pi\mathsf{N}_{\beta}&\overset{(\mathsf{f})}{\approx} k\left(z_{\tq}+\frac{y_{\mathsf{r};{\beta}}^2}{2z_{\tq}}\right),\label{eq:OOO}
\end{empheq}
 \end{subequations}
where we used the approximation $\sqrt{1+x}\approx1+\frac{x}{2}$ for $x=\frac{y_{\mathsf{r};{\kappa}}^2}{z_{\tq}^2}$ in $(\mathsf{f})$. Since $\frac{y_{\mathsf{r};{\kappa}}^2}{z_{\tq}^2}<\frac{D_{\rq}^2}{\left(2.4D_{\rq}\right)^2}\approx 0.174$, this approximation error is smaller than $3.5\times 10^{-3}$. Plugging \eqref{eq:NNN} and \eqref{eq:NNNN} into \eqref{eq:OO} and \eqref{eq:OOO}, we can solve $z_{\tq}$ (denoted as $\mathsf{z}_{\textit{II}1}$):
\begin{subequations}
 \begin{empheq}[left={\mathsf{z}_{\textit{II}1}=}\empheqlbrace]{align}
 &\frac{k\left(y_{\mathsf{r};{\beta}}^2-y_{\mathsf{r};{\alpha}}^2\right)}{2\left(\Theta_{\beta}-\Theta_{\alpha}\right)},~~~~~~\text{if}~\Theta_{\beta}>\Theta_{\alpha}, \label{eq:34a}\\
&\frac{k\left(y_{\mathsf{r};{\beta}}^2-y_{\mathsf{r};{\alpha}}^2\right)}{2\left(\Theta_{\beta}-\Theta_{\alpha}+2\pi\right)},~~\text{if}~\Theta_{\beta}\leq\Theta_{\alpha}\label{eq:34b}.
 \end{empheq}
 \end{subequations}
Then, we solve $t_z$ (denoted as $\mathsf{t}_{\textit{II}1}$) via \eqref{eq:tsolve1}, where $\mathsf{z}_{\textit{I}}=\mathsf{z}_{\textit{II}1}$. The closed-form solution in \eqref{eq:34a} and \eqref{eq:34b} holds for $z_{\tq}\geq d_{\mathsf{PA}}$. If $d_{\mathsf{SC}}\leq z_{\tq}<d_{\mathsf{PA}}$, we leverage elements with index $1$ and $2$. Then, similar to \eqref{eq:OO}--\eqref{eq:34b}, we can solve $z_{\tq}$ (denoted as $\mathsf{z}_{\textit{II}2}$) as
\begin{subequations}
 \begin{empheq}[left={\mathsf{z}_{\textit{II}2}=}\empheqlbrace]{align}
 &\frac{kl_{\mathsf{s}}^2}{\Theta_{2}-\Theta_{1}},~~~~~~~~~\text{if}~\Theta_{2}>\Theta_{1}, \label{eq:aaaa}\\
&\frac{kl_{\mathsf{s}}^2}{\Theta_{2}-\Theta_{1}+2\pi},~~\text{if}~\Theta_{2}\leq\Theta_{1}\label{eq:aa}.
 \end{empheq}
 \end{subequations}
Then, we solve $t_z$ (denoted as $\mathsf{t}_{\textit{II}2}$) via \eqref{eq:tsolve1}, where $\mathsf{z}_{\textit{I}}=\mathsf{z}_{\textit{II}2}$. 

\begin{remark}[Summary of closed-form solutions] \label{re1} Based on the \textit{prior} distributions given by \textit{Assumption} \assumptionref{assum:p}, we have: 
\begin{itemize}[leftmargin=*]
    \item If $\sqrt{\mathsf{H}_2^2+y_{\mathsf{r};{\kappa}}^2}<\lambda$, the closed-form solution of $z_\tq$ is given by \eqref{eq:ztsslove}.
    \item If $\sqrt{\mathsf{H}_2^2+y_{\mathsf{r};{\kappa}}^2}\geq\lambda$ and $\mathsf{H}_1 \geq d_{\mathsf{PA}}$, the closed-form solution of $z_\tq$ is given by \eqref{eq:34a} and \eqref{eq:34b}.
    \item If $\sqrt{\mathsf{H}_2^2+y_{\mathsf{r};{\kappa}}^2}\geq\lambda$ and $d_{\mathsf{SC}}\leq\mathsf{H}_1 < d_{\mathsf{PA}}$, the closed-form solution of $z_\tq$ is given by \eqref{eq:aaaa} and \eqref{eq:aa}.
    \item The closed-form solution of $t_z$ is given by \eqref{eq:tsolve1}, where $\mathsf{z}_{\textit{I}}$ is replaced by the corresponding closed-form solution of $z_\tq$.
    {\item The position and attitude parameters can be directly computed, resulting in a time complexity of $\mathcal{O}\left(1\right)$. Besides, the closed-form solutions can be used as initial estimates of the unknown parameters \cite{10366272}.}
\end{itemize}
\end{remark}

\section{Performance Bounds of Joint Estimation} \label{sec:method1}
In Section \ref{sec:method}, we have derived the closed-form solutions for the noise-free case. In practice, the observed voltages in \eqref{eq:esmxny} are inevitably affected by noise. We will discuss the impact of noise on joint estimation performance in this section.

For the noisy case, the goal of joint {position} and {attitude} estimation is to estimate $z_{\tq}$ and $t_z$ based on the \textit{observations} at the receiving array. The \textit{observation equation} for the $n_{\mathsf{y}}$-th element can be written as
\begin{equation}
\tilde{{\mathrm{v}}}_{n_{\mathsf{y}}}={\mathrm{v}}_{n_{\mathsf{y}}}+n_{n_{\mathsf{y}}},\label{eq:observationV}
\end{equation}
where $\tilde{{\mathrm{v}}}_{n_{\mathsf{y}}}$ is the observation voltage and $n_{n_{\mathsf{y}}}$ is the random noise associated with the $n_{\mathsf{y}}$-th element modeled as follows.
\begin{assumption}[Random noise field modeling]\assumptionlabel{assum:3} The random noise field can be modeled as a spatially uncorrelated circularly symmetric zero-mean
complex-Gaussian process with the {correlation function}: $\mathbb{E}\left\{{n}_{n_{\mathsf{y}}}{n}^{*}_{n_{\mathsf{y}}^{\prime}}\right\}=\sigma^{2}\delta\left(n_{\mathsf{y}}-n_{\mathsf{y}}^{\prime}\right)$, where $n_{\mathsf{y}}^{\prime}$ is an arbitrary index different from $n_{\mathsf{y}}$, and $\sigma^{2}$ is the variance measured in ${\mathsf{V}_{\mathsf{olt}}^{2}}$. 
\end{assumption}

By stacking the \textit{observation equations} from \eqref{eq:observationV}, the \textit{observation equation vector} for the observation array is obtained as
\begin{equation}
\tilde{\bm{{\mathrm{v}}}}=\bm{{\mathrm{v}}}+\bm{n}=\mathcal{E}_{\mathsf{in}}l_{\mathsf{s}}{\bm{h}}_{\mathsf{y}}\left(\bm{\xi}\right)+\bm{n},\label{eq:vector_observation}
\end{equation}
where $\tilde{\bm{{\mathrm{v}}}}=\left[\tilde{{\mathrm{v}}}_{1},\cdots,\tilde{{\mathrm{v}}}_{\mathsf{N}}\right]^{\mathsf{T}}$ and ${\bm{{\mathrm{v}}}}=\left[{{\mathrm{v}}}_{1}\,\cdots,{{\mathrm{v}}}_{\mathsf{N}}\right]^{\mathsf{T}}$ are the observation voltage vector and the noise-free voltage vector, respectively.  ${\bm{h}}_{\mathsf{y}}\left(\bm{\xi}\right)=
\left[{{h}}_{\mathsf{y}}\left(\bm{\xi};y_{\mathsf{r};1}\right),\cdots,{{h}}_{\mathsf{y}}\left(\bm{\xi};y_{\mathsf{r};\mathsf{N}}\right)\right]^{\mathsf{T}}$ is the channel vector, whose element is provided in continuous form in \eqref{eq:hy}. $\bm{n}=\left[n_{1},\cdots,n_{\mathsf{N}}\right]^{\mathsf{T}}$ is the random noise vector.

The mean squared error (MSE) is a well-accepted metric to assess the estimation performance and can be expressed as
\begin{equation}
\mathrm{MSE}\left(\bm{\xi}_{\imath}\right)=\mathbb{E}\left\{\left(\bm{\xi}_{\imath}-\hat{\bm{\xi}}_{\imath}\right)^{2}\right\},\imath=1,2.\label{eq:MSE}
\end{equation}
where $\hat{\bm{\xi}}\triangleq\left(\hat{z}_{\tq},\hat{t}_{z}\right)^{\mathsf{T}}$ is the estimate of $\bm{\xi}$. As a challenging non-linear vector parameter estimation problem, there is no exact closed-form expression of the minimum MSE for the joint {position} and {attitude} estimation, which motivates us to find the closed-form lower bounds to MSE. As discussed in Section \ref{sec:ZZB} and \ref{sec:ECRB}, we will develop the closed-form  ZZB and ECRB to provide the lower bounds to MSE. Besides, the MSE of the maximum a
posteriori (MAP) estimator is evaluated in Section \ref{sec:simulation} for reference.

\subsection{ZZB for Joint Position and Attitude Estimation}\label{sec:ZZB}
The ZZB is a \textit{globally tight} bound applicable in all SNR regimes, which relates the MSE to the probability of error in a binary hypothesis testing problem. The preliminaries of ZZB are given in \textit{Lemma} \ref{lemm:ZZB1}.
\begin{lemma}
[ZZB preliminaries for $\bm{\xi}$]\label{lemm:ZZB1} 
Denoting the estimate of $\bm{\xi}$ by $\hat{\bm{\xi}}$, the MSE matrix is defined as
\begin{equation}
\bm{R}_{\bm{\epsilon}}=\mathbb{E}\left\{\bm{\epsilon}\bm{\epsilon}^{\mathsf{T}}\right\}=\mathbb{E}\left\{\left(\hat{\bm{\xi}}-\bm{\xi}\right)\left(\hat{\bm{\xi}}-\bm{\xi}\right)^{\mathsf{T}}\right\}.\label{eq:MSEmatrix}
\end{equation}
The ZZB for estimating $\bm{\xi}$ to lower
bound the MSE matrix is
\begin{align}
\notag\bm{u}^{\mathsf{T}}\bm{R}_{\bm{\epsilon}}\bm{u} \geq\frac{1}{2} \int_0^{\infty} \mathcal{V}\bigg\{&\max_{\bm{\delta}:\bm{u}^{\mathsf{T}}\bm{\delta}=h}\iint_{\bm{\Xi}}\left[f_{\bm{\xi}}\left(\bm{\vartheta}\right)+f_{\bm{\xi}}\left(\bm{\vartheta}+\bm{\delta}\right)\right]\times\\
&\mathrm{P}_{\min}\left(\bm{\vartheta},\bm{\vartheta}+\bm{\delta}\right)d\bm{\vartheta}\bigg\} h d h,\label{eq:ZZBURU}
\end{align}
where $\bm{u}\in\mathbb{R}^{2}$ is any normalized weight vector, $\mathcal{V}\left\{\cdot\right\}$ is the valley-filling function, i.e., $\Vc \{ f(t) \}= \sup_{u: u\geq t} f(u)$, $\bm{\Xi}$ denotes the region where $\bm{\xi}$ is defined, $f_{\bm{\xi}}\left(\bm{\vartheta}\right)\triangleq f_{\bm{\xi}}\left(\bm{\xi}\right)|_{\bm{\xi}=\bm{\vartheta}}$ with $f_{\bm{\xi}}\left(\bm{\xi}\right)$ denoting the \textit{prior} probability density function (PDF) of the unknown vector $\bm{\xi}$ , and $\mathrm{P}_{\min}\left(\bm{\vartheta},\bm{\vartheta}+\bm{\delta}\right)$ is the minimum error probability of the binary hypothesis testing problem, deciding between $\mathcal{H}_{0}: \bm{\xi}=\bm{\vartheta}$ and $\mathcal{H}_{1}: \bm{\xi}=\bm{\vartheta}+\bm{\delta}$. The hypothesis testing problem is
\begin{align}
& \notag \mathcal{H}_{0}: \bm{\xi}=\bm{\vartheta}; \tilde{\bm{\mathrm{v}}}\sim f_{\tilde{\bm{\mathrm{v}}}|\bm{\xi}}\left(\tilde{\bm{\mathrm{v}}}|\bm{\vartheta}\right), \\
& \mathcal{H}_{1}: \bm{\xi}=\bm{\vartheta}+\bm{\delta}; \tilde{\bm{\mathrm{v}}}\sim f_{\tilde{\bm{\mathrm{v}}}|\bm{\xi}}\left(\tilde{\bm{\mathrm{v}}}| \bm{\vartheta}+\bm{\delta}\right), \label{eq:problem1}
    \end{align}
where $f_{\tilde{\bm{\mathrm{v}}}| \bm{\xi}}\left(\tilde{\bm{\mathrm{v}}}| \bm{\vartheta}\right)\triangleq f_{\tilde{\bm{\mathrm{v}}}| \bm{\xi}}\left(\tilde{\bm{\mathrm{v}}}| \bm{\xi}\right)|_{\bm{\xi}=\bm{\vartheta}}$ with $f_{\tilde{\bm{\mathrm{v}}}| \bm{\xi}}\left(\tilde{\bm{\mathrm{v}}}| \bm{\xi}\right)$ representing the conditional PDF (CPDF) of $\tilde{\bm{\mathrm{v}}}$ given the vector $\bm{\xi}$. The \textit{prior} probabilities of \eqref{eq:problem1} can be written as
\begin{align}
&\mathrm{Pr}\left\{\mathcal{H}_{0}\right\}=\frac{f_{\bm{\xi}}\left(\bm{\vartheta}\right)}{f_{\bm{\xi}}\left(\bm{\vartheta}\right)+f_{\bm{\xi}}\left(\bm{\vartheta}+\bm{\delta}\right)},\label{eq:PrH0}\\
&\mathrm{Pr}\left\{\mathcal{H}_{1}\right\}=\frac{f_{\bm{\xi}}\left(\bm{\vartheta}+\bm{\delta}\right)}{f_{\bm{\xi}}\left(\bm{\vartheta}\right)+f_{\bm{\xi}}\left(\bm{\vartheta}+\bm{\delta}\right)}\label{eq:PrH1}.
\end{align}
\end{lemma}
\begin{IEEEproof}
    The proof can be referred to \cite{556118,7116574,542439}.
\end{IEEEproof}

It is worth noting that lower bounding the
quadratic form of the MSE matrix in \textit{Lemma} \ref{lemm:ZZB1} offers an effective way to bound either the total error (sum of the diagonal elements of $\bm{R}_{\bm{\epsilon}}$) or errors of specific components of $\bm{\xi}$. To compute the
estimation error of a specific parameter, the element of $\bm{u}$ associated with the parameter of
interest is set to $1$, and the rest is set to $0$.

To obtain the explicit expressions of the ZZBs, we derive $\mathrm{P}_{\min}\left(\bm{\vartheta},\bm{\vartheta}+\bm{\delta}\right)$ in \eqref{eq:ZZBURU} by computing the error probability of the likelihood ratio test for the problem in \eqref{eq:problem1} using the MAP criterion, which is given below. 
\begin{lemma}
[Minimum error probability]\label{lemm:ZZB2} 
The expression of $\mathrm{P}_{\min}\left(\bm{\vartheta},\bm{\vartheta}+\bm{\delta}\right)$ in \eqref{eq:ZZBURU} can be computed as
\begin{align}
\notag\mathrm{P}_{\min}\left(\bm{\vartheta},\bm{\vartheta}+\bm{\delta}\right)=&\mathrm{Pr}\left\{\mathcal{H}_{0}\right\}\mathcal{Q}\left(\frac{\mu_{\mathcal{L}_{\bm{\vartheta}}}-\ln \frac{\mathrm{Pr}\left\{\mathcal{H}_{1}\right\}}{\mathrm{Pr}\left\{\mathcal{H}_{0}\right\}}}{\sqrt{2 \mu_{\mathcal{L}_{\bm{\vartheta}}}}}\right)\\
&+\mathrm{Pr}\left\{\mathcal{H}_{1}\right\}\mathcal{Q}\left(\frac{\mu_{\mathcal{L}_{\bm{\vartheta}}}+\ln \frac{\mathrm{Pr}\left\{\mathcal{H}_{1}\right\}}{\mathrm{Pr}\left\{\mathcal{H}_{0}\right\}}}{\sqrt{2 \mu_{\mathcal{L}_{\bm{\vartheta}}}}}\right),\label{eq:lemma2}
\end{align}
where $\mathrm{Pr}\left\{\mathcal{H}_{0}\right\}$ and $\mathrm{Pr}\left\{\mathcal{H}_{1}\right\}$ are given in \eqref{eq:PrH0} and \eqref{eq:PrH1}, $\mathcal{Q}\left(\cdot\right)$ denotes the tail distribution function of the standard normal distribution, i.e., $\mathcal{Q}\left(x\right)=\frac{1}{\sqrt{2 \pi}} \int_x^{\infty} \mathrm{e}^{-\frac{v^2}{2}} d v$, and $\mu_{\mathcal{L}_{\bm{\vartheta}}}$ is 
\begin{equation}
\mu_{\mathcal{L}_{\bm{\vartheta}}}=\text{SNR}l_{\mathsf{s}}\int_{0}^{{D_{\rq}}}\mathcal{A}\left(\bm{\vartheta},\bm{\delta},y_{\rq}\right)dy_{\rq},\label{eq:mumu} 
\end{equation}
where $\text{SNR}\triangleq\frac{{\mathcal{E}_{\mathsf{in}}^{2}}}{\sigma^{2}}$ is the SNR and $\mathcal{A}\left(\bm{\vartheta},\bm{\delta},y_{\rq}\right)$ is the \textit{ambiguity function} (AF) defined as
\begin{align}
\notag\mathcal{A}&\left(\bm{\vartheta},\bm{\delta},y_{\rq}\right)=\left|{{h}}_{\mathsf{y}}\left(\bm{\vartheta}+\bm{\delta};y_{\mathsf{r}}\right)\right|^{2}+\left|{{h}}_{\mathsf{y}}\left(\bm{\vartheta};y_{\mathsf{r}}\right)\right|^{2}\\
&\quad\quad-2\left|{{h}}_{\mathsf{y}}\left(\bm{\vartheta}+\bm{\delta};y_{\mathsf{r}}\right)\right|\left|{{h}}_{\mathsf{y}}\left(\bm{\vartheta};y_{\mathsf{r}}\right)\right|\cos {\left(k{\angle{\mathcal{A}}}\right)},\label{eq:AF}
\end{align}
where ${h}_{\mathsf{y}}\left(\bm{\vartheta};y_{\mathsf{r}}\right)\triangleq {{h}}_{\mathsf{y}}\left(\bm{\xi};y_{\rq}\right)|_{\bm{\xi}=\bm{\vartheta}}$ with ${{h}}_{\mathsf{y}}\left(\bm{\xi};y_{\rq}\right)$ denoting the \textit{near-field electromagnetic (NF-EM) channel} given in \eqref{eq:hy}, and 
\begin{align}
{\angle{\mathcal{A}}}\triangleq r_{\mathsf{ry}}|_{z_{\tq}={\vartheta}_{z}+\delta_{z}}-{r_{\mathsf{ry}}|_{z_{\tq}={\vartheta}_{z}}}
\end{align}
with $\bm{\vartheta}\triangleq\left(\vartheta_{z},\vartheta_{t}\right)^{\mathsf{T}}$ and $\bm{\delta}\triangleq\left(\delta_{z},\delta_{t}\right)^{\mathsf{T}}$.
\end{lemma}
\begin{IEEEproof}
    The proof is provided in Appendix~\ref{app:lemma2}.
\end{IEEEproof}

\textit{Lemma}~\ref{lemm:ZZB2} clearly shows that the minimum error probability $\mathrm{P}_{\min}\left(\bm{\vartheta},\bm{\vartheta}+\bm{\delta}\right)$ is the function of two \textit{prior} conditional PDFs $f_{\bm{\xi}}\left(\bm{\vartheta}\right)$ and $f_{\bm{\xi}}\left(\bm{\vartheta}+\bm{\delta}\right)$, and the AF $\mathcal{A}\left(\bm{\vartheta},\bm{\delta},y_{\rq}\right)$ given in \eqref{eq:AF}. Besides, $\mathcal{A}\left(\bm{\vartheta},\bm{\delta},y_{\rq}\right)$ captures the difference between two NF-EM channels ${h}_{\mathsf{y}}\left(\bm{\vartheta};y_{\mathsf{r}}\right)$ and ${h}_{\mathsf{y}}\left(\bm{\vartheta}+\bm{\delta};y_{\mathsf{r}}\right)$.

Based on the \textit{prior} distributions in \textit{Assumption} \assumptionref{assum:p}, we derive the explicit expressions of ZZB in the following proposition.

\begin{proposition}[ZZB expressions for $\bm{\xi}$]\propositionlabel{prop:ZZBexpression} The ZZB expressions for estimating $\bm{\xi}$ can be obtained as
\begin{align}
&\mathrm{ZZB}\left(z_{\tq}\right)= \frac{1}{\mathsf{H}_{\tq}} \int_{0}^{\mathsf{H}_{\mathsf{t}}}\left[\max _{\delta_{t}} \iint_{\bm{\Xi}} \mathcal{Q}\left(\sqrt{\frac{\mu_{\mathcal{L}_{\bm{\vartheta}}}}{2}}\right)d\bm{\vartheta}\right] \delta_z d \delta_z,\label{eq:ZZBz}\\
&\mathrm{ZZB}\left(t_{z}\right)=\frac{1}{\mathsf{H}_{\tq}} \int_{0}^{1}\left[\max _{\delta_z} \iint_{\bm{\Xi}} \mathcal{Q}\left(\sqrt{\frac{\mu_{\mathcal{L}_{\bm{\vartheta}}}}{2}}\right)d\bm{\vartheta}\right] \delta_{t} d \delta_{t},\label{eq:ZZBtz}
\end{align}
where $\mu_{\mathcal{L}_{\bm{\vartheta}}}$ is given in \eqref{eq:mumu}, and 
\begin{align}
  \bm{\Xi}=\big\{\vartheta_{z},\vartheta_{t}:
    \vartheta_{z}\in[\mathsf{H}_1, \mathsf{H}_{2}-\delta_{z}],
    \vartheta_{t}\in[0, 1-\delta_{t})\big\},\label{eq:space}
\end{align}
where $\delta_{z}\in[0,\mathsf{H}_{\mathsf{t}}]$ and $\delta_{t}\in[0,1)$.
\end{proposition}
\begin{IEEEproof}
We have $f_{\bm{\xi}}\left(\bm{\vartheta}\right)=f_{\bm{\xi}}\left(\bm{\vartheta}+\bm{\delta}\right)=\frac{1}{\mathsf{H}_{\tq}}$ and
\begin{equation}
\mathrm{P}_{\min}\left(\bm{\vartheta},\bm{\vartheta}+\bm{\delta}\right)=\mathcal{Q}\left(\sqrt{\frac{\mu_{\mathcal{L}_{\bm{\vartheta}}}}{2}}\right).\label{eq:Pminsim} 
\end{equation}
Ignoring the $\mathcal{V}\left\{\cdot\right\}$ function for simplicity, \eqref{eq:ZZBURU} reduces to
\begin{equation}
\bm{u}^{\mathsf{T}}\bm{R}_{\bm{\epsilon}}\bm{u} \geq\frac{1}{\mathsf{H}_{\tq}} \int_0^{\infty} \left[\max_{\bm{\delta}:\bm{u}^{\mathsf{T}}\bm{\delta}=h}\iint_{\bm{\Xi}}\mathcal{Q}\left(\sqrt{\frac{\mu_{\mathcal{L}_{\bm{\vartheta}}}}{2}}\right)d\bm{\vartheta}\right] h d h,\label{eq:ZZBpro1}
\end{equation}
where $\bm{\Xi}$ is provided in \eqref{eq:space}. Substituting $\bm{u}$ in \eqref{eq:ZZBpro1} with $\bm{u}_{z}\triangleq\left(1,0\right)^{\mathsf{T}}$ and
$\bm{u}_{t}\triangleq\left(0,1\right)^{\mathsf{T}}$, we have
\begin{align}
&\mathbb{E}\left\{|z_{\tq}-\hat{z}_{\tq}|^2\right\}=\bm{u}_z^{\mathsf{T}} \bm{R}_{\bm{\epsilon}} \bm{u}_z=\left[\bm{R}_{\bm{\epsilon}}\right]_{11} \geq  \mathrm{ZZB}\left(z_{\tq}\right),\\
&\mathbb{E}\left\{|t_{z}-\hat{t}_{z}|^2\right\}=\bm{u}_t^{\mathsf{T}} \bm{R}_{\bm{\epsilon}} \bm{u}_t=\left[\bm{R}_{\bm{\epsilon}}\right]_{22} \geq  \mathrm{ZZB}\left(t_{z}\right),
\end{align}
where the ZZBs are provided in \eqref{eq:ZZBz} and \eqref{eq:ZZBtz}.
\end{IEEEproof}\begin{discussion}[Computational difficulty and potential optimization strategies for the ZZBs]{\label{definition22444} \eqref{eq:ZZBz} and \eqref{eq:ZZBtz} have similar forms to the ZZBs for common joint estimation problems \cite{7116574}. The computational difficulty lies in \textit{multiple integrals} and \textit{maximum value search}, which is the common difficulty in most multi-parameter ZZB computations. We provide the following potential optimization strategies. Firstly, we can utilize coarse-grained search and the \textit{rectangular numerical integration} with low accuracy of interval division to compute the preliminary value of ZZB, then refine the granularity and interval division according to the required accuracy. Secondly, we can avoid the maximum search through single parameter estimation, as discussed next in \textit{Corollary}~\ref{coro:ZZBonly}. Finally, it should be pointed out that using appropriate optimization algorithms to reduce the complexity of computations deserves further exploration.}
\end{discussion}

In the asymptotic case where the SNR is small, or the long side length $D_{\rq}$ is small, or the hypothesis increment $\bm{\xi}$ is small, the observation vector
in \eqref{eq:problem1} is indistinguishable under the two hypotheses, leading to the simplified ZZBs given below.

\begin{corollary}[Asymptotic analysis of ZZB]\label{coro:ZZBCase3} When $\text{SNR}\to 0$, or $D_{\rq}\to 0$, or $\bm{\delta}\to\bm{0}$,  the ZZBs in \eqref{eq:ZZBz} and \eqref{eq:ZZBtz} reduce to
 \begin{align}
&\mathrm{ZZB}\left(z_{\tq}\right)=\frac{\mathsf{H}^{2}_{\tq}}{12},~~\mathrm{ZZB}\left(t_{z}\right)=\frac{1}{12}\label{eq:zzbcase3t}.
 \end{align}
\end{corollary}
\begin{IEEEproof}
If the $\text{SNR}$ or $D_{\rq}$ approaches zero, $\mathcal{L}_{\bm{\vartheta}}$ and $\mathcal{L}_{\bm{\vartheta}+\bm{\delta}}$ in \eqref{eq:LXI} can be seen as the determined value $0$. Thus, we can obtain $\mathrm{P}_{\min}=\frac{1}{2}$ under \textit{Assumption} \assumptionref{assum:p}, where we use $\mathrm{P}_{\min}$ to represent $\mathrm{P}_{\min}\left(\bm{\vartheta},\bm{\vartheta}+\bm{\delta}\right)$. Besides, if $\bm{\delta}$ approaches $\bm{0}$, the observation  voltage vector $\tilde{\mathbf{v}}$
in \eqref{eq:problem1} is indistinguishable under the two hypotheses. Thus, we have $\mathrm{P}_{\min}=\frac{f_{\bm{\xi}}\left(\bm{\vartheta}\right)}{2f_{\bm{\xi}}\left(\bm{\vartheta}\right)}=\frac{1}{2}$. Consequently, $\mathrm{P}_{\min}$ in \eqref{eq:Pminsim} reduces to $\mathrm{P}_{\min}=\frac{1}{2}$, which can be substituted into \eqref{eq:ZZBz} and \eqref{eq:ZZBtz} to obtain \eqref{eq:zzbcase3t}.
\end{IEEEproof}

\textit{Corollary} \ref{coro:ZZBCase3} shows that ZZBs are the prior variances of the {position} and {attitude} parameters if $\text{SNR}\to 0$, or $D_{\rq}\to 0$, or $\bm{\delta}\to\bm{0}$. In other words, the ZZB uses the \textit{prior} knowledge if the observation vector $\tilde{\mathbf{v}}$ is indistinguishable, which enables it to provide a \textit{globally tight} bound for the MSE. 

Further, to simplify ZZB expressions, we consider the non-joint estimation, namely, attitude-only estimation\footnote{Another non-joint estimation is the position-only estimation, where the {attitude} parameter $t_z$ is assumed known and the unknown parameter is $z_{\tq}$. The position-only estimation is often discussed in the classic near-field positioning literature. For example, \cite{hu2018beyond2} and \cite{alegria2019cramer} assume that UE is a mass point, \cite{de2021cramer} and \cite{angchen} assume that $t_z=0$. Thus, we only discuss the attitude-only estimation.}, where the {position }parameter $z_{\tq}$ is known and the unknown parameter is $t_{z}$. The attitude-only estimation is a special case of our joint estimation and has not been studied. We make \textit{Assumption}~\assumptionref{assum:4.1} for the attitude-only estimation to focus on the original structure of the non-joint estimation case without being disturbed by the value of the known parameter.
\begin{assumption}[Attitude-only estimation]\assumptionlabel{assum:4.1}
$z_{\tq}$ is not assumed to take a deterministic value. In particular, it is still considered a random variable with the same probability distribution as the corresponding random variable in \textit{Assumption}~\assumptionref{assum:p}. 
\end{assumption}
\begin{corollary}[ZZB for the attitude-only estimation]\label{coro:ZZBonly}  
The expression of ZZB for the attitude-only estimation is 
\begin{equation}
\mathrm{ZZB}_{\mathrm{AO}}\left(t_{z}\right)=\int_{0}^{1} \delta_{t} \int_{0}^{1-\delta_{t}}\mathbb{E}_{z_{\tq}}\left\{\mathcal{Q}\left(\sqrt{\frac{\mu_{\mathcal{L}_{\bm{\vartheta}_{a}}}}{2}}\right)\right\}d\vartheta_{t} d \delta_{t},\label{eq:mulva}
\end{equation}
where
{\begin{align}
\frac{{\mu_{\mathcal{L}_{\bm{\vartheta}_{a}}}}}{\text{SNR}l_{\mathsf{s}}}
=\frac{\delta_{t}^2\tau^3+2\delta_{t}\mathcal{F}_{t}+\mathcal{F}_{t}^2\left(2\tau^2+3\right)\tau}{3z_{\tq}\left(\tau^2+1\right)^{3/2}}-\frac{2\delta_{t}\mathcal{F}_{t}}{3z_{\tq}},\label{eq:mulvaclose}
\end{align}
and $\mathcal{F}_{t}\triangleq\sqrt{1-\vartheta_{t}^2}-\sqrt{1-\left(\vartheta_{t}+\delta_{t}\right)^2}$. In \eqref{eq:mulvaclose}, the parameter $\tau\triangleq D_{\rq}/z_{\tq}$ measures the maximum geometric dimension $D_{\rq}$ of the receiving surface
normalized by the distance  from the UE to the
receiver.} Further, when $\mathcal{R}_{\rq}$ is infinite in the $\mathsf{Y}$-dimension, i.e., ${D_{\mathsf{r}} \to \infty}$ ($\tau \to \infty$), \eqref{eq:mulvaclose} can be written as
\begin{equation}
    \lim_{D_{\mathsf{r}} \to \infty} \mu_{\mathcal{L}_{\bm{\vartheta}_{a}}}=\text{SNR}\frac{l_{\mathsf{s}}}{3z_{\tq}}\left[\left(\delta_{t}-\mathcal{F}_{t}\right)^2+\mathcal{F}_{t}^2\right].\label{eq:mulvaclose11}
\end{equation}
\end{corollary}

\begin{IEEEproof}
The proof is provided in Appendix~\ref{app:coro:ZZBonly}.
\end{IEEEproof}

The ZZB in \eqref{eq:mulva} is greatly simplified compared to \eqref{eq:ZZBtz} since the dimension of the integral is reduced and there is no 1D maximum search process. Through \textit{Corollary}~\ref{prop:singleVS} and Section \ref{sec:jointvsnu}, we indicate that $\mathrm{ZZB}_{\mathrm{AO}}\left(t_z\right)$ can be used to approximate $\mathrm{ZZB}\left(t_z\right)$ with very small error. Also, it can be observed that the wavelength $\lambda$ does not affect the ZZB for $t_z$, as the phase term of the channel ${{h}}_{\mathsf{y}}\left(\bm{\xi};y_{\rq}\right)$ in
\eqref{eq:hy} contains only $z_{\tq}$, not $t_{z}$. 

\subsection{ECRB for Joint Position and Attitude Estimation}\label{sec:ECRB}
The CRB offers a lower bound to the MSE of any unbiased estimate, stating that the MSE of any such estimator is at least as high as the inverse of the Fisher information. Compared with the ZZB in Section \ref{sec:ZZB}, the CRB is a \textit{locally bound} and only \textit{asymptotically tight} in small error estimation scenarios (say, high SNR). However, as the CRB is easier to compute and widely used, we will develop the closed-form CRB for the joint estimation to gain more insights in high SNR regimes.

Since the parameter vector $\bm{\xi}$ is random with a known \textit{prior} distribution $f_{\bm{\xi}}\left(\bm{\xi}\right)$, we consider the Bayesian version of the CRB, i.e., expected CRB (ECRB)  \cite{van2007bayesian}. The preliminaries of ECRB are given in \textit{Lemma} \ref{lemma:ECRB1}.
\begin{lemma}[ECRB preliminaries for $\bm{\xi}$]\label{lemma:ECRB1} Based on estimation theory, the ECRB for estimating the $\imath$-th entry of ${\bm{\xi}}$ is 
\begin{equation}
\mathrm{ECRB}\left(\bm{\xi}_{\imath} \right)=\mathbb{E}_{\bm{\xi}}\left\{\left[\mathbf{F}^{-1}\left(\bm{\xi}\right)\right]_{\imath\imath}\right\},\imath=1,2,
\label{eq:ECRB}
\end{equation}
where $\mathbf{F}\left(\bm{\xi}\right)$ is the deterministic Fisher information matrix (FIM) for a specific vector $\bm{\xi}$, whose element on the $\imath$-th row and $\jmath$-th column is given by
\begin{equation}
\left[\mathbf{F}\left(\bm{\xi}\right)\right]_{\imath \jmath}=2\text{SNR}l_{\mathsf{s}}\int_{0}^{D_{\mathsf{r}}}\Re\left\{\frac{\partial {{h}}_{\mathsf{y}}^{*}\left(\bm{\xi};y_{\mathsf{r}}\right)}{\partial \bm{\xi}_{\imath}}\frac{\partial {{h}}_{\mathsf{y}}\left(\bm{\xi};y_{\mathsf{r}}\right)}{\partial \bm{\xi}_{\jmath}}\right\}d{y_{\rq}}.
\end{equation}
\end{lemma} 
\begin{IEEEproof}
The proof is provided in Appendix~\ref{app:CRB1}.
\end{IEEEproof}\begin{discussion}[Bayesian CRB]\label{definition22} {Another Bayesian version of the CRB is the Bayesian CRB (BCRB). The BCRB for estimating the $\imath$-th entry of ${\bm{\xi}}$ is given by $\mathrm{BCRB}\left(\bm{\xi}_{\imath} \right)=\left[\mathbf{B}^{-1}\right]_{\imath\imath}$, $\imath=1,2$, where $\mathbf{B}$ is a $2\times 2$ positive definite matrix, whose element on the $\imath$-th row and $\jmath$-th column is given by 
\begin{equation}
    \left[\mathbf{B}\right]_{\imath  \jmath}=\underbrace{\mathbb{E}_{\bm{\xi}}\left\{\left[\mathbf{F}\left(\bm{\xi}\right)\right]_{\imath \jmath}\right\}}_{\triangleq \left[{\mathbf{F}_{\mathsf{O}}}\right]_{\imath \jmath}}+\underbrace{\mathbb{E}_{\bm{\xi}}\left\{\frac{\partial \ln f_{\bm{\xi}}\left(\bm{\xi}\right)}{\partial \bm{\xi}_{\imath}}\frac{\partial \ln f_{\bm{\xi}}\left(\bm{\xi}\right)}{\partial \bm{\xi}_{\jmath}}\right\}}_{\triangleq\left[\mathbf{B}_{\mathsf{P}}\right]_{\imath \jmath}},
\end{equation}
where $\mathbf{F}_{\mathsf{O}}$ and $\mathbf{B}_{\mathsf{P}}$ denote the information gained from the observation data and the \textit{prior} knowledge, respectively. Under \textit{Assumption}~\assumptionref{assum:p}, there is not
\textit{prior} information of $\bm{\xi}$ contributing to $\mathbf{B}$, i.e.,  $\mathbf{B}_{\mathsf{P}}=\mathbf{0}$. Note that the BCRB is hard to assess since the expectation of $\mathbf{F}\left(\bm{\xi}\right)$ over $\bm{\xi}$ is not analytically tractable. Also, the
ECRB is a tighter bound of the MSE in the high SNR regimes than the BCRB, i.e., $\mathrm{BCRB}\left(\bm{\xi}_{\imath} \right){<}\mathrm{ECRB}\left(\bm{\xi}_{\imath} \right)\leq \mathrm{MSE}\left(\bm{\xi}_{\imath} \right)$\footnote{{We have $\mathbf{B}^{-1}=\mathbf{F}_{\mathsf{O}}^{-1}-\mathbf{F}_{\mathsf{O}}^{-1}\mathbf{K}^{-1}\mathbf{F}_{\mathsf{O}}^{-1}$ with $\mathbf{K}\triangleq \mathbf{B}_{\mathsf{P}}^{-1}+\mathbf{F}_{\mathsf{O}}^{-1}$ based on Woodbury's identity. Since $\mathbf{F}_{\mathsf{O}}$ and $\mathbf{B}_{\mathsf{P}}$ are positive definite, $\mathbf{F}_{\mathsf{O}}^{-1}\mathbf{K}^{-1}\mathbf{F}_{\mathsf{O}}^{-1}$ is also positive definite. Thus, it follows that $\mathrm{BCRB}\left(\bm{\xi}_{\imath} \right)<\big[\mathbf{F}_{\mathsf{O}}^{-1}\big]_{\imath \imath}$. Further, we have $\big[\mathbf{F}_{\mathsf{O}}^{-1}\big]_{\imath \imath}<\mathrm{ECRB}\left(\bm{\xi}_{\imath} \right)$ by using Jensen's inequality.}} \cite{7359173CP}. Thus, we adopt the ECRB in this paper.}
\end{discussion}

Now, we provide the expressions of the ECRB as follows.
\begin{proposition}[ECRB expressions for $\bm{\xi}$]\label{coro:ECRB} The ECRB expressions for estimating $\bm{\xi}$ are given by
\begin{align}
&\mathrm{ECRB}\left(z_{\tq}\right)=\frac{\text{SNR}^{-1}}{2l_{\mathsf{s}}}\mathbb{E}_{\bm{\xi}}\left\{\frac{\mathcal{I}_{\mathsf{tt}}}{\mathcal{I}_{\mathsf{zz}}\mathcal{I}_{\mathsf{tt}}-\mathcal{I}_{\mathsf{zt}}^2}\right\},\label{eq:ECRBYT}\\
&\mathrm{ECRB}\left(t_{z}\right)=\frac{\text{SNR}^{-1}}{2l_{\mathsf{s}}}\mathbb{E}_{\bm{\xi}}\left\{\frac{\mathcal{I}_{\mathsf{zz}}}{\mathcal{I}_{\mathsf{zz}}\mathcal{I}_{\mathsf{tt}}-\mathcal{I}_{\mathsf{zt}}^2}\right\}\label{eq:ECRBT},
\end{align}
where $\mathcal{I}_{\mathsf{zz}}\triangleq \mathcal{I}_{\mathsf{zz}1}+k^{2}\mathcal{I}_{\mathsf{zz}2}$,
\begin{align}
\mathcal{I}_{\mathsf{zz}1}&\triangleq \frac{1}{z_{\tq}}\int_{0}^{D_{\mathsf{r}}}\frac{\left[t_{z}y_{\rq}\left(y_{\rq}^2-4z_{\tq}^2\right)+t_{y}\ell_{\mathsf{yz}}z_{\tq}\right]^2}{4r_{\mathsf{ry}}^{9}}dy_{\rq},\label{eq:Izz1FIM}\\
\mathcal{I}_{\mathsf{zz}2}&\triangleq z_{\tq}^3\int_{0}^{D_{\mathsf{r}}}\frac{\left(t_{z}y_{\rq}+t_{y}z_{\tq}\right)^2}{r_{\mathsf{ry}}^{7}}dy_{\rq},\\
\mathcal{I}_{\mathsf{tt}}&\triangleq z_{\tq}\int_{0}^{D_{\mathsf{r}}}\frac{\left(y_{\rq}-z_{\tq}t_{z}/{t_{y}}\right)^2}{r_{\mathsf{ry}}^{5}}dy_{\rq},\label{eq:Ittapp}\\
\mathcal{I}_{\mathsf{zt}}&\triangleq \int_{0}^{D_{\mathsf{r}}}\frac{t_{z}y_{\rq}\left(y_{\rq}^2-4z_{\tq}^2\right)+t_{y}z_{\tq}\ell_{\mathsf{yz}}}{2r_{\mathsf{ry}}^{7}\left(y_{\rq}-{z_{\tq}t_{z}}/{t_{y}}\right)^{-1}}dy_{\rq},\label{eq:IztFIM}
\end{align}
with $\ell_{\mathsf{yz}}\triangleq3y_{\rq}^2-2z_{\tq}^2$. {We provide the closed-form expressions of $\mathcal{I}_{\mathsf{zz}1}$, $\mathcal{I}_{\mathsf{zz}2}$, $\mathcal{I}_{\mathsf{tt}}$, and $\mathcal{I}_{\mathsf{zt}}$ in \eqref{eq:Izz1FIMclose}--\eqref{eq:IztFIMclose} of Appendix~\ref{app:ECRB1}.}
\end{proposition}
\begin{IEEEproof}
The proof is provided in Appendix~\ref{app:ECRB1}.
\end{IEEEproof}

\textit{Proposition}~\ref{coro:ECRB} shows that 
the ECRBs are \textit{inversely proportional} to the SNR. In particular, the ECRBs will decrease by a factor of $10$ if the SNR increases by $10$ whether in \textit{high} SNR regimes or \textit{low} SNR regimes. Note that the ECRB is an \textit{asymptotically tight} lower bound for MSE only in \textit{high} SNR regimes. Besides, the ECRBs are \textit{inversely proportional} to $l_{\mathsf{s}}$, which is the length of the short side of the {observation region}. If $l_{\mathsf{s}}\to 0$, the CRB will increase infinitely. This is because the \textit{near-field electromagnetic signal} and \textit{channel} in \eqref{eq:12} and \eqref{eq:hy} are the 3D models in the physical world, while the reduction of $l_{\mathsf{s}}$ to $0$ will cause the noise-free observation signal in \eqref{eq:esmxny} to be reduced to $0$, i.e., no effective information can be captured and the \textit{prior} knowledge cannot be exploited either.

In the asymptotic case where the SNR is small or the long side length $D_{\rq}$ is small, namely, $\text{SNR}\to 0$ or $D_{\rq}\to 0$, we have $\mathrm{ECRB}\left(\bm{\xi}_{\imath}\right)=\infty$, which means that no observed information and \textit{prior} knowledge can be used. 

In the asymptotic case where the observation region $\mathcal{R}_{\rq}$ is infinite in the $\mathsf{Y}$-dimension and the dipole points to the
positive direction of $\mathsf{Y}$-axis, the ECRBs can be simplified as follows.
\begin{corollary}[Asymptotic analysis of ECRB]\label{coro:ECRBCase2} When $D_{\rq} \to \infty$ and $t_{z} \to 0$,  the ECRBs in \eqref{eq:ECRBYT} and \eqref{eq:ECRBT} are simplified to
    \begin{align}
&\mathrm{ECRB}\left(z_{\tq}\right)=\frac{\text{SNR}^{-1}}{2l_{\mathsf{s}}}\mathbb{E}_{z_{\tq}}\left\{\frac{210z_{\tq}^3}{112k^2z_{\tq}^2+75}\right\},\label{eq:ECRBzcase2}\\
        &\mathrm{ECRB}\left(t_{z}\right)=\frac{3\text{SNR}^{-1}}{2l_{\mathsf{s}}}\mathbb{E}_{z_{\tq}}\left\{z_{\tq}\right\}\label{eq:ECRBtcase2}.
    \end{align}
Further, when the component of the position coordinates along $\mathsf{Z}$ direction is much larger than the wavelength, i.e., $z_{\tq}\gg \lambda$\footnote{Since $z_{\tq}\ll 2\mathsf{D}^{2}/\lambda$ is always satisfied when $z_{\tq}\gg \lambda$ and $\mathsf{D}$ is not very small, we know that $z_{\tq}\gg \lambda$ corresponds to the near-field region when the size of the \textit{observation region} $\mathcal{R}_{\rq}$ is on the order of meters.}, which generally holds in the systems with carrier frequencies in the range of GHz or above, \eqref{eq:ECRBzcase2} can be  simplified to
   \begin{align}
\mathrm{ECRB}\left(z_{\tq}\right)\approx\frac{15\text{SNR}^{-1}}{64\pi^2l_{\mathsf{s}}}\lambda^{2}\mathbb{E}_{z_{\tq}}\left\{z_{\tq}\right\}.\label{eq:limmmmecrbz}
    \end{align}
\end{corollary}
\begin{IEEEproof}
The proof is provided in Appendix~\ref{app:ECRBcase2}.
\end{IEEEproof}

\textit{Corollary}~\ref{coro:ECRBCase2} shows that we no longer need to use the Monte Carlo simulation to evaluate \eqref{eq:ECRBtcase2} and \eqref{eq:limmmmecrbz} since $\mathbb{E}_{z_{\tq}}\left\{z_{\tq}\right\}$ can be obtained directly from the \textit{prior} knowledge. Besides, \eqref{eq:limmmmecrbz} reveals that the ECRB for $z_\tq$ depends on $\lambda$ \textit{quadratically}.  Indeed, reducing $\lambda$ by a factor of $\varrho$ will reduce it by $\varrho^2$. The same conclusion does not hold for the ECRB for $t_{z}$. 

Similar to \textit{Corollary}~\ref{coro:ZZBonly}, we simplify the ECRB expression for the case with attitude-only estimation.
\begin{corollary}[ECRB for the attitude-only estimation]\label{coro:ECRBonly}  
The expression of ECRB for the attitude-only estimation is 
\begin{align}
\mathrm{ECRB}_{\mathrm{AO}}\left(t_{z}\right)=\frac{\textrm{SNR}^{-1}}{2l_{\mathsf{s}}}\mathbb{E}_{\bm{\xi}}\left\{{\mathcal{I}_{\mathsf{tt}}^{-1}}\right\}\label{eq:ECRBAO},
\end{align}
where $\mathcal{I}_{\mathsf{tt}}$ is given in \eqref{eq:Ittapp} and its closed-form expression is computed in \eqref{eq:Ittappclose}. Further, if ${D_{\mathsf{r}} \to \infty}$, \eqref{eq:Ittappclose} is written as
\begin{equation}
    \lim_{D_{\mathsf{r}} \to \infty} \mathcal{I}_{\mathsf{tt}}=\frac{1}{3z_{\tq}}\frac{1+t_{z}^2-2t_{z}^2\sqrt{1-t_{z}^2}}{1-t_{z}^2}.\label{eq:711}
\end{equation}    
\end{corollary}
\begin{IEEEproof}
The FIM for the attitude-only estimation degenerates to a scalar Fisher information parameter. Then, we can carry out the similar proof process of \textit{Proposition}~\ref{coro:ECRB}.
\end{IEEEproof}

Having obtained the ZZBs and ECRBs for joint estimation and attitude-only estimation, we discuss their relation.
\begin{corollary}
[ZZB, ECRB, joint vs. attitude-only estimation]\label{prop:singleVS} For the joint estimation and attitude-only estimation, we have 
\begin{align}
&\mathrm{ZZB}\left(t_z\right)\geq \mathrm{ZZB}_{\mathrm{AO}}\left(t_z\right)\label{eq:vs1},\\
&\mathrm{ECRB}\left(t_{z}\right)\geq \mathrm{ECRB}_{\mathrm{AO}}\left(t_{z}\right)\label{eq:vs2}.
\end{align}
\end{corollary}
\begin{IEEEproof}
The proof is provided in Appendix~\ref{app:VS}.
\end{IEEEproof}

\textit{Corollary}~\ref{prop:singleVS} indicates that the achievable lower bounds of the 
ZZB and ECRB for the joint estimation are respectively the 
ZZB and ECRB for the attitude-only estimation. This is reasonable since estimating a larger number of parameters will lead to larger errors due to uncertainty from other parameters.

\begin{table*}[!tb]
\caption{Accuracy of the closed-form solutions with $\alpha=1$ and $\beta=\mathsf{N}/2$. The parameter settings for each case are specified in the table.}
\label{tab:performance}
\resizebox{0.99\linewidth}{!}{
\begin{tabular}{c|c c||c c|c c|c c|c c|c c|c c||c c|c c}
\toprule
\toprule
& \multicolumn{4}{c||}{\textit{Case I}  \eqref{eq:ztsslove}, \eqref{eq:tsolve1} (\textit{Unit is} $[\textrm{m}]$)} & \multicolumn{6}{c||}{\textit{Case II} with $z_{\tq}\geq d_{\mathsf{PA}}$ \eqref{eq:34a}, \eqref{eq:34b}, \eqref{eq:tsolve1} (\textit{Unit is} $[\textrm{m}]$)}& \multicolumn{8}{c}{\textit{Case II} with $d_{\mathsf{SC}}\leq z_{\tq}<d_{\mathsf{PA}}$ \eqref{eq:aaaa}, \eqref{eq:aa}, \eqref{eq:tsolve1} (\textit{Unit is} $[\textrm{m}]$)}\\
\cline{2-7}
\cline{8-13}
\cline{14-19}
& \multicolumn{2}{c|}{$\lambda=1$} & \multicolumn{2}{c||}{$\lambda=0.5$}& \multicolumn{2}{c|}{$\lambda=0.1$}& \multicolumn{2}{c|}{$\lambda=0.1$} & \multicolumn{2}{c||}{$\lambda=0.01$}& \multicolumn{2}{c|}{$\lambda=0.1$}& \multicolumn{2}{c|}{$\lambda=0.1$} & \multicolumn{2}{c|}{$\lambda=0.01$} & \multicolumn{2}{c}{$\lambda=0.01$}\\
& \multicolumn{2}{c|}{$D_{\rq}=0.5$} & \multicolumn{2}{c||}{$D_{\rq}=1$} & \multicolumn{2}{c|}{$D_{\rq}=1$}& \multicolumn{2}{c|}{$D_{\rq}=2$} & \multicolumn{2}{c||}{$D_{\rq}=2$}& \multicolumn{2}{c|}{$D_{\rq}=1$} & \multicolumn{2}{c|}{$D_{\rq}=1$}& \multicolumn{2}{c|}{$D_{\rq}=2$} & \multicolumn{2}{c}{$D_{\rq}=2$}\\
& \multicolumn{2}{c|}{$l_{\mathsf{s}}=0.05$} &\multicolumn{2}{c||}{$l_{\mathsf{s}}=0.05$}& \multicolumn{2}{c|}{$l_{\mathsf{s}}=0.05$} & \multicolumn{2}{c|}{$l_{\mathsf{s}}=0.05$}&\multicolumn{2}{c||}{$l_{\mathsf{s}}=0.05$}& \multicolumn{2}{c|}{$l_{\mathsf{s}}=0.05$} & \multicolumn{2}{c|}{$l_{\mathsf{s}}=0.1$} & \multicolumn{2}{c|}{$l_{\mathsf{s}}=0.05$}& \multicolumn{2}{c}{$l_{\mathsf{s}}=0.005$}\\
& \multicolumn{2}{c|}{$y_{\mathsf{r};{1}}=0.025$} & \multicolumn{2}{c||}{$y_{\mathsf{r};{1}}=0.025$}& \multicolumn{2}{c|}{$y_{\mathsf{r};{1}}=0.025$} & \multicolumn{2}{c|}{$y_{\mathsf{r};{1}}=0.025$}& \multicolumn{2}{c||}{$y_{\mathsf{r};{1}}=0.025$}&  \multicolumn{2}{c|}{$y_{\mathsf{r};{1}}=0.025$} &  \multicolumn{2}{c|}{$y_{\mathsf{r};{1}}=0.05$}&  \multicolumn{2}{c|}{$y_{\mathsf{r};{1}}=0.025$} &  \multicolumn{2}{c}{$y_{\mathsf{r};{1}}=0.0025$}\\
& \multicolumn{2}{c|}{$y_{\mathsf{r};{\mathsf{N}}/2}=0.225$} & \multicolumn{2}{c||}{$y_{\mathsf{r};{\mathsf{N}}/2}=0.475$} & \multicolumn{2}{c|}{$y_{\mathsf{r};{\mathsf{N}}/2}=0.475$}& \multicolumn{2}{c|}{$y_{\mathsf{r};{\mathsf{N}}/2}=0.975$}& \multicolumn{2}{c||}{$y_{\mathsf{r};{\mathsf{N}}/2}=0.975$}& \multicolumn{2}{c|}{$y_{\mathsf{r};{2}}=0.075$} & \multicolumn{2}{c|}{$y_{\mathsf{r};{2}}=0.15$} & \multicolumn{2}{c|}{$y_{\mathsf{r};{2}}=0.075$}& \multicolumn{2}{c}{$y_{\mathsf{r};{2}}=0.0075$}\\
& \multicolumn{2}{c|}{$\mathsf{H}_1=0.5$} & \multicolumn{2}{c||}{$\mathsf{H}_1=0.1$}& \multicolumn{2}{c|}{$\mathsf{H}_1=5$} & \multicolumn{2}{c|}{$\mathsf{H}_1=20$}& \multicolumn{2}{c||}{$\mathsf{H}_1=200$}& \multicolumn{2}{c|}{$\mathsf{H}_1=0.18$} & \multicolumn{2}{c|}{$\mathsf{H}_1=0.36$} & \multicolumn{2}{c|}{$\mathsf{H}_1=0.25$}& \multicolumn{2}{c}{$\mathsf{H}_1=0.018$} \\
& \multicolumn{2}{c|}{$\mathsf{H}_2=0.9$} & \multicolumn{2}{c||}{$\mathsf{H}_2=0.45$} & \multicolumn{2}{c|}{$\mathsf{H}_2=20$}& \multicolumn{2}{c|}{$\mathsf{H}_2=80$} & \multicolumn{2}{c||}{$\mathsf{H}_2=400$}& \multicolumn{2}{c|}{$\mathsf{H}_2=4.98$} & \multicolumn{2}{c|}{$\mathsf{H}_2=4.98$} & \multicolumn{2}{c|}{$\mathsf{H}_2=199$}& \multicolumn{2}{c}{$\mathsf{H}_2=199$}\\
& \multicolumn{2}{c|}{$d_{\mathsf{PA}}= 0.125$ \eqref{eq:dPA}} & \multicolumn{2}{c||}{$d_{\mathsf{PA}}= 1$} & \multicolumn{2}{c|}{$d_{\mathsf{PA}}= 5$} & \multicolumn{2}{c|}{$d_{\mathsf{PA}}= 20$}& \multicolumn{2}{c||}{$d_{\mathsf{PA}}= 200$}& \multicolumn{2}{c|}{$d_{\mathsf{PA}}=5$} & \multicolumn{2}{c|}{$d_{\mathsf{PA}}=5$} & \multicolumn{2}{c|}{$d_{\mathsf{PA}}=200$}& \multicolumn{2}{c}{$d_{\mathsf{PA}}=200$}\\
& \multicolumn{2}{c|}{$d_{\mathsf{F}}= 0.5$ \eqref{eq:dPA}} & \multicolumn{2}{c||}{$d_{\mathsf{F}}= 4$} & \multicolumn{2}{c|}{$d_{\mathsf{F}}= 20$} & \multicolumn{2}{c|}{$d_{\mathsf{F}}= 80$}& \multicolumn{2}{c||}{$d_{\mathsf{F}}= 800$}& \multicolumn{2}{c|}{$d_{\mathsf{SC}}=0.18$ \eqref{eq:dSC}} & \multicolumn{2}{c|}{$d_{\mathsf{SC}}=0.36$} & \multicolumn{2}{c|}{$d_{\mathsf{SC}}=0.25$}& \multicolumn{2}{c}{$d_{\mathsf{SC}}=0.018$} \\
\midrule
\midrule
$\mathrm{RMSE}\left(z_{\tq}\right)~[\textrm{m}]$ (\Checkmark)&\multicolumn{2}{c|}{$0$} &\multicolumn{2}{c||}{$0$}&\multicolumn{2}{c|}{$5.70\times10^{-3}$}&\multicolumn{2}{c|}{$5.90\times10^{-3}$}&\multicolumn{2}{c||}{$8.41\times10^{-4}$} &\multicolumn{2}{c|}{$1.60\times10^{-3}$} &\multicolumn{2}{c|}{$4.60\times10^{-3}$} &\multicolumn{2}{c|}{$2.22\times10^{-4}$}&\multicolumn{2}{c}{$2.19\times10^{-5}$}\\
{$\mathrm{RMSE}_{\mathrm{SIM}}\left(z_{\tq}\right)~[\textrm{m}]$ (\Checkmark)}&\multicolumn{2}{c|}{{$0$}}&\multicolumn{2}{c||}{{$0$}} &\multicolumn{2}{c|}{{$5.70\times 10^{-3}$}}&\multicolumn{2}{c|}{{$5.90\times 10^{-3}$}}  &\multicolumn{2}{c||}{{$8.41\times 10^{-4}$}} &\multicolumn{2}{c|}{{$1.60\times 10^{-3}$}} &\multicolumn{2}{c|}{{$4.60\times 10^{-3}$}} &\multicolumn{2}{c|}{{$2.22\times 10^{-4}$}}& \multicolumn{2}{c}{{$2.19\times 10^{-5}$}}\\
$\mathrm{RMSE}_{\mathrm{mis}}\left(z_{\tq}\right)~[\textrm{m}]$ (\XSolidBrush)&\multicolumn{2}{c|}{$-$}&\multicolumn{2}{c||}{$-$} &\multicolumn{2}{c|}{$13.2-\imagunit0.0236$}&\multicolumn{2}{c|}{$52.9-\imagunit0.0236$}  &\multicolumn{2}{c||}{$306-\imagunit0.0245$} &\multicolumn{2}{c|}{$33.1$} &\multicolumn{2}{c|}{$27.2$} &\multicolumn{2}{c|}{$2.74\times 10^{3}$}& \multicolumn{2}{c}{$2.19\times10^{3}$}\\
\midrule
$\mathrm{RMSE}\left(t_{z}\right)~[1]$ (\Checkmark) &\multicolumn{2}{c|}{$0$}&\multicolumn{2}{c||}{$0$} &\multicolumn{2}{c|}{$7.72\times10^{-4}$} &\multicolumn{2}{c|}{$2.15\times10^{-4}$}&\multicolumn{2}{c||}{$3.66\times10^{-6}$}& \multicolumn{2}{c|}{$3.80\times10^{-3}$}& \multicolumn{2}{c|}{$5.50\times10^{-3}$} & \multicolumn{2}{c|}{$3.99\times10^{-4}$}& \multicolumn{2}{c}{$7.29\times10^{-4}$}\\
{$\mathrm{RMSE}_{\mathrm{SIM}}\left(t_z\right)~[1]$ (\Checkmark)} &\multicolumn{2}{c|}{{$3.57\times 10^{-2}$}}&\multicolumn{2}{c||}{{$9.85\times 10^{-2}$}} &\multicolumn{2}{c|}{{$7.74\times 10^{-4}$}} &\multicolumn{2}{c|}{{$2.15\times 10^{-4}$}}&\multicolumn{2}{c||}{{$3.66\times 10^{-6}$}} &\multicolumn{2}{c|}{{$4.30\times 10^{-3}$}} &\multicolumn{2}{c|}{{$5.60\times 10^{-3}$}}  &\multicolumn{2}{c|}{{$4.00\times 10^{-4}$}}& \multicolumn{2}{c}{{$8.22\times 10^{-4}$}}\\
$\mathrm{RMSE}_{\mathrm{mis}}\left(t_z\right)~[1]$ (\XSolidBrush) &\multicolumn{2}{c|}{$-$}&\multicolumn{2}{c||}{$-$} &\multicolumn{2}{c|}{$0.440+\imagunit0.0686$} &\multicolumn{2}{c|}{$0.469+\imagunit 0.0529$}&\multicolumn{2}{c||}{$0.525+\imagunit 0.00910$} &\multicolumn{2}{c|}{$1.82\times10^{4}$} &\multicolumn{2}{c|}{$1.36\times10^{4}$}  &\multicolumn{2}{c|}{$1.14\times10^{7}$}& \multicolumn{2}{c}{$4.85\times10^{4}$}\\
\bottomrule
\bottomrule
\end{tabular}
}
\end{table*}

\section{Numerical Results and Discussions}\label{sec:simulation}
In this section, we will provide numerical results to verify the analytical results. First, we will analyze the closed-form solutions derived in Section \ref{sec:method}. Then, we will evaluate the ZZB and ECRB given in Section \ref{sec:ZZB} and \ref{sec:ECRB} to show the effect of the various system parameters on the joint estimation performance. Additionally, the MSE of the MAP estimator is given for benchmarking.
\subsection{Closed-form Solutions for Joint estimation}
In order to evaluate the accuracy of the closed-form solutions, we compute the root mean squared error (RMSE), i.e., $\mathrm{RMSE}\left(z_{\tq}\right)= \sqrt{\frac{1}{\mathsf{UV}}\sum_{u=1}^{\mathsf{U}}\sum_{v=1}^{\mathsf{V}}\left(z_{\tq;u}-\mathsf{z}_{{\chi};u,v}\right)^2}$,
where $z_{\tq;u}$ is the $u$-th uniform {sampling point} in \textit{prior} region $[\mathsf{H}_{1}, \mathsf{H}_{2}]$, $\mathsf{z}_{{\chi};u,v}$ is the closed-form solution provided in \eqref{eq:ztsslove}, \eqref{eq:34a}, \eqref{eq:34b}, \eqref{eq:aaaa}, \eqref{eq:aa} corresponding to different conditions ($\chi=\textit{I},\textit{II}1,\textit{II}2$), the $u$-th uniform sampling point of $z_{\tq}$, and the $v$-th uniform sampling point of $t_{z}$, $\mathsf{U}$ and $\mathsf{V}$ are the number of sampling points, which are set to $\mathsf{U}=10,000$ and $\mathsf{V}=10,000$. Similarly, we define $\mathrm{RMSE}\left(t_{z}\right)$, omitted here due to space constraints. $\mathrm{RMSE}\left(z_{\tq}\right)$ is measured in meters $[\textrm{m}]$ and $\mathrm{RMSE}\left(t_{z}\right)$ is dimensionless whose unit is the constant $[1]$. 

{Then, we study the accuracy of closed-form solutions for the EM-SIMP channel. We define the simulated RMSE as $\mathrm{RMSE}_{\mathrm{SIM}}\left(z_{\tq}\right)= \sqrt{\frac{1}{\mathsf{UV}}\sum_{u=1}^{\mathsf{U}}\sum_{v=1}^{\mathsf{V}}\left(z_{\tq;u}-\widetilde{\mathsf{z}}_{{\chi};u,v}\right)^2}$. $\widetilde{\mathsf{z}}_{{\chi};u,v}$ is the simulated closed-form solution for $z_{\tq}$, where the known quantities $\Theta_{\kappa}$ and $\Psi_{\kappa}$ in \eqref{eq:ztsslove}, \eqref{eq:34a}, \eqref{eq:34b}, \eqref{eq:aaaa}, and \eqref{eq:aa} are generated by $h_{\mathrm{SIM}}\left(\bm{\xi};x_{\rq},y_{\rq}\right)$ in \textit{Discussion} \ref{disminor1}.} 

Further, to illustrate the importance of dividing the distance domain of the UE based on the wavelength $\lambda$, \textit{Phase ambiguity distance} $d_{\mathsf{PA}}$ in \eqref{eq:dPA}, and \textit{Spacing constraint distance} $d_{\mathsf{SC}}$ in \eqref{eq:dSC}, we define the mismatched RMSE as $\mathrm{RMSE}_{\mathrm{mis}}\left(t_z\right)= \sqrt{\frac{1}{\mathsf{UV}}\sum_{u=1}^{\mathsf{U}}\sum_{v=1}^{\mathsf{V}}\left(t_{z;v}-\breve{\mathsf{t}}_{{\chi};u,v}\right)^2}$, where $\breve{\mathsf{t}}_{{\chi};u,v}$ is the mismatched closed-form solution for $t_{z}$. In particular, for \textit{Case II} with $z_{\tq}\geq d_{\mathsf{PA}}$, we use the solutions of \textit{Case I} in \eqref{eq:ztsslove} and \eqref{eq:tsolve1}, and for \textit{Case II} with $d_{\mathsf{SC}}\leq z_{\tq}<d_{\mathsf{PA}}$, we utilize the solutions of \textit{Case II} with $z_{\tq}\geq d_{\mathsf{PA}}$ in \eqref{eq:34a}, \eqref{eq:34b}, and \eqref{eq:tsolve1}.

Extensive numerical evaluation results are provided in Table \ref{tab:performance}. It can be observed that the RMSEs of the closed-form solutions are close to $0$ in all cases, which reveals the accuracy of our closed-form solutions. The reason why the RMSEs of \textit{Case II} are not $0$ is that the Taylor approximation in \eqref{eq:OO} and \eqref{eq:OOO} introduces a tolerable error. {Besides, $\mathrm{RMSE}_{\mathrm{SIM}}\left(z_{\tq}\right)$ and $\mathrm{RMSE}\left(z_{\tq}\right)$ are equal since the phase of the NF-EM channel is completely accurate. $\mathrm{RMSE}_{\mathrm{SIM}}\left(t_z\right)$ has a very tiny increment relative to $\mathrm{RMSE}\left(t_z\right)$. Thus, our closed-form solutions are also accurate for the
electromagnetic simulation platform channel.} Moreover, it can be observed that the mismatched RMSEs of \textit{Case II} are abnormal complex numbers or very large real values. This is because the mismatch will make the value inside the root of \eqref{eq:ztsslove} negative or $\mathsf{N}_{\beta}-\mathsf{N}_{\alpha}\gg1$. Thus, it is important to finely divide the distance domain of the UE based on $\lambda$, $d_{\mathsf{PA}}$, and $d_{\mathsf{SC}}$.

\subsection{ZZB and ECRB for Joint estimation}
Then, we numerically evaluate the ZZBs given in  \eqref{eq:ZZBz} and \eqref{eq:ZZBtz}, the ECRBs given in \eqref{eq:ECRBYT} and \eqref{eq:ECRBT}.
\subsubsection{Impact of SNR}
\begin{figure}
\centering
\includegraphics[scale=0.44]{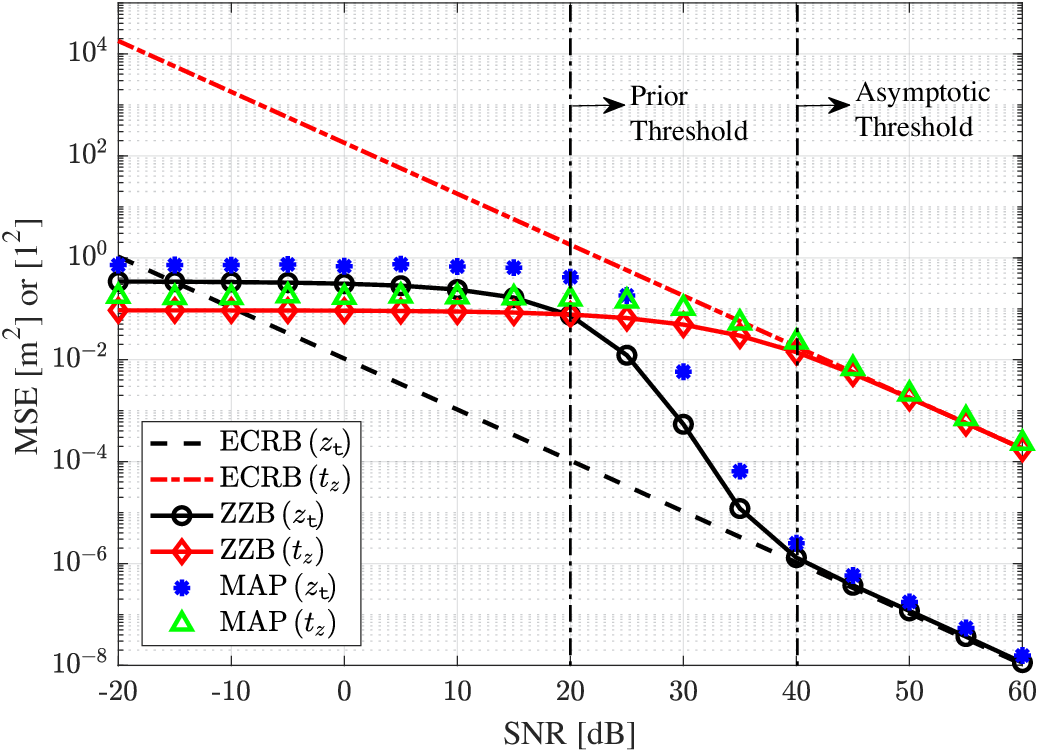}
\caption{MSEs versus SNR  for the joint estimates of $z_{\tq}$ and $t_z$, with $z_{\tq} \sim \mathcal{U}[3~\textrm{m}, 5~\textrm{m}]$, $t_{z} \sim \mathcal{U}[0, 1)$, $\lambda=0.1~\textrm{m}$, $D_{\rq}=5~\textrm{m}$, and $l_{\mathsf{s}}=0.1~\textrm{m}$. $\mathrm{ZZB}\left(z_{\tq}\right)$ and $\mathrm{ZZB}\left(t_{z}\right)$ are computed based on \eqref{eq:ZZBz} and \eqref{eq:ZZBtz}, $\mathrm{ECRB}\left(z_{\tq}\right)$ and $\mathrm{ECRB}\left(t_{z}\right)$ are computed based on \eqref{eq:ECRBYT} and \eqref{eq:ECRBT}, and $\mathrm{MAP}\left(z_{\tq}\right)$ and $\mathrm{MAP}\left(t_{z}\right)$ are obtained from Monte Carlo simulations.}
\label{fig:sim1}
\end{figure}

Fig. \ref{fig:sim1}  performs a simulation to illustrate the ZZBs, ECRBs, and MSEs of the MAP
estimates obtained from Monte Carlo simulations with
respect to the SNR when $z_{\tq} \sim \mathcal{U}[3~\textrm{m}, 5~\textrm{m}]$, $t_{z} \sim \mathcal{U}[0, 1)$, $\lambda=0.1~\textrm{m}$ (i.e., $f=3~\textrm{GHz}$), $D_{\rq}=5~\textrm{m}$, and $l_{\mathsf{s}}=0.1~\textrm{m}$. The following interesting conclusions
can be drawn from the results in Fig. \ref{fig:sim1}:
\begin{itemize}[leftmargin=*]
    \item \textit{The relationship between ZZB, ECRB, and MSE:} The MSEs of the MAP estimators are globally bounded by the corresponding ZZBs. Since the
ECRBs are \textit{locally tight} bounds while the ZZBs are \textit{globally tight} bounds, it is
not surprising that the ZZBs provide bounds at least as tight or
tighter than the corresponding ECRBs in all SNR regimes.
\item \textit{The asymptotic threshold and prior
threshold:} Fig. \ref{fig:sim1} shows that the prior threshold is $20~\textrm{dB}$ and asymptotic threshold is $40~\textrm{dB}$. When the SNR is less than the prior threshold (\textit{prior region}), the ZZBs are dominated by prior information and approach the prior variances (see \textit{Corollary} \ref{coro:ZZBCase3}), which can be achieved by ignoring the observations and estimating by the prior mean. When the SNR is greater than the asymptotic threshold (\textit{asymptotic region}), the ZZBs are consistent with the ECRBs and dominated by the inverse of the Fisher information. When the SNR is between these two thresholds (\textit{transition region}), the role of observation becomes larger, making ZZB and ECRB gradually approach each other.
\item {\textit{The difference between the curves of $\mathrm{ZZB}\left(z_{\tq}\right)$ and $\mathrm{ZZB}\left(t_{z}\right)$:} For the curve of $\mathrm{ZZB}\left(t_z\right)$, the absolute value of its slope gradually enlarges in the \textit{transition region} and subsequently stabilizes in the \textit{asymptotic region}. For the curve of $\mathrm{ZZB}\left(z_{\mathsf{t}}\right)$, the absolute value of its slope experiences a gradual increase, followed by a decrease, ultimately reaching a constant value. In fact, the physical parameter settings of the sensing system cause the distinct characteristics of these two curves, which
will lead to $\mathrm{ECRB}\left(z_{\mathsf{t}}\right)$ and $\mathrm{ECRB}\left(t_z\right)$ having different \textit{deviations} (manifesting as different positive and negative signs, along with diverse absolute values) relative to the corresponding ZZBs in the \textit{prior region} as the ECRBs exhibit linear changes according to the slopes of the ZZBs in the \textit{asymptotic region}. This also reveals that the ECRBs cannot provide effective performance predictions outside the \textit{asymptotic region}.} 
\item \textit{The accuracy of the joint estimation:} When the receiving linear array size is several meters, the frequency is around $3~\textrm{GHz}$, and the SNR is high, the position estimation can reach millimeter-level accuracy and 0.1-level accuracy can be achieved for the {attitude} estimation.
\end{itemize}
\begin{figure}
\centering
\includegraphics[scale=0.44]{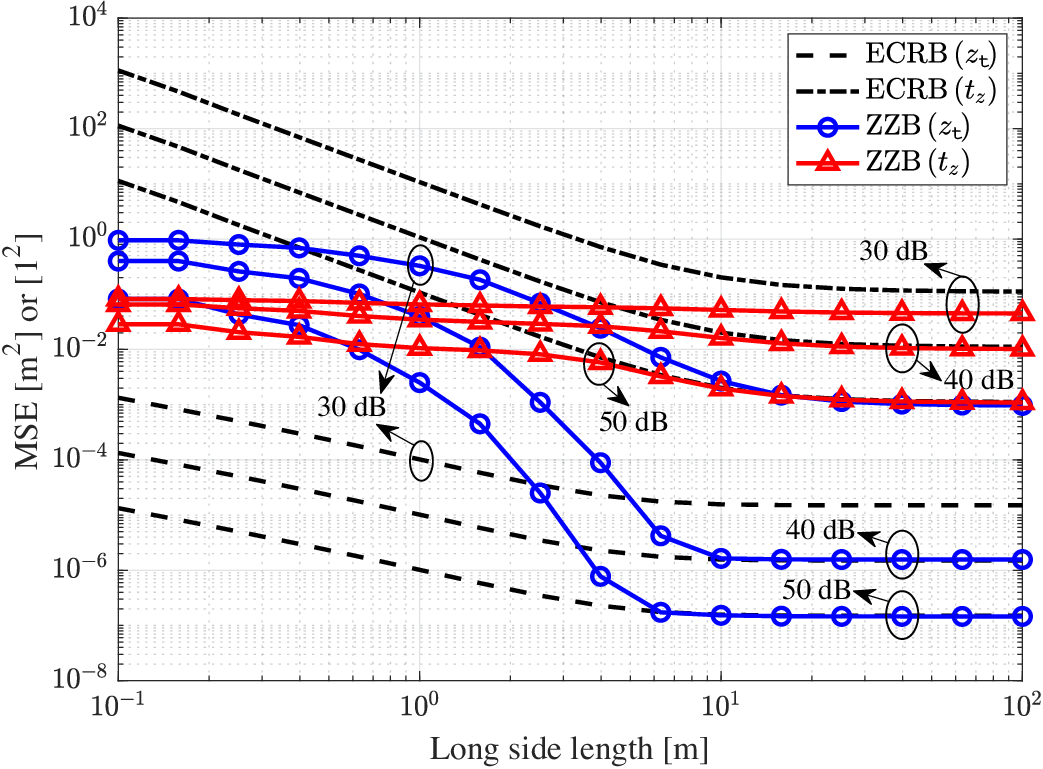}
\caption{ECRBs and ZZBs versus the long side length $D_{\rq}$  for the joint estimates of $z_{\tq}$ and $t_z$, with $z_{\tq} \sim \mathcal{U}[4~\textrm{m}, 8~\textrm{m}]$, $t_{z} \sim \mathcal{U}[0, 1)$, $\lambda=0.1~\textrm{m}$, $l_{\mathsf{s}}=0.1~\textrm{m}$, and $\textrm{SNR}=30~\textrm{dB}$, $40~\textrm{dB}$, or $50~\textrm{dB}$.}
\label{fig:sim2}
\end{figure}\subsubsection{Impact of the Long Side Length \texorpdfstring{$D_{\rq}$}{zt}}
Fig. \ref{fig:sim2}  illustrates the ECRBs and ZZBs versus the long side length $D_{\rq}$ when $z_{\tq} \sim \mathcal{U}[4~\textrm{m}, 8~\textrm{m}]$, $t_{z} \sim \mathcal{U}[0, 1)$, $\lambda=0.1~\textrm{m}$, $l_{\mathsf{s}}=0.1~\textrm{m}$, and $\textrm{SNR}=30~\textrm{dB}$, $40~\textrm{dB}$, or $50~\textrm{dB}$. It can be seen that all the ECRBs and ZZBs decrease to varying degrees with the increase of $D_{\rq}$. Specifically, we have the following insights:
\begin{itemize}[leftmargin=*]
    \item \textit{The ECRBs versus $D_{\rq}$:} All the ECRBs decrease dramatically as $D_{\rq}$ increases in $0.1~\textrm{m} \sim  10~\textrm{m}$ and they approach the asymptotic limits from $D_{\rq}\approx 10~\textrm{m}$. The reason is that an increase in $D_{\rq}$ will greatly facilitate the performance of the observation for values
of $D_{\rq}$ of practical interest, while this increase no longer enhances the observation if $D_{\rq}$ exceeds the critical point $10~\textrm{m}$.
\item \textit{The ZZBs versus $D_{\rq}$:} The ZZB for $t_z$ decreases slowly with the increase of $D_{\rq}$. This indicates that increasing $D_{\rq}$ has a limited effect on improving the performance of estimating $t_z$. Besides, the ZZB for $z_{\tq}$ decreases rapidly as $D_{\rq}$ increases in the range $1~\textrm{m} \sim  10~\textrm{m}$, since observations greatly improve the performance of estimating $z_{\tq}$. In addition, we also find that $\mathrm{ZZB}\left(z_{\tq}\right)$ decreases slowly or approaches the limit value when $D_{\rq}$ is less than $1~\textrm{m}$ or greater than $10~\textrm{m}$.
\item \textit{Convergence of ZZB and ECRB:} When SNR is $30~\textrm{dB}$, the ZZBs fail to converge to the corresponding ECRBs, even though the differences between them gradually decrease as $D_{\rq}$ increases. When SNR is $40~\textrm{dB}$ and $D_{\rq}$ is greater than $10~\textrm{m}$, or SNR is $50~\textrm{dB}$ and $D_{\rq}$ is greater than $6~\textrm{m}$, the ZZBs converge to the corresponding ECRBs. This reveals that for high SNR and satisfactory observation (the long side length $D_{\rq}$ of the observation region is in the range of several meters), the ZZB converges to the ECRB.
\end{itemize}
\begin{figure}
\centering
\includegraphics[scale=0.44]{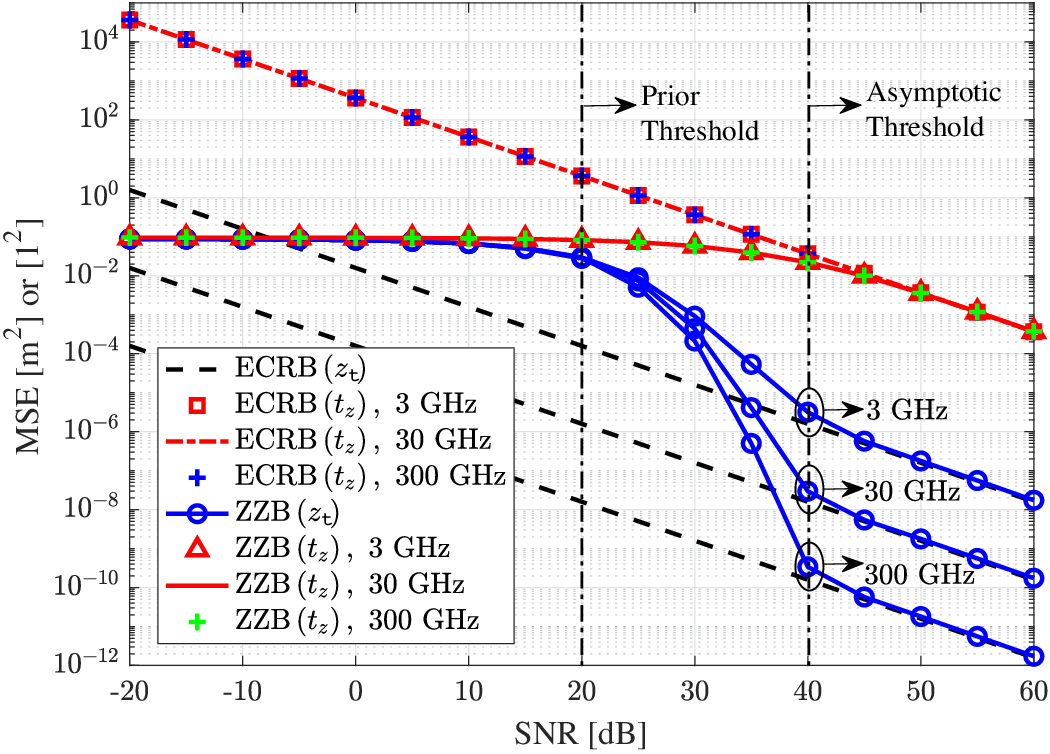}
\caption{ECRBs and ZZBs versus SNR for the joint estimates of $z_{\tq}$ and $t_z$, with $z_{\tq} \sim \mathcal{U}[5~\textrm{m}, 6~\textrm{m}]$, $t_{z} \sim \mathcal{U}[0, 1)$, $D_{\rq}=5~\textrm{m}$, $l_{\mathsf{s}}=0.1~\textrm{m}$, and three different carrier frequencies $f=3~\textrm{GHz}$,~$30~\textrm{GHz}$,~$\textrm{or}~300~\textrm{GHz}$ (i.e., $\lambda=0.1~\textrm{m}$,~$0.01~\textrm{m}$,~$\textrm{or}~0.001~\textrm{m}$).}
\label{fig:sim3}
\end{figure}
\subsubsection{Impact of Carrier Frequency}
Fig. \ref{fig:sim3} shows the ECRBs and ZZBs as the functions of the SNR for three different values of the wavelength, namely, $\lambda=0.1~\textrm{m}$ (corresponding to
$f = 3~\textrm{GHz}$), $\lambda=0.01~\textrm{m}$ (i.e., $f = 30~\textrm{GHz}$), and $\lambda=0.001~\textrm{m}$ (i.e., $f = 300~\textrm{GHz}$) with $z_{\tq} \sim \mathcal{U}[5~\textrm{m}, 6~\textrm{m}]$, $t_{z} \sim \mathcal{U}[0, 1)$, $D_{\rq}=5~\textrm{m}$, and $l_{\mathsf{s}}=0.1~\textrm{m}$. As expected, the
estimation accuracy improves as the SNR increases, and the asymptotic threshold and prior
threshold appear at $40~\textrm{dB}$ and $20~\textrm{dB}$, respectively. Notice that the ECRB for $z_{\mathsf{t}}$ depends linearly on the square of the wavelength (i.e., $\lambda^2$). Indeed,
reducing the wavelength by a factor of $10$ reduces $\mathrm{ECRB}\left(z_{\tq}\right)$ by the factor of $10^2$. This can be derived analytically by considering the results in \textit{Corollary}~\ref{coro:ECRBCase2}. Besides, the ZZB for $z_{\mathsf{t}}$ also depends linearly on $\lambda^2$ when the SNR is greater than the asymptotic threshold. The same dependence on $\lambda^2$ no longer holds for $\mathrm{ZZB}\left(z_{\tq}\right)$ when the SNR is less than the asymptotic threshold. In particular, the three ZZBs for $z_{\tq}$ approach the prior variance of $z_\tq$ when the SNR is less than the prior
threshold, and the $\mathrm{ZZB}\left(z_{\tq}\right)$ with a higher carrier frequency will decrease faster to converge to the corresponding smaller ECRB when the SNR is in the \textit{transition
region}. Notice that both ECRB and ZZB for $t_z$ are independent of $\lambda$. This is because the phase term of the channel ${{h}}_{\mathsf{y}}\left(\bm{\xi};y_{\rq}\right)$ in
\eqref{eq:hy} contains only $z_{\tq}$, not $t_{z}$. 
\begin{figure}
\centering
\includegraphics[scale=0.44]{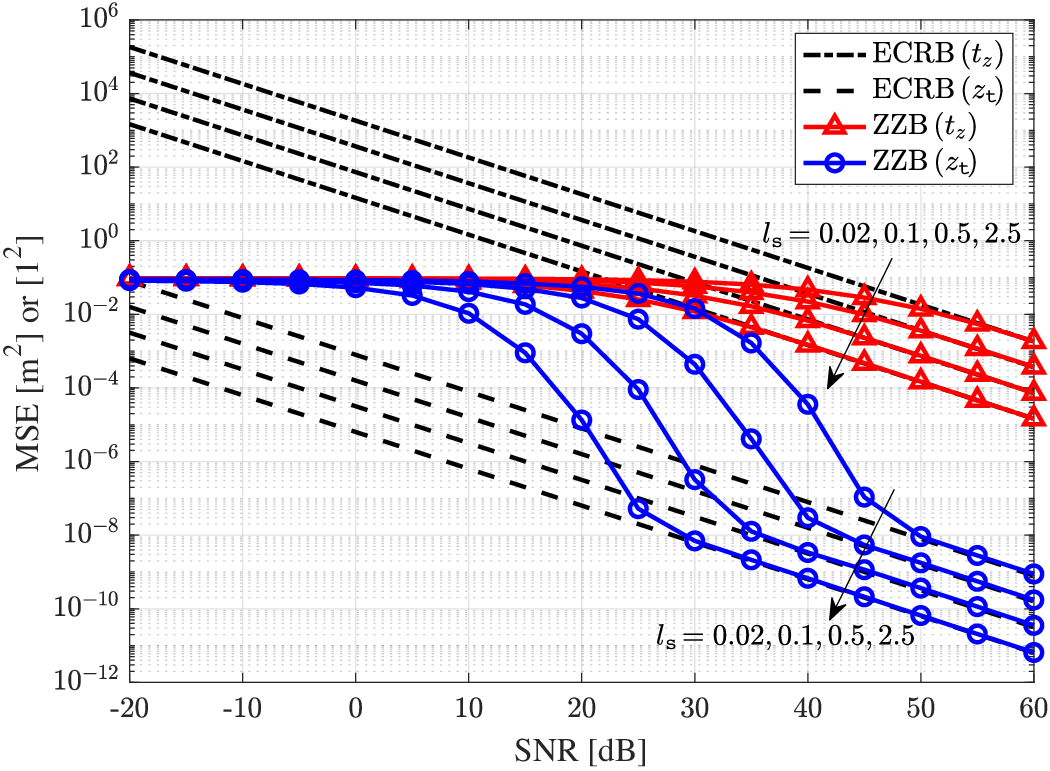}
\caption{ECRBs and ZZBs versus SNR for the joint estimates of $z_{\tq}$ and $t_z$, with $z_{\tq} \sim \mathcal{U}[5~\textrm{m}, 6~\textrm{m}]$, $t_{z} \sim \mathcal{U}[0, 1)$, $D_{\rq}=5~\textrm{m}$, $\lambda=0.01~\textrm{m}$, and four different short side lengths $l_{\mathsf{s}}=0.02~\textrm{m}$,~$0.1~\textrm{m}$,~$0.5~\textrm{m}$,~$\textrm{and}~2.5~\textrm{m}$.}
\label{fig:sim4}
\end{figure}
\begin{figure}[!t]
	\centering
	\subfloat[$\mathrm{ECRB}\left(z_{\tq} \right)$ versus $D_{\rq}$.]{
	\includegraphics[scale=0.44]{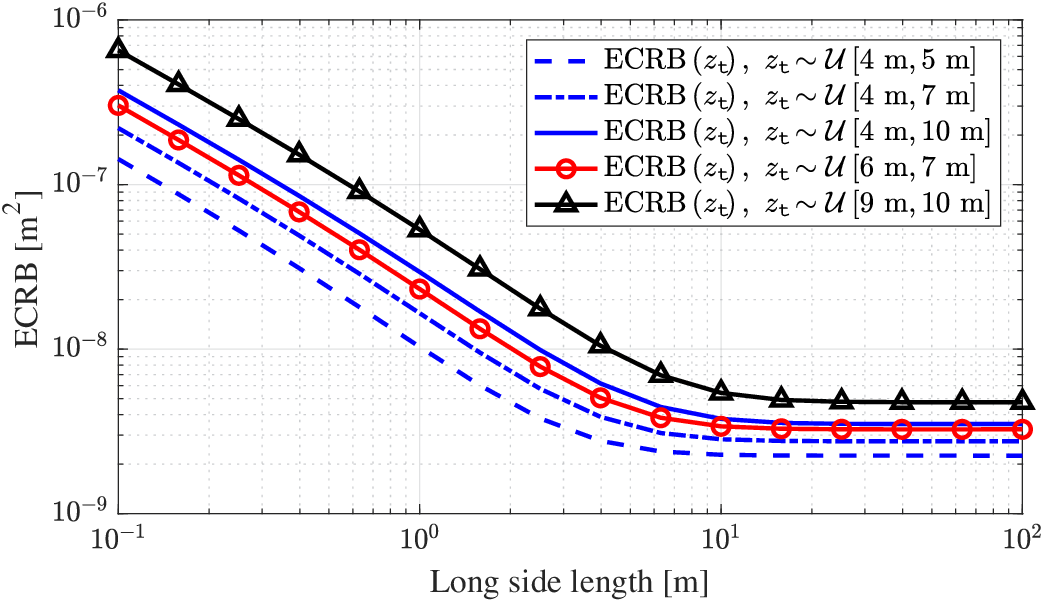}
		\label{fig:sim51}
	}
	\vfill
	\subfloat[$\mathrm{ECRB}\left(t_z\right)$ versus $D_{\rq}$.]{
		\includegraphics[scale=0.44]{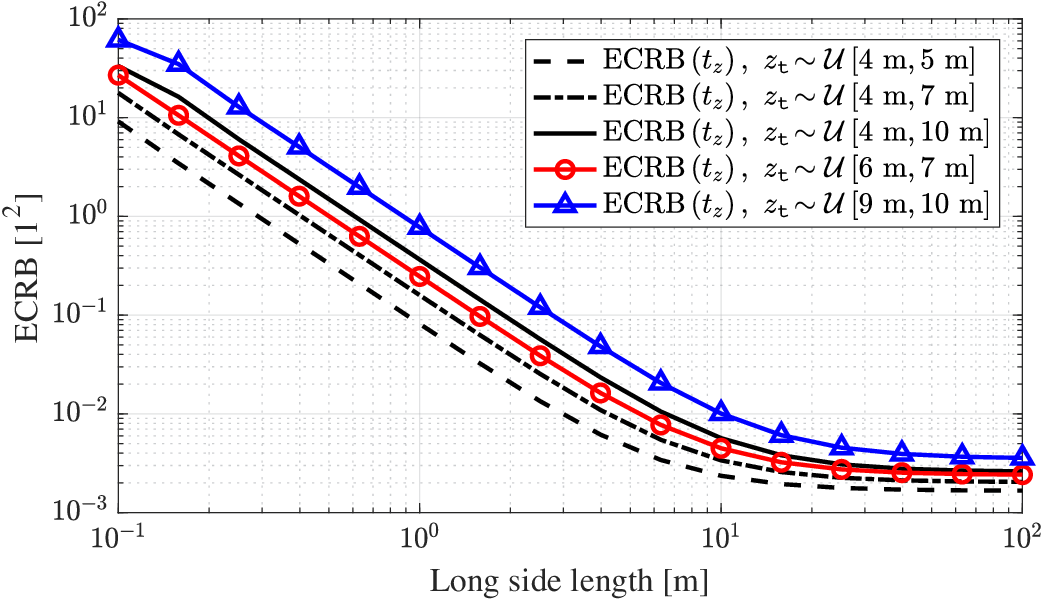}
		\label{fig:sim52}
	}
 \caption{ECRBs versus the long side length $D_{\rq}$  for the joint estimates of $z_{\tq}$ and $t_z$, with $t_{z} \sim \mathcal{U}[0, 1)$, $\textrm{SNR}=40~\textrm{dB}$, $\lambda=0.01~\textrm{m}$, $l_{\mathsf{s}}=0.5~\textrm{m}$, and five different prior distributions for $z_{\tq}$, i.e., $z_{\tq} \sim \mathcal{U}[4~\textrm{m}, 5~\textrm{m}]$, $z_{\tq} \sim \mathcal{U}[4~\textrm{m}, 7~\textrm{m}]$, $z_{\tq} \sim \mathcal{U}[4~\textrm{m}, 10~\textrm{m}]$, $z_{\tq} \sim \mathcal{U}[6~\textrm{m}, 7~\textrm{m}]$, and $z_{\tq} \sim \mathcal{U}[9~\textrm{m}, 10~\textrm{m}]$.}
   \label{fig:simo}
\end{figure}
\subsubsection{Impact of the Short Side Length \texorpdfstring{$l_{\mathsf{s}}$}{zt}}
Fig. \ref{fig:sim4}  illustrates the ECRBs and ZZBs versus the short side length $l_{\mathsf{s}}$ for four different values of $l_{\mathsf{s}}$, namely, $l_{\mathsf{s}}=0.02~\textrm{m}$, $0.1~\textrm{m}$, $0.5~\textrm{m}$, and $2.5~\textrm{m}$ with $z_{\tq} \sim \mathcal{U}[5~\textrm{m}, 6~\textrm{m}]$, $t_{z} \sim \mathcal{U}[0, 1)$, $D_{\rq}=5~\textrm{m}$, and $\lambda=0.01~\textrm{m}$. To illustrate the impact of the short side length $l_{\mathsf{s}}$ along the $\mathsf{X}$-dimension on the joint estimation performance, the interval $l_{\mathsf{s}}$ of the sub-elements along the $\mathsf{Y}$-dimension is still set to $l_{\mathsf{s}}=0.1~\textrm{m}$, and only the $l_{\mathsf{s}}$ along the $\mathsf{X}$-dimension is changed. It can be seen that the ECRBs for $z_{\mathsf{t}}$ and  $t_{z}$ depend linearly on the reciprocal of $l_{\mathsf{s}}$ (i.e., $l_{\mathsf{s}}^{-1}$). Indeed,
increasing $l_{\mathsf{s}}$ by a factor of $5$ reduces $\mathrm{ECRB}\left(z_{\tq}\right)$ and $\mathrm{ECRB}\left(t_z\right)$ of the same factor. Since the ZZBs converge to the corresponding ECRBs in the high SNR regions, $\mathrm{ZZB}\left(z_{\tq}\right)$ and $\mathrm{ZZB}\left(t_z\right)$ also depend linearly on the reciprocal of $l_{\mathsf{s}}$ (i.e., $l_{\mathsf{s}}^{-1}$) when the SNR is greater than the asymptotic threshold. In addition, Fig. \ref{fig:sim4} also illustrates that when $l_{\mathsf{s}}=0.02~\textrm{m}$, $0.1~\textrm{m}$, $0.5~\textrm{m}$, and $2.5~\textrm{m}$, the asymptotic threshold is around $47~\textrm{dB}$, $40~\textrm{dB}$, $33~\textrm{dB}$, and $26~\textrm{dB}$ respectively, and the prior
threshold is around $27~\textrm{dB}$, $20~\textrm{dB}$, $13~\textrm{dB}$, and  $6~\textrm{dB}$ respectively. Hence, we find that 
$l_{\mathsf{s}}$ affects the values of the asymptotic threshold and prior
threshold, and this effect is the same as the effect of SNR. Specifically, the larger $l_{\mathsf{s}}$ or SNR results in a smaller asymptotic threshold and prior
threshold. In fact, based on the results in \eqref{eq:mumu}, we know that the asymptotic threshold appears at $\text{SNR}l_{\mathsf{s}}\approx 10^3~\textrm{m}$ and the prior
threshold appears at $\text{SNR}l_{\mathsf{s}}\approx 10~\textrm{m}$.
\subsubsection{Impact of the 
Parameters \texorpdfstring{$\mathsf{H}_{1}$ and $\mathsf{H}_{2}$ of the Prior Region of $z_\tq$}{zt}}
Fig. \ref{fig:sim51} and Fig. \ref{fig:sim52} demonstrate the ECRBs versus the long side length $D_{\rq}$ for five different prior distributions for $z_{\tq}$, i.e., $z_{\tq} \sim \mathcal{U}[4~\textrm{m}, 5~\textrm{m}]$, $z_{\tq} \sim \mathcal{U}[4~\textrm{m}, 7~\textrm{m}]$, $z_{\tq} \sim \mathcal{U}[4~\textrm{m}, 10~\textrm{m}]$, $z_{\tq} \sim \mathcal{U}[6~\textrm{m}, 7~\textrm{m}]$, and $z_{\tq} \sim \mathcal{U}[9~\textrm{m}, 10~\textrm{m}]$ with $t_{z} \sim \mathcal{U}[0, 1)$, $\textrm{SNR}=40~\textrm{dB}$, $\lambda=0.01~\textrm{m}$, and $l_{\mathsf{s}}=0.5~\textrm{m}$. It is observed that both $\mathrm{ECRB}\left(z_{\tq}\right)$ and $\mathrm{ECRB}\left(t_z\right)$ become larger with the increase of $\mathsf{H}_{\tq}$ when $\mathsf{H}_1$ is fixed. This illustrates that increasing the range of the prior region of the position parameter $z_{\tq}$ will reduce the accuracy of the joint estimation. Besides, both $\mathrm{ECRB}\left(z_{\tq}\right)$ and $\mathrm{ECRB}\left(t_z\right)$ become larger as the prior mean of $z_{\tq}$ increases when $\mathsf{H}_{\tq}$ is fixed, which reveals that an increase in the position parameter $z_{\tq}$ will also reduce the accuracy.
\subsubsection{Joint vs. Attitude-Only Estimation}\label{sec:jointvsnu} 
\begin{figure}
\centering
\includegraphics[scale=0.44]{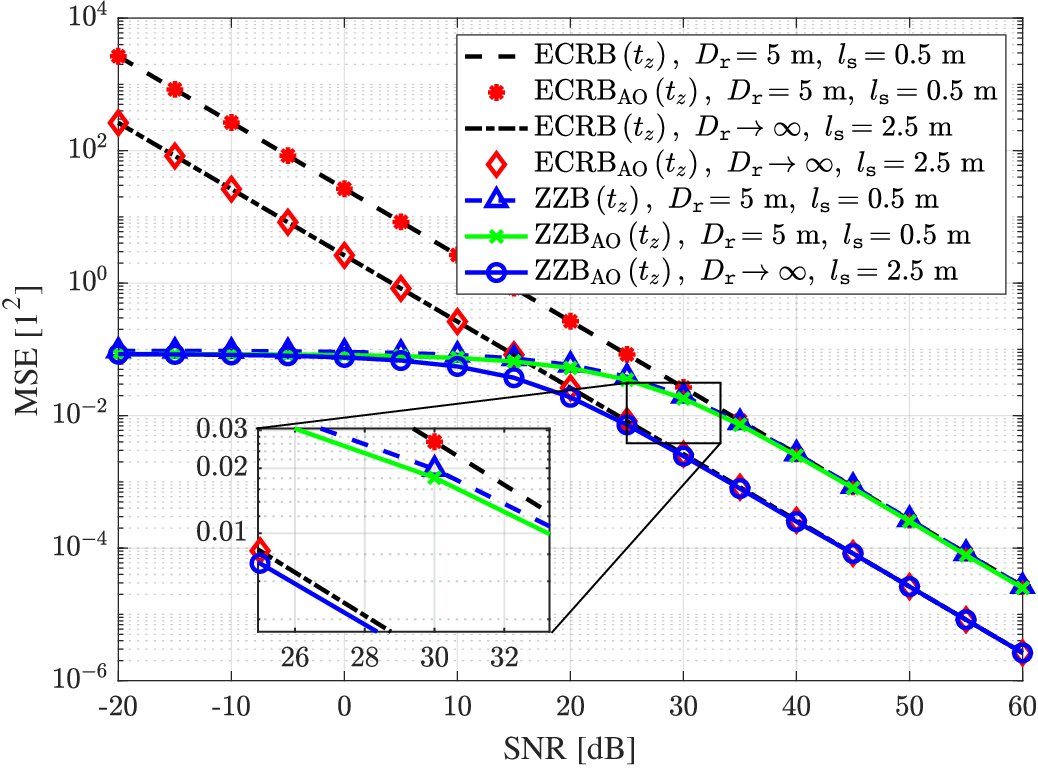}
\caption{ECRBs and ZZBs versus SNR for the joint estimation and attitude-only estimation, with $z_{\tq} \sim \mathcal{U}[3~\textrm{m}, 4~\textrm{m}]$, $t_{z} \sim \mathcal{U}[0, 1)$, $\lambda=0.1~\textrm{m}$, $D_{\rq}=5~\textrm{m}$ or $D_{\rq}\to \infty$, and $l_{\mathsf{s}}=0.5~\textrm{m}$ or $l_{\mathsf{s}}=2.5~\textrm{m}$. $\mathrm{ZZB}_{\mathrm{AO}}\left(t_z\right)$ is computed based on \eqref{eq:mulva}, \eqref{eq:mulvaclose}, \eqref{eq:mulvaclose11}. $\mathrm{ECRB}_{\mathrm{AO}}\left(t_z\right)$ is computed based on \eqref{eq:ECRBAO}, \eqref{eq:711}, \eqref{eq:Ittappclose}.}
\label{fig:sim8}
\end{figure}

Fig. \ref{fig:sim8} demonstrates the ECRBs and ZZBs versus the SNR for the joint estimation and attitude-only estimation, where we set $z_{\tq} \sim \mathcal{U}[3~\textrm{m}, 4~\textrm{m}]$, $t_{z} \sim \mathcal{U}[0, 1)$, $\lambda=0.1~\textrm{m}$, $D_{\rq}=5~\textrm{m}$ or $D_{\rq}\to \infty$, and $l_{\mathsf{s}}=0.5~\textrm{m}$ or $l_{\mathsf{s}}=2.5~\textrm{m}$. It is observed that 
the values of $\mathrm{ECRB}\left(t_z\right)$ and $\mathrm{ECRB}_{\mathrm{AO}}\left(t_z\right)$ are consistent with no visible difference. The reason is that $\mathcal{I}_{\mathsf{zz}}$ and $\mathcal{I}_{\mathsf{tt}}$ in \eqref{eq:ECRBYT} and \eqref{eq:ECRBT} are much larger than $\mathcal{I}_{\mathsf{zt}}$,
making the FIM approximate a diagonal matrix. In addition, we find that $\mathrm{ZZB}_{\mathrm{AO}}\left(t_z\right)$ is less than $\mathrm{ZZB}\left(t_z\right)$ and the difference between them is small enough to be neglected. It is worth mentioning is that \textit{Assumption}~\assumptionref{assum:4.1} allows us to focus on the original structure of the attitude-only estimation without being disturbed by the deterministic value of $z_\tq$, and we have proved in \textit{Corollary}~\ref{prop:singleVS} that $\mathrm{ZZB}_{\mathrm{AO}}\left(t_z\right)$ and $\mathrm{ECRB}_{\mathrm{AO}}\left(t_z\right)$ are the achievable lower bounds of $\mathrm{ZZB}\left(t_z\right)$ and $\mathrm{ECRB}\left(t_z\right)$, respectively. The results in Fig. \ref{fig:sim8} reveal that $\mathrm{ZZB}_{\mathrm{AO}}\left(t_z\right)$ derived in \eqref{eq:mulva}, \eqref{eq:mulvaclose} and \eqref{eq:mulvaclose11}, and $\mathrm{ECRB}_{\mathrm{AO}}\left(t_z\right)$ derived in \eqref{eq:ECRBAO}, \eqref{eq:711}, and \eqref{eq:Ittappclose} can be leveraged as the closed-form   approximations of $\mathrm{ZZB}\left(t_z\right)$ and $\mathrm{ECRB}\left(t_z\right)$, respectively.
\section{Conclusions}\label{sec:con}
In this paper, we have developed the \textit{electromagnetic propagation model} (EPM) to accurately
characterize the functional dependence of the observed signals on the position and attitude parameters of UE. Efficient methods have been proposed to jointly estimate the position and attitude of UE based on the EPM. Specifically, starting from the noise-free case, we have proposed the \textit{Phase ambiguity distance} and \textit{Spacing constraint distance}  to finely divide the distance domain of UE into three regions, and then provided the closed-form solutions to the position and attitude parameters for each region. To elucidate the impact of noise on joint estimation performance, we have derived the ZZB for the joint estimation for the first time. Further, we have given the closed-form expression of ECRB for comparison. We have revealed that the derived ZZB is tighter than the ECRB at low and moderate
SNR regimes and it converges to the ECRB at the high SNR regime. Besides, we have found that the {position} estimation can reach the millimeter-level accuracy and the 0.1-level accuracy can be achieved for the {attitude} estimation when the SNR is sufficiently large. The effects of different system configurations on performance have been demonstrated by numerical studies.

{We conclude the paper with some interesting directions to inspire future research. Firstly, further investigation into \textit{low-complexity algorithms} that approach the derived ZZBs and CRBs for near-field positioning and attitude sensing is imperative to fulfill the requirements of diverse real-time sensing applications. Secondly, although electromagnetic signals and sensing algorithms can be simulated and verified using electromagnetic-compliant models on common numerical computing platforms, true confidence in the technology is obtained through
\textit{real-world testbeds} and \textit{measurements}. This paper establishes a robust theoretical foundation and provides a clear evaluation baseline, setting the stage for the necessary real-world testing and validation that are to follow. Thirdly, for the near-field boundary, the \textit{Fraunhofer
distance} has established the boundary between the UPW and the SWM by making the phase difference no larger than $\pi/8$. Due to the accuracy and complexity of the EPM, establishing the boundary between the EPM and the SWM is necessary to meet the requirements of high-precision and low-complexity sensing. Finally, the \textit{implementation} and \textit{refinement} of near-field positioning and attitude sensing techniques across multiple sectors, including IoT environments like smart homes and factories, autonomous vehicle systems such as smart parking and roadblock detection, and sensing-assisted communication within mobile communications, warrant further investigation.}

\appendix
\subsection{\texorpdfstring{Proof of Eq.~\eqref{eq:dis1}}{zt} }\label{app:eq:dis1} 
{The description of the general system is given in \textit{Discussion} \ref{definition21} and Fig. \ref{systemapp}. Due to the existence of three components of $\hat{\mathbf{t}}$, \eqref{eq:exftheta}--\eqref{eq:ezftheta} are rewritten as
\begin{align}
&e_{x}\left({\mathbf{r}}_{\mathsf{ar}}\right)=\imagunit \mathcal{E}_{\mathsf{in}}\frac{{\Gamma}_1 t_{x}+\Gamma_2 t_{y}+\Gamma_3 t_{z}}{\|\mathbf{r}_{\mathsf{ar}}\|}\mathrm{e}^{\imagunit k \|\mathbf{r}_{\mathsf{ar}}\|}\label{eq:appa1},
\\
&e_{y}\left({\mathbf{r}}_{\mathsf{ar}}\right)=\imagunit \mathcal{E}_{\mathsf{in}}\frac{\Gamma_2 t_{x}+\Gamma_4 t_{y}+\Gamma_5 t_{z}}{\|\mathbf{r}_{\mathsf{ar}}\|}\mathrm{e}^{\imagunit k \|\mathbf{r}_{\mathsf{ar}}\|},\\
&e_{z}\left({\mathbf{r}}_{\mathsf{ar}}\right)=\imagunit \mathcal{E}_{\mathsf{in}}\frac{\Gamma_3 t_{x}+\Gamma_5 t_{y}+\Gamma_6 t_{z}}{\|\mathbf{r}_{\mathsf{ar}}\|}\mathrm{e}^{\imagunit k \|\mathbf{r}_{\mathsf{ar}}\|}\label{eq:appa2},
\end{align}
where ${\Gamma}_1\triangleq 1-\sin^{2}{\theta}\cos^{2}{\phi}$, ${\Gamma}_2\triangleq -\sin{\theta}\cos{\theta}\cos{\phi}$, ${\Gamma}_3\triangleq \sin^{2}{\theta}\sin{\phi}\cos{\phi}$, $\Gamma_4\triangleq 1-\cos^{2}{\theta}$, $\Gamma_5\triangleq \sin{\theta}\cos{\theta}\sin{\phi}$, and $\Gamma_6\triangleq 1-\sin^{2}{\theta}\sin^{2}{\phi}$. 
Since $\|\mathbf{r}_{\mathsf{ar}}\|=\sqrt{x_{\mathsf{r},\mathsf{t}}^2+y_{\mathsf{r},\mathsf{t}}^{2}+z_{\mathsf{t}}^{2}}$, $\cos \theta=\frac{y_{\mathsf{r},\mathsf{t}}}{\|\mathbf{r}_{\mathsf{ar}}\|}$, $\tan \phi=\frac{z_{\mathsf{t}}}{{x_{\mathsf{r},\mathsf{t}}}}$, $x_{\mathsf{r},\mathsf{t}}\triangleq x_{\mathsf{r}}-x_{\mathsf{t}}$, and $y_{\mathsf{r},\mathsf{t}}\triangleq y_{\mathsf{r}}-y_{\mathsf{t}}$, we have $\Gamma_1=1-\frac{x_{\rq,\tq}^{2}}{\|\mathbf{r}_{\mathsf{ar}}\|^{2}}$, $\Gamma_2=-\frac{{x_{\rq,\tq}y_{\rq,\tq}}}{\|\mathbf{r}_{\mathsf{ar}}\|^{2}}$, $\Gamma_3=\frac{x_{\rq,\tq}z_{\tq}}{\|\mathbf{r}_{\mathsf{ar}}\|^{2}}$, $\Gamma_4=1-\frac{y_{\rq,\tq}^{2}}{\|\mathbf{r}_{\mathsf{ar}}\|^{2}}$, $\Gamma_5=\frac{y_{\rq,\tq}z_{\tq}}{\|\mathbf{r}_{\mathsf{ar}}\|^{2}}$, and $\Gamma_6=1-\frac{z_{\tq}^{2}}{\|\mathbf{r}_{\mathsf{ar}}\|^{2}}$. Plugging above conversions into \eqref{eq:appa1}--\eqref{eq:appa2} yields the explicit expressions of $e_{x}\left({\mathbf{r}}_{\mathsf{ar}}\right)$, $e_{y}\left({\mathbf{r}}_{\mathsf{ar}}\right)$, and $e_{z}\left({\mathbf{r}}_{\mathsf{ar}}\right)$ about the unknown position and attitude
parameters $\bm{\xi}_{\mathsf{ar}}$:
\begin{align}
&e_{x}\left({\mathbf{r}}_{\mathsf{ar}}\right)=\imagunit \mathcal{E}_{\mathsf{in}}\frac{\left(y_{\rq,\tq}^{2}+z_{\tq}^{2}\right)t_{x}-x_{\rq,\tq}y_{\rq,\tq}t_{y}+x_{\rq,\tq}z_{\tq}t_{z}}{\|\mathbf{r}_{\mathsf{ar}}\|^3 \mathrm{e}^{-\imagunit k \|\mathbf{r}_{\mathsf{ar}}\|}}\label{eq:appa3},
\\
&e_{y}\left({\mathbf{r}}_{\mathsf{ar}}\right)=\imagunit \mathcal{E}_{\mathsf{in}}\frac{-x_{\rq,\tq}y_{\rq,\tq}t_{x}+\left(x_{\rq,\tq}^{2}+z_{\tq}^{2}\right)t_{y}+y_{\rq,\tq}z_{\tq}t_{z}}{\|\mathbf{r}_{\mathsf{ar}}\|^3 \mathrm{e}^{-\imagunit k \|\mathbf{r}_{\mathsf{ar}}\|}},\\
&e_{z}\left({\mathbf{r}}_{\mathsf{ar}}\right)=\imagunit \mathcal{E}_{\mathsf{in}}\frac{x_{\rq,\tq}z_{\tq}t_{x}+y_{\rq,\tq}z_{\tq}t_{y}+\left(x_{\rq,\tq}^{2}+y_{\rq,\tq}^{2}\right)t_{z}}{\|\mathbf{r}_{\mathsf{ar}}\|^3 \mathrm{e}^{-\imagunit k \|\mathbf{r}_{\mathsf{ar}}\|}}.\label{eq:appa4}
\end{align}
Then, plugging \eqref{eq:appa3}--\eqref{eq:appa4} into 
\begin{equation}
h_{\mathsf{ar}}\left(\bm{\xi}_{\mathsf{ar}};\mathbf{p}_{\mathsf{r}}\right)=\frac{\mathrm{e}^{\imagunit k \|\mathbf{r}_{\mathsf{ar}}\|}}{\mathcal{E}_{\mathsf{in}}}  \sqrt{\sum_{\tilde{\epsilon}\in \left\{x,y,z\right\}}|e_{\tilde{\epsilon}}\left({\mathbf{r}}_{\mathsf{ar}}\right)|^2\left(- {\hat{\mathbf{r}}_{\mathsf{ar}}}\cdot{\hat{\mathbf{z}}}\right)}
\end{equation}
yields \eqref{eq:dis1}. So far, we have given the general channel model.
\begin{figure}[!t]
\centering
\includegraphics[scale=0.59]{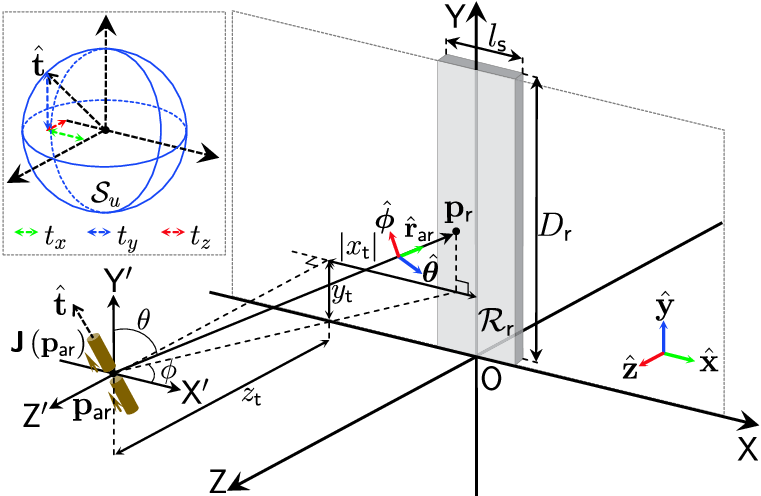}
\caption{{Illustration of the general system introduced in \textit{Discussion} \ref{definition21}. The UE is located at an arbitrary point $\mathbf{p}_{\mathsf{ar}}=\left(x_{\tq},y_{\tq},z_{\tq}\right)^{\mathsf{T}}$ in front of  $\mathcal{R}_{\mathsf{r}}$. Meanwhile, the UE has a general spatial orientation $\hat{\mathbf{t}}=t_{x}\hat{\mathbf{x}}+t_{y}\hat{\mathbf{y}}+t_{z}\hat{\mathbf{z}}$, whose endpoint is located on the surface of the \textit{unit spherical shell $\mathcal{S}_{u}$}. The current distribution of the UE is denoted as ${\bm{\mathsf{J}}}\left(\mathbf{p}_{\mathsf{ar}}\right)$. As in Fig. \ref{system2}, we establish the $\mathsf{PX^{\prime}Y^{\prime}Z^{\prime}}$ system with $\mathbf{p}_{\mathsf{ar}}$ as its origin, which has a pure translational relationship with $\mathsf{OXYZ}$ system. Then, we establish a spherical coordinate system $\left(r_{\mathsf{ar}},\theta,\phi\right)$ of point $\mathbf{p}_{\mathsf{ar}}$ with respect to $\mathsf{PX^{\prime}Y^{\prime}Z^{\prime}}$ system, where ${\hat{\mathbf{r}}_{\mathsf{ar}}}=\frac{\mathbf{r}_{\mathsf{ar}}}{\|\mathbf{r}_{\mathsf{ar}}\|}=\frac{\mathbf{p}_{\mathsf{r}}-\mathbf{p}_{\mathsf{ar}}}{\|\mathbf{p}_{\mathsf{r}}-\mathbf{p}_{\mathsf{ar}}\|}$.}}
\label{systemapp}
\end{figure}}

\subsection{\texorpdfstring{Calculation Process of $h_{\mathrm{SIM}}\left(\bm{\xi};x_{\rq},y_{\rq}\right)$}{zt} }\label{app:eq:disminor1} 
{Substituting \eqref{eq:green} into \eqref{eq:eff}, the vector electric
field ${\mathbf{e}}^{\mathrm{SIM}}\left({\mathbf{r}}\right)$ of the EM-SIMP channel can be derived as
\begin{align}
\notag{\mathbf{e}}^{\mathrm{SIM}}\left({\mathbf{r}}\right)&=\left(1+\frac{\imagunit 3}{k r}-\frac{3}{k^2 r^2}\right){\mathbf{e}}\left({\mathbf{r}}\right)\\
&=\underbrace{{G_{s}}(r) {\bm{\mathsf{J}}}_{\mathcal{R}_{\mathsf{t}}}^{\perp}}_{\text{propagating }}+\underbrace{\left(\frac{\imagunit 3}{k r}-\frac{3}{k^2 r^2}\right){G_{s}}(r){\bm{\mathsf{J}}}_{\mathcal{R}_{\mathsf{t}}}^{\perp}}_{\text {non-propagating}}.\label{eq:appminor1}
    \end{align}
It is observed that ${\mathbf{e}}^{\mathrm{SIM}}\left({\mathbf{r}}\right)$ includes a propagating component whose power decays as $r^{-2}$, as well as two non-propagating
components whose power decay as $r^{-4}$ and $r^{-6}$, respectively. Substituting \eqref{eq:appminor1} into \eqref{eq:esrrt}, the scalar electric
field ${{e}}_{\mathsf{s}}^{\mathrm{SIM}}\left({\mathbf{r}}\right)$ of the EM-SIMP channel can be derived as
\begin{align}
\notag e_{\mathsf{s}}^{\mathrm{SIM}}\left(\mathbf{r}\right)&=\left|1+\frac{\imagunit 3}{k r}-\frac{3}{k^2 r^2}\right|\sqrt{\frac{z_{\tq}}{r}}\|\mathbf{e}\left(\mathbf{r} \right)\|\mathrm{e}^{\imagunit kr}\\
&=\mathcal{E}_{\mathsf{in}}\underbrace{\sqrt{1+\frac{3}{k^2r^2}+\frac{9}{k^4r^4}}{h}\left(\bm{\xi};x_{\rq},y_{\rq}\right)}_{\triangleq h_{\mathrm{SIM}}\left(\bm{\xi};x_{\rq},y_{\rq}\right)},
\end{align}
where $h_{\mathrm{SIM}}\left(\bm{\xi};x_{\rq},y_{\rq}\right)$ is the EM-SIMP channel.}

\subsection{Proof of \textit{Lemma}~\ref{lemm:ZZB2} }\label{app:lemma2} 
The minimum error probability related to the MAP criterion can be expressed as 
\begin{align} \notag\mathrm{P}_{\min}\left(\bm{\vartheta},\bm{\vartheta}+\bm{\delta}\right)&= \mathrm{Pr}\left\{\mathcal{H}_0\right\} \mathrm{Pr}\left\{\mathcal{L}_{\bm{\xi}} \leq \ln\frac{\mathrm{Pr}\left\{\mathcal{H}_1\right\}}{\mathrm{Pr}\left\{\mathcal{H}_0\right\}} \bigg|\bm{\xi}=\bm{\vartheta}\right\}\\
\notag+&\mathrm{Pr}\left\{\mathcal{H}_1\right\} \mathrm{Pr}\left\{\mathcal{L}_{\bm{\xi}} > \ln\frac{\mathrm{Pr}\left\{\mathcal{H}_1\right\}}{\mathrm{Pr}\left\{\mathcal{H}_0\right\}} \bigg|\bm{\xi}=\bm{\vartheta}+\bm{\delta}\right\}\\
\notag& \triangleq \mathrm{Pr}\left\{\mathcal{H}_0\right\} \mathrm{Pr}\left\{\mathcal{L}_{\bm{\vartheta}} \leq \ln\frac{\mathrm{Pr}\left\{\mathcal{H}_1\right\}}{\mathrm{Pr}\left\{\mathcal{H}_0\right\}}\right\}\\
+&\mathrm{Pr}\left\{\mathcal{H}_1\right\} \mathrm{Pr}\left\{\mathcal{L}_{\bm{\vartheta}+\bm{\delta}}>\ln\frac{\mathrm{Pr}\left\{\mathcal{H}_1\right\}}{\mathrm{Pr}\left\{\mathcal{H}_0\right\}}\right\},
\label{eq:Pmin1}
\end{align}
where $\mathcal{L}_{\bm{\xi}}$ is the log-likelihood ratio (LLR) test, given as
\begin{equation}
    \mathcal{L}_{\bm{\xi}}=\ln \frac{f_{\tilde{\bm{\mathrm{v}}}| \bm{\xi}}\left(\tilde{\bm{\mathrm{v}}}| \bm{\vartheta}\right)}{f_{\tilde{\bm{\mathrm{v}}}| \bm{\xi}}\left(\tilde{\bm{\mathrm{v}}}| \bm{\vartheta}+\bm{\delta}\right)}=\Lambda\left(\bm{\vartheta}|\tilde{\bm{\mathrm{v}}}\right)-\Lambda\left(\bm{\vartheta}+\bm{\delta}|\tilde{\bm{\mathrm{v}}}\right),\label{eq:LXI}
\end{equation}
and $\Lambda\left(\bm{\vartheta}|\tilde{\bm{\mathrm{v}}}\right)\triangleq \Lambda\left(\bm{\xi}|\tilde{\bm{\mathrm{v}}}\right)|_{\bm{\xi}=\bm{\vartheta}}$ with $\Lambda\left(\bm{\xi}|\tilde{\bm{\mathrm{v}}}\right)$ denoting the log-likelihood function of \eqref{eq:vector_observation}. Specifically, for a given vector $\bm{\xi}$, the observation vector $\tilde{\bm{\mathrm{v}}}$ follows a complex Gaussian distribution $\mathcal{CN}\left(\bar{\bm{\mathrm{v}}}|{\bm{\xi}},\mathbf{C}_{\tilde{\bm{\mathrm{v}}}}\right)$ with  the mean vector $\bar{{\bm{{{\mathrm{v}}}}}}|{\bm{\xi}}=\mathcal{E}_{\mathsf{in}}l_{\mathsf{s}}{\bm{h}}_{\mathsf{y}}\left(\bm{\xi}\right)$ and the covariance matrix $\mathbf{C}_{\tilde{\bm{\mathrm{v}}}}=\mathbb{E}_{\tilde{\bm{{\mathrm{v}}}}|\bm{\xi}}\left\{\left(\tilde{\bm{\mathrm{v}}}-\bar{\bm{\mathrm{v}}}{|{\bm{\xi}}}\right)\left(\tilde{\bm{\mathrm{v}}}-\bar{\bm{\mathrm{v}}}{|{\bm{\xi}}}\right)^{\dagger}\right\}=\sigma^{2}\mathbf{I}_{\mathsf{N}}$. Thus, the conditional PDF of the observation vector $\tilde{\bm{\mathrm{v}}}$ given the vector $\bm{\xi}$ is expressed as
\begin{equation}
    f_{\tilde{\bm{\mathrm{v}}}| \bm{\xi}}\left(\tilde{\bm{\mathrm{v}}}| \bm{\xi}\right)=\frac{1}{\pi^{\mathsf{N}}\mathrm{det}\left\{\mathbf{C}_{\tilde{\bm{\mathrm{v}}}}\right\}}\mathrm{e}^{-\left(\tilde{\bm{\mathrm{v}}}-\bar{{\bm{\mathrm{v}}}}{|\bm{\xi}}\right)^{\dagger}\mathbf{C}_{\tilde{\bm{\mathrm{v}}}}^{-1}\left(\tilde{\bm{\mathrm{v}}}-\bar{{\bm{\mathrm{v}}}}{|\bm{\xi}}\right)},\label{eq:CPDF}
\end{equation}
where $\mathrm{det}\left\{\cdot\right\}$ denotes the determinant of a matrix. Then, for the given observation vector $\tilde{\bm{\mathrm{v}}}$, the log-likelihood function is
\begin{equation}
\Lambda\left({\bm{\xi}}|\tilde{\bm{\mathrm{v}}}\right)= -\frac{1}{\sigma^{2}}\sum_{n_{\mathsf{y}}=1}^{\mathsf{N}}
\left|\tilde{\mathrm{v}}_{n_{\mathsf{y}}}-\mathcal{E}_{\mathsf{in}}l_{\mathsf{s}}{{h}}_{\mathsf{y}}\left({\bm{\xi}};y_{\mathsf{r};{n_{\mathsf{y}}}}\right)\right|^{2}.\label{eq:Lambda}
\end{equation}
Substituting \eqref{eq:Lambda} into \eqref{eq:LXI}, we can further compute $\mathcal{L}_{\bm{\xi}}$ as
\begin{align}
\notag\mathcal{L}_{\bm{\xi}}=&\frac{\mathcal{E}_{\mathsf{in}}^{2}l_{\mathsf{s}}^{2}}{\sigma^{2}}\left[\left\|{\bm{h}}_{\mathsf{y}}\left(\bm{\vartheta}+\bm{\delta}\right)\right\|^{2}-\left\|{\bm{h}}_{\mathsf{y}}\left(\bm{\vartheta}\right)\right\|^{2}\right]\\
    &-\frac{2\mathcal{E}_{\mathsf{in}}l_{\mathsf{s}}}{\sigma^{2}}\Re\left\{\tilde{\bm{\mathrm{v}}}^{\dagger}{\bm{h}}_{\mathsf{y}}\left(\bm{\vartheta}+\bm{\delta}\right)-\tilde{\bm{\mathrm{v}}}^{\dagger}{\bm{h}}_{\mathsf{y}}\left(\bm{\vartheta}\right)\right\},\label{eq:LXI2}
\end{align}
\begin{figure*}[htbp]
\begin{align}
\mathcal{L}_{\bm{\vartheta}}=\underbrace{\frac{\mathcal{E}_{\mathsf{in}}^{2}l_{\mathsf{s}}^{2}}{\sigma^{2}}\left\{\left\|{\bm{h}}_{\mathsf{y}}\left(\bm{\vartheta}+\bm{\delta}\right)\right\|^{2}+\left\|{\bm{h}}_{\mathsf{y}}\left(\bm{\vartheta}\right)\right\|^{2}-2\Re\left\{{\bm{h}}_{\mathsf{y}}^{\dagger}\left(\bm{\vartheta}\right){\bm{h}}_{\mathsf{y}}\left(\bm{\vartheta}+\bm{\delta}\right)\right\}\right\}}_{\mu_{\mathcal{L}_{\bm{\vartheta}}}}+\underbrace{\frac{2\mathcal{E}_{\mathsf{in}}l_{\mathsf{s}}}{\sigma^{2}}\Re\left\{\bm{n}^{\dagger}\left[{\bm{h}}_{\mathsf{y}}\left(\bm{\vartheta}\right)-{\bm{h}}_{\mathsf{y}}\left(\bm{\vartheta}+\bm{\delta}\right)\right]\right\}}_{\text{random term via $\bm{n}$}}.\label{eq:lva}
\end{align}
\hrule
\begin{align}
\notag\mu_{\mathcal{L}_{\bm{\vartheta}}}
&=\frac{\mathcal{E}_{\mathsf{in}}^{2}l_{\mathsf{s}}^{2}}{\sigma^{2}}\sum_{n_{\mathsf{y}}=1}^{\mathsf{N}}\left\{\left|h_{\mathsf{y}}\left(\bm{\vartheta}+\bm{\delta};y_{\mathsf{r};{n_{\mathsf{y}}}}\right)\right|^{2}+\left|h_{\mathsf{y}}\left(\bm{\vartheta};y_{\mathsf{r};{n_{\mathsf{y}}}}\right)\right|^{2}-2\Re\left\{h_{\mathsf{y}}^{*}\left(\bm{\vartheta};y_{\mathsf{r};{n_{\mathsf{y}}}}\right)h_{\mathsf{y}}\left(\bm{\vartheta}+\bm{\delta};y_{\mathsf{r};{n_{\mathsf{y}}}}\right)\right\}\right\}\\
&\overset{(\mathsf{g})}{\simeq}\frac{\mathcal{E}_{\mathsf{in}}^{2}l_{\mathsf{s}}}{\sigma^{2}}\int_{0}^{{D_{\rq}}}\left|{{h}}_{\mathsf{y}}\left(\bm{\vartheta}+\bm{\delta};y_{\mathsf{r}}\right)\right|^{2}+\left|{{h}}_{\mathsf{y}}\left(\bm{\vartheta};y_{\mathsf{r}}\right)\right|^{2}-2\left|{{h}}_{\mathsf{y}}\left(\bm{\vartheta}+\bm{\delta};y_{\mathsf{r}}\right)\right|\left|{{h}}_{\mathsf{y}}\left(\bm{\vartheta};y_{\mathsf{r}}\right)\right|\cos \left[{k\left({r_{\mathsf{ry}}|_{z_{\tq}={\vartheta}_{z}+\delta_{z}}}-r_{\mathsf{ry}}|_{z_{\tq}={\vartheta}_{z}}\right)}\right]dy_{\rq},\label{eq:mul1}
\end{align}
\hrule
\end{figure*}where ${\bm{h}}_{\mathsf{y}}\left(\bm{\vartheta}\right)\triangleq {\bm{h}}_{\mathsf{y}}\left(\bm{\xi}\right)|_{\bm{\xi}=\bm{\vartheta}}$, $\tilde{\bm{\mathrm{v}}}$ and ${\bm{h}}_{\mathsf{y}}\left(\bm{\xi}\right)$ are given in \eqref{eq:vector_observation}.  From \eqref{eq:LXI2}, it is easy to find that $\mathcal{L}_{\bm{\xi}}$ is a function of
unknown parameter vector $\bm{\xi}$ via $\tilde{\bm{\mathrm{v}}}$. To obtain the expressions of the conditional probabilities in \eqref{eq:Pmin1}, it is necessary to find the distributions of $\mathcal{L}_{\bm{\vartheta}}$ and $\mathcal{L}_{\bm{\vartheta}+\bm{\delta}}$. Given $\bm{\xi}=\bm{\vartheta}$, we derive $\mathcal{L}_{\bm{\vartheta}}$ in \eqref{eq:lva}. Obviously, since the only
random variable in $\mathcal{L}_{\bm{\vartheta}}$ is the Gaussian noise vector $\bm{n}$, $\mathcal{L}_{\bm{\vartheta}}$ follows a
Gaussian distributions $\mathcal{N}\left(\mu_{\mathcal{L}_{\bm{\vartheta}}}, \sigma_{\mathcal{L}_{\bm{\vartheta}}}^2\right)$ with $\mu_{\mathcal{L}_{\bm{\vartheta}}}$ and $\sigma_{\mathcal{L}_{\bm{\vartheta}}}^2$ given in \eqref{eq:mul1} and \eqref{eq:v1}.
    \begin{align}
\notag\sigma_{\mathcal{L}_{\bm{\vartheta}}}^2&=\frac{4\mathcal{E}_{\mathsf{in}}^{2}l_{\mathsf{s}}^{2}}{\sigma^{4}}\mathbb{V}_{\bm{n}}\left\{\Re\left\{\bm{n}^{\dagger}\left[{\bm{h}}_{\mathsf{y}}\left(\bm{\vartheta}\right)-{\bm{h}}_{\mathsf{y}}\left(\bm{\vartheta}+\bm{\delta}\right)\right]\right\}\right\}\\
\notag&=\frac{2\mathcal{E}_{\mathsf{in}}^{2}l_{\mathsf{s}}^{2}}{\sigma^{2}}\left[{\bm{h}}_{\mathsf{y}}\left(\bm{\vartheta}\right)-{\bm{h}}_{\mathsf{y}}\left(\bm{\vartheta}+\bm{\delta}\right)\right]^{\dagger}\left[{\bm{h}}_{\mathsf{y}}\left(\bm{\vartheta}\right)-{\bm{h}}_{\mathsf{y}}\left(\bm{\vartheta}+\bm{\delta}\right)\right]\\
&=2\mu_{\mathcal{L}_{\bm{\vartheta}}}.\label{eq:v1}    \end{align}
In $(\mathsf{g})$ of \eqref{eq:mul1}, we approximate the summation as an integral by using the \textit{Riemann integral method}, and if each continuous point on the $\mathsf{Y}$-axis can observe the voltage independently, the equality in $(\mathsf{g})$ holds. Besides, we utilize the Euler's formula to eliminate the operation of taking the real part $\Re\left\{\cdot\right\}$, where we denote $\bm{\vartheta}$ and $\bm{\delta}$ as $\bm{\vartheta}\triangleq\left(\vartheta_{z},\vartheta_{t}\right)^{\mathsf{T}}$ and $\bm{\delta}\triangleq\left(\delta_{z},\delta_{t}\right)^{\mathsf{T}}$. Therefore, we have $\mathcal{L}_{\bm{\vartheta}}\sim\mathcal{N}\left(\mu_{\mathcal{L}_{\bm{\vartheta}}}, 2\mu_{\mathcal{L}_{\bm{\vartheta}}}\right)$.  Similarly, given $\bm{\xi}=\bm{\vartheta}+\bm{\delta}$, then $\mathcal{L}_{\bm{\vartheta}+\bm{\delta}}=-\mu_{\mathcal{L}_{\bm{\vartheta}}}-\frac{2\mathcal{E}_{\mathsf{in}}l_{\mathsf{s}}}{\sigma^{2}}\Re\left\{\bm{n}^{\dagger}\left[{\bm{h}}_{\mathsf{y}}\left(\bm{\vartheta}+\bm{\delta}\right)-{\bm{h}}_{\mathsf{y}}\left(\bm{\vartheta}\right)\right]\right\}$
follows $\mathcal{L}_{\bm{\vartheta}+\bm{\delta}}\sim\mathcal{N}\left(-\mu_{\mathcal{L}_{\bm{\vartheta}}}, 2\mu_{\mathcal{L}_{\bm{\vartheta}}}\right)$. Hence, the probabilities in \eqref{eq:Pmin1} are obtained by
\begin{align}
    &\mathrm{Pr}\left\{\mathcal{L}_{\bm{\vartheta}} \leq \ln \frac{\mathrm{Pr}\left\{\mathcal{H}_{1}\right\}}{\mathrm{Pr}\left\{\mathcal{H}_{0}\right\}}\right\}=\mathcal{Q}\left(\frac{\mu_{\mathcal{L}_{\bm{\vartheta}}}-\ln \frac{\mathrm{Pr}\left\{\mathcal{H}_{1}\right\}}{\mathrm{Pr}\left\{\mathcal{H}_{0}\right\}}}{\sqrt{2 \mu_{\mathcal{L}_{\bm{\vartheta}}}}}\right),\label{eq:Q1}\\
    &\mathrm{Pr}\left\{\mathcal{L}_{\bm{\vartheta}+\bm{\delta}}>\ln \frac{\mathrm{Pr}\left\{\mathcal{H}_{1}\right\}}{\mathrm{Pr}\left\{\mathcal{H}_{0}\right\}}\right\}=\mathcal{Q}\left(\frac{\mu_{\mathcal{L}_{\bm{\vartheta}}}+\ln \frac{\mathrm{Pr}\left\{\mathcal{H}_{1}\right\}}{\mathrm{Pr}\left\{\mathcal{H}_{0}\right\}}}{\sqrt{2 \mu_{\mathcal{L}_{\bm{\vartheta}}}}}\right)\label{eq:Q2},
\end{align}
where $\mathcal{Q}\left(x\right)=\frac{1}{\sqrt{2 \pi}} \int_x^{\infty} \mathrm{e}^{-\frac{v^2}{2}} d v$ is the tail distribution function of the standard normal distribution. Plugging \eqref{eq:Q1} and  \eqref{eq:Q2} into \eqref{eq:Pmin1} yields \eqref{eq:lemma2}. Thus, \textit{Lemma}~\ref{lemm:ZZB2} holds.

\subsection{Proof of \textit{Corollary}~\ref{coro:ZZBonly} }\label{app:coro:ZZBonly} 
For the attitude-only estimation, the vectors $\bm{\vartheta}$ and $\bm{\delta}$ reduce into two scalar parameters ${\vartheta}_{t}$ and ${\delta}_{t}$, and \eqref{eq:ZZBURU} reduces to
\begin{align}
\notag\mathbb{E} \left\{|t_{z}-\hat{t}_{z}|^2\right\} \geq\frac{1}{2} &\int_0^{1} \int_{0}^{1-\delta_t}\left[f_{t_{z}}\left({\vartheta}_{t}\right)+f_{t_{z}}\left({\vartheta}_{t}+{\delta}_{t}\right)\right]\\
&\times\mathrm{P}_{\min}^{\mathrm{AO}}\left({\vartheta}_{t},{\vartheta}_{t}+{\delta}_{t}\right)d{\vartheta}_{t} \delta_{t} d \delta_{t}.\label{eq:ZZBttAO}
\end{align}
Under \textit{Assumption}~\assumptionref{assum:p} and \assumptionref{assum:4.1}, we obtain the \textit{prior} PDF: 
\begin{equation}
f_{t_z}\left({\vartheta}_{t}\right)=f_{t_z}\left({\vartheta}_{t}+{\delta}_{t}\right)=1,\label{eqftz1}
\end{equation}
and the conditional minimum error probability for a deterministic value of $z_{\tq}$:
\begin{equation}
\mathrm{CP}_{\min}^{\mathrm{AO}}\left({\vartheta}_{t},{\vartheta}_{t}+{\delta}_{t}\right)=\mathcal{Q}\left(\sqrt{\frac{\mu_{\mathcal{L}_{\bm{\vartheta}_{a}}}}{2}}\right)\label{eq:CPminAO},
\end{equation}
where $\mu_{\mathcal{L}_{\bm{\vartheta}_{a}}}$ can be derived by replacing $\bm{\vartheta}$ and $\bm{\delta}$ of $\mu_{\mathcal{L}_{\bm{\vartheta}}}$ in \eqref{eq:mumu} with $\left(z_{\tq},{\vartheta}_{t}\right)^{\mathsf{T}}$ and $\left(0,\delta_t\right)^{\mathsf{T}}$ respectively, i.e., 
\begin{align}
\mu_{\mathcal{L}_{\bm{\vartheta }_{a}}}=\text{SNR}l_{\mathsf{s}}z_{\tq}\int_{0}^{D_{\mathsf{r}}}\frac{\left(y_{\rq}\delta_{t}+z_{\tq}\vartheta_{\delta \mathsf{s}}-z_{\tq}\vartheta_{t\mathsf{s}}\right)^2}{\left(y_{\rq}^{2}+z_{\tq}^{2}\right)^{5/2}}dy_{\rq},\label{eq:mulva1}
\end{align}
where $\vartheta_{t\mathsf{s}}\triangleq\sqrt{1-\vartheta_{t}^2}$, and $\vartheta_{\delta\mathsf{s}}\triangleq\sqrt{1-\left(\vartheta_{t}+\delta_{t}\right)^2}$. The integrand in \eqref{eq:mulva1} no longer contains the cosine function, thus we can provide a closed-form expression for $\mu_{\mathcal{L}_{\bm{\vartheta }_{a}}}$ in \eqref{eq:mulvaclose}.

By taking the expectation of \eqref{eq:CPminAO} over $z_{\tq}$, the  minimum error probability can be obtained as
\begin{equation}
\mathrm{P}_{\min}^{\mathrm{AO}}\left({\vartheta}_{t},{\vartheta}_{t}+{\delta}_{t}\right)=\mathbb{E}_{z_{\tq}}\left\{\mathcal{Q}\left(\sqrt{\frac{\mu_{\mathcal{L}_{\bm{\vartheta}_{a}}}}{2}}\right)\right\}.\label{eq:PminAO}
\end{equation}
Plugging \eqref{eqftz1} and \eqref{eq:PminAO} into \eqref{eq:ZZBttAO} yields \eqref{eq:mulva}.

\subsection{Proof of \textit{Lemma}~\ref{lemma:ECRB1} }\label{app:CRB1} 
The proof can be completed based on \cite{angchen}. we mainly provide the derivation of $\mathbf{F}\left(\bm{\xi}\right)$. Differentiating the log-likelihood function in \eqref{eq:Lambda} once produces
\begin{equation}
        \frac{\partial \ln \Lambda\left({\bm{\xi}}|\tilde{\bm{\mathrm{v}}}\right)}{\partial \bm{\xi}_{\imath}}=\frac{2\mathcal{E}_{\mathsf{in}}l_{\mathsf{s}}}{\sigma^{2}}\Re\left\{\left[\tilde{\bm{\mathrm{v}}}-\mathcal{E}_{\mathsf{in}}l_{\mathsf{s}}{\bm{h}}_{\mathsf{y}}\left(\bm{\xi}\right)\right]^{\dagger}\frac{\partial {\bm{h}}_{\mathsf{y}}\left(\bm{\xi}\right)}{\partial \bm{\xi}_{\imath}}\right\},
\end{equation}
where the required regularity condition $\mathbb{E}_{\tilde{\bm{{\mathrm{v}}}}|\bm{\xi}}\left\{\frac{\partial \ln \Lambda\left({\bm{\xi}}|\tilde{\bm{\mathrm{v}}}\right)}{\partial \bm{\xi}_{\imath}}\right\}=0$ is satisfied 
as $\mathbb{E}_{\tilde{\bm{{\mathrm{v}}}}|\bm{\xi}}\left\{\tilde{\bm{\mathrm{v}}}\right\}=\mathcal{E}_{\mathsf{in}}l_{\mathsf{s}}{\bm{h}}_{\mathsf{y}}\left(\bm{\xi}\right)$. Then, we compute the element on the $\imath$-th row and $\jmath$-th column of $\mathbf{F}\left(\bm{\xi}\right)$ as
\begin{align}
&\notag\left[\mathbf{F}\left(\bm{\xi}\right)\right]_{\imath \jmath}=\mathbb{E}_{\tilde{\bm{{\mathrm{v}}}}|\bm{\xi}}\left\{\frac{\partial \ln \Lambda\left({\bm{\xi}}|\tilde{\bm{\mathrm{v}}}\right)}{\partial \bm{\xi}_{\imath}}\frac{\partial \ln \Lambda\left({\bm{\xi}}|\tilde{\bm{\mathrm{v}}}\right)}{\partial \bm{\xi}_{\jmath}}\right\}\\
\notag&\quad\quad\overset{(\mathsf{h})}{=}\frac{\mathcal{E}_{\mathsf{in}}^{2}l_{\mathsf{s}}^{2}}{\sigma^{4}}\left[\frac{\partial {\bm{h}}_{\mathsf{y}}^{\dagger}\left(\bm{\xi}\right)}{\partial \bm{\xi}_{\imath}}\mathbf{C}_{\tilde{\bm{\mathrm{v}}}}\frac{\partial {\bm{h}}_{\mathsf{y}}\left(\bm{\xi}\right)}{\partial \bm{\xi}_{\jmath}}+\frac{\partial {\bm{h}}_{\mathsf{y}}^{\dagger}\left(\bm{\xi}\right)}{\partial \bm{\xi}_{\jmath}}\mathbf{C}_{\tilde{\bm{\mathrm{v}}}}\frac{\partial {\bm{h}}_{\mathsf{y}}\left(\bm{\xi}\right)}{\partial \bm{\xi}_{\imath}}\right]\\
\notag&\quad\quad=\frac{2\mathcal{E}_{\mathsf{in}}^{2}l_{\mathsf{s}}^{2}}{\sigma^{2}}\Re\left\{\sum_{n_{\mathsf{y}}=1}^{\mathsf{N}}\frac{\partial {{h}}_{\mathsf{y}}^{*}\left(\bm{\xi};y_{\mathsf{r};{n_{\mathsf{y}}}}\right)}{\partial \bm{\xi}_{\imath}}\frac{\partial {{h}}_{\mathsf{y}}\left(\bm{\xi};y_{\mathsf{r};{n_{\mathsf{y}}}}\right)}{\partial \bm{\xi}_{\jmath}}\right\}\\
&\quad\quad\overset{(\mathsf{i})}{\simeq}2\text{SNR}l_{\mathsf{s}}\int_{0}^{D_{\mathsf{r}}}\Re\left\{\frac{\partial {{h}}_{\mathsf{y}}^{*}\left(\bm{\xi};y_{\mathsf{r}}\right)}{\partial \bm{\xi}_{\imath}}\frac{\partial {{h}}_{\mathsf{y}}\left(\bm{\xi};y_{\mathsf{r}}\right)}{\partial \bm{\xi}_{\jmath}}\right\}d{y_{\rq}},\label{FIMequation}
\end{align}
where $(\mathsf{h})$ follows that $\mathbb{E}_{\tilde{\bm{{\mathrm{v}}}}|\bm{\xi}}\left\{\left(\tilde{\bm{\mathrm{v}}}-\bar{\bm{\mathrm{v}}}{|{\bm{\xi}}}\right)\left(\tilde{\bm{\mathrm{v}}}-\bar{\bm{\mathrm{v}}}{|{\bm{\xi}}}\right)^{\mathsf{T}}\right\}=\mathbf{0}$ and $\mathbb{E}_{\tilde{\bm{{\mathrm{v}}}}|\bm{\xi}}\left\{\left(\tilde{\bm{\mathrm{v}}}-\bar{\bm{\mathrm{v}}}{|{\bm{\xi}}}\right)^{*}\left(\tilde{\bm{\mathrm{v}}}-\bar{\bm{\mathrm{v}}}{|{\bm{\xi}}}\right)^{\mathsf{\dagger}}\right\}=\mathbf{0}$. Besides, in $(\mathsf{i})$,  we use the \textit{Riemann integral method} like $(\mathsf{g})$ in \eqref{eq:mul1}.

\subsection{Proof of \textit{Proposition}~\ref{coro:ECRB}  }\label{app:ECRB1} 
According to \eqref{eq:hy} and \textit{Lemma} \ref{lemma:ECRB1}, the first-order derivatives ${\partial h_{\mathsf{y}}\left(\bm{\xi};y_{\rq}\right)}/{\partial \bm{\xi}_{\imath}},\imath=1,2$ in FIM are first computed as
\begin{align}
 &\frac{\partial {{h}}_{\mathsf{y}}\left(\bm{\xi};y_{\mathsf{r}}\right)}{\partial z_{\tq}}=\frac{t_{z}y_{\rq}\left(y_{\rq}^2-\ell_{1}\right)+{t}_{y}z_{\tq}\left(3y_{\rq}^2-\ell_{2}\right)}{2z_{\tq}^{1/2}r_{\mathsf{ry}}^{9/2}}\mathrm{e}^{\imagunit k r_{\mathsf{ry}}},\label{eq:dhz}\\
 &\frac{\partial {{h}}_{\mathsf{y}}\left(\bm{\xi};y_{\mathsf{r}}\right)}{\partial t_{z}}=\frac{y_{\rq}-{z_{\tq}t_{z}}/{t_{y}}}{z_{\tq}^{-1/2}r_{\mathsf{ry}}^{5/2}}\mathrm{e}^{\imagunit k r_{\mathsf{ry}}}\label{eq:dht},
\end{align}
where $\ell_{1}\triangleq 2z_{\tq}^2\left(2-\imagunit kr_{\mathsf{ry}}\right)$, $\ell_{2}\triangleq 2z_{\tq}^2\left(1-\imagunit kr_{\mathsf{ry}}\right)$, and $t_{y}\triangleq \sqrt{1-t_{z}^2}$.
Substituting \eqref{eq:dhz} and \eqref{eq:dht} into \eqref{FIMequation}, we derive the FIM as $\mathsf{F}_{\mathsf{zz}}\triangleq\left[\mathbf{F}\left(\bm{\xi}\right)\right]_{11}
=2\text{SNR}l_{\mathsf{s}}\left(\mathcal{I}_{\mathsf{zz}1}+ k^2 \mathcal{I}_{\mathsf{zz}2}\right)$, $\mathsf{F}_{\mathsf{tt}}\triangleq\left[\mathbf{F}\left(\bm{\xi}\right)\right]_{22}
=2\text{SNR}l_{\mathsf{s}}\mathcal{I}_{\mathsf{tt}}$, and $\mathsf{F}_{\mathsf{zt}}\triangleq\left[\mathbf{F}\left(\bm{\xi}\right)\right]_{23}
=2\text{SNR}l_{\mathsf{s}}\mathcal{I}_{\mathsf{zt}}$, where $\mathcal{I}_{\mathsf{zz}1}$, $\mathcal{I}_{\mathsf{zz}2}$, $\mathcal{I}_{\mathsf{tt}}$, and $\mathcal{I}_{\mathsf{zt}}$ are provided in \eqref{eq:Izz1FIM}--\eqref{eq:IztFIM}. {By defining $\tau\triangleq D_{\rq}/z_{\tq}$, the closed-form expressions of \eqref{eq:Izz1FIM}--\eqref{eq:IztFIM} can be derived as follows.
\begin{align}
\mathcal{I}_{\mathsf{zz}1}
&=\frac{1}{z_{\tq}^3}\left(t_z t_y\mathcal{F}_{\tau1}+t_z^2 \mathcal{F}_{\tau2}+t_y^2 \mathcal{F}_{\tau3}\right),\label{eq:Izz1FIMclose}\\
\mathcal{I}_{\mathsf{zz}2}
&=\frac{1}{z_{\tq}}\left(t_z t_y\mathcal{F}_{\tau4}+t_z^2 \mathcal{F}_{\tau5}+t_y^2 \mathcal{F}_{\tau6}\right),\\
\mathcal{I}_{\mathsf{tt}}
&=\frac{1}{z_{\tq}}\left(\frac{t_z^2}{t_y^2}\mathcal{F}_{\tau7}+\frac{t_z}{t_y}\mathcal{F}_{\tau8}+\frac{1}{t_y^2}\mathcal{F}_{\tau 9}\right),\label{eq:Ittappclose}\\
\mathcal{I}_{\mathsf{zt}}&=\frac{1}{z_{\tq}^2}\left(\frac{t_z^2}{t_y}\mathcal{F}_{\tau10}+t_z\mathcal{F}_{\tau11}+t_y\mathcal{F}_{\tau 12}\right),\label{eq:IztFIMclose}
\end{align}
where $\mathcal{F}_{\tau1}\triangleq \frac{2\tau^8+8\tau^6+12\tau^4+8\tau^2+2}{7\left(\tau^2+1\right)^4}- \frac{7\tau^4-14\tau^2+4}{14\left(\tau^2+1\right)^{7/2}}$, $\mathcal{F}_{\tau2}\triangleq \frac{19\tau^7+56\tau^5+112\tau^3}{84\left(\tau^2+1\right)^{7/2}}$, $\mathcal{F}_{\tau 3}\triangleq \frac{10\tau^7+35\tau^5+28\tau^3+28\tau}{28\left(\tau^2+1\right)^{7/2}}$, $\mathcal{F}_{\tau 4}\triangleq \frac{2}{5}-\frac{2}{5\left(\tau^2+1\right)^{5/2}}$, $\mathcal{F}_{\tau5}\triangleq \frac{\tau^3\left(2\tau^2+5\right)}{15\left(\tau^2+1\right)^{5/2}}$, $\mathcal{F}_{\tau6}\triangleq \frac{8\tau^5+20\tau^3+15\tau}{15\left(\tau^2+1\right)^{5/2}}$, $\mathcal{F}_{\tau 7}\triangleq \frac{\tau^3+3\tau}{3\left(\tau^2+1\right)^{3/2}}$, $\mathcal{F}_{\tau8}\triangleq \frac{2}{3\left(\tau^2+1\right)^{3/2}}-\frac{2}{3}$, $\mathcal{F}_{\tau9}\triangleq \frac{\tau^3}{3\left(\tau^2+1\right)^{3/2}}$, $\mathcal{F}_{\tau10}\triangleq \frac{\tau^2-2}{6\left(\tau^2+1\right)^{5/2}}+\frac{\tau^4+2\tau^2+1}{3\left(\tau^2+1\right)^2}$, $\mathcal{F}_{\tau11}\triangleq \frac{\tau^5+\tau^3+6\tau}{6\left(\tau^2+1\right)^{5/2}}$, and $\mathcal{F}_{\tau12}\triangleq \frac{-\tau^2}{2\left(\tau^2+1\right)^{5/2}}$.} Then, by applying the matrix inversion lemma, we obtain the inverse of $\mathbf{F}\left(\bm{\xi}\right)$, denoted as $\mathbf{F}^{-1}\left(\bm{\xi}\right)$. Taking the expectation of $\mathbf{F}^{-1}\left(\bm{\xi}\right)$ over $\bm{\xi}$ leads to \eqref{eq:ECRBYT} and \eqref{eq:ECRBT} and completes the proof of \textit{Proposition}~\ref{coro:ECRB}.

\subsection{Proof of \textit{Corollary}~\ref{coro:ECRBCase2} }\label{app:ECRBcase2}
When $D_{\rq} \to \infty$ and $t_{z} \to 0$,  \eqref{eq:Izz1FIM}--\eqref{eq:IztFIM}  reduce to
\begin{align}
    &\mathcal{I}_{\mathsf{zz}1}=\frac{z_{\tq}}{4}\int_{0}^{\infty}\frac{\left(3y_{\rq}^2-2z_{\tq}^2\right)^2}{\left(y_{\rq}^2+z_{\tq}^2\right)^{9/2}}dy_{\rq}=\frac{5}{14z_{\tq}^3},\label{eq:Izz1lim}\\
        &  \mathcal{I}_{\mathsf{zz}2}=z_{\tq}^5\int_{0}^{\infty}\frac{1}{\left(y_{\rq}^2+z_{\tq}^2\right)^{7/2}}dy_{\rq}=\frac{8}{15z_{\tq}},\\
        &\mathcal{I}_{\mathsf{tt}}=z_{\tq}\int_{0}^{\infty}\frac{y_{\rq}^2}{\left(y_{\rq}^2+z_{\tq}^2\right)^{5/2}}dy_{\rq}=\frac{1}{{3z_{\tq}}},\\
        &\mathcal{I}_{\mathsf{zt}}=\frac{z_{\tq}}{2}\int_{0}^{\infty}\frac{y_{\rq}\left(3y_{\rq}^2-2z_{\tq}^2\right)}{\left(y_{\rq}^2+z_{\tq}^2\right)^{7/2}}dy_{\rq}=0\label{eq:Iztlim}.
\end{align}
We can also use the closed-form expressions in \eqref{eq:Izz1FIMclose}--\eqref{eq:IztFIMclose}. Specifically, when $D_{\rq} \to \infty$ (i.e., $\tau \to \infty$) and $t_{z} \to 0$, \eqref{eq:Izz1FIMclose}--\eqref{eq:IztFIMclose} degenerate to $\mathcal{I}_{\mathsf{zz}1}=\lim\frac{1}{z_{\tq}^3}\frac{10\tau^7+35\tau^5+28\tau^3+28\tau}{28\left(\tau^2+1\right)^{7/2}}=\frac{5}{14z_{\tq}^3}$, $\mathcal{I}_{\mathsf{zz}2}
=\lim\frac{1}{z_{\tq}}\frac{8\tau^5+20\tau^3+15\tau}{15\left(\tau^2+1\right)^{5/2}}=\frac{8}{15z_{\tq}}$, $\mathcal{I}_{\mathsf{tt}}
=\lim\frac{1}{z_{\tq}}\frac{\tau^3}{3\left(\tau^2+1\right)^{3/2}}=\frac{1}{{3z_{\tq}}}$, and $\mathcal{I}_{\mathsf{zt}}=\lim\frac{1}{z_{\tq}^2}\frac{-\tau^2}{2\left(\tau^2+1\right)^{5/2}}=0$, where we use $\lim$ to represent $\lim_{\tau\to \infty}$. Plugging \eqref{eq:Izz1lim}--\eqref{eq:Iztlim} into \eqref{eq:ECRBYT} and \eqref{eq:ECRBT} yields \eqref{eq:ECRBzcase2} and \eqref{eq:ECRBtcase2}.
If $z_{\tq}\gg \lambda$, we have $k^2z_{\tq}^2 \gg 4\pi^2$. Thus, $\frac{210z_{\tq}^3}{112k^2z_{\tq}^2+75}\approx \frac{15z_{\tq}}{8k^2}$ and \eqref{eq:limmmmecrbz} can be proved.

\subsection{Proof of \textit{Corollary}~\ref{prop:singleVS} }\label{app:VS}
\eqref{eq:ECRBT} can be written as $\mathrm{ECRB}\left(t_{z}\right)=\frac{\text{SNR}^{-1}}{2l_{\mathsf{s}}}\mathbb{E}_{\bm{\xi}}\left\{\frac{1}{\mathcal{I}_{\mathsf{tt}}-\mathcal{I}_{\mathsf{zz}}^{-1}\mathcal{I}_{\mathsf{zt}}^2}\right\}$, where $\mathcal{I}_{\mathsf{zz}}^{-1}\mathcal{I}_{\mathsf{zt}}^2\geq0$. So we have $\mathrm{ECRB}\left(t_{z}\right)\geq \mathrm{ECRB}_{\mathrm{AO}}\left(t_{z}\right)$. Moreover, from \eqref{eq:ECRBtcase2}, \eqref{eq:ECRBAO}, and \eqref{eq:711}, we have $\mathrm{ECRB}\left(t_{z}\right)= \mathrm{ECRB}_{\mathrm{AO}}\left(t_{z}\right)=\frac{3\text{SNR}^{-1}}{2l_{\mathsf{s}}}\mathbb{E}_{z_{\tq}}\left\{z_{\tq}\right\}$ when $D_{\rq} \to \infty$ and $t_{z} \to 0$, which shows that the lower bound is achievable. Besides, \eqref{eq:ZZBtz} and \eqref{eq:mulva} can be rewritten as $\mathrm{ZZB}\left(t_{z}\right)= \int_{0}^{1}\mathsf{W}_{t} \delta_{t} d \delta_{t}$ and $\mathrm{ZZB}_{\mathrm{AO}}\left(t_{z}\right)=\int_{0}^{1} \mathsf{W}_{{\mathrm{AO}};t}\delta_{t} d \delta_{t}$, where
\begin{align}
\notag\mathsf{W}_{t}&\triangleq\frac{1}{\mathsf{H}_{\tq}}\max _{\delta_{z}} \int_{0}^{1-\delta_{t}}\int_{\mathsf{H}_1}^{\mathsf{H}_{2}-\delta_{z}}\mathcal{Q}\left(\sqrt{\frac{\mu_{\mathcal{L}_{\bm{\vartheta}}}}{2}}\right)d\vartheta_{z}d\vartheta_{t} \\
\notag&\overset{(\mathsf{j})}{\geq} \frac{1}{\mathsf{H}_{\tq}} \max\limits_{\delta_{z}=0} \int_{0}^{1-\delta_{t}}\int_{\mathsf{H}_1}^{\mathsf{H}_{2}-\delta_{z}}\mathcal{Q}\left(\sqrt{\frac{\mu_{\mathcal{L}_{\bm{\vartheta}}}}{2}}\right)d\vartheta_{z}d\vartheta_{t}\\
\notag&= \frac{1}{\mathsf{H}_{\tq}} \int_{0}^{1-\delta_{t}}\int_{\mathsf{H}_1}^{\mathsf{H}_{2}}\mathcal{Q}\left(\sqrt{\frac{\mu_{\mathcal{L}_{\bm{\vartheta}_{a}}}}{2}}\right)dz_{\tq}d\vartheta_{t}\\
\notag&\overset{(\mathsf{k})}{=} \int_{0}^{1-\delta_{t}}\mathbb{E}_{z_{\tq}}\left\{\mathcal{Q}\left(\sqrt{\frac{\mu_{\mathcal{L}_{\bm{\vartheta}_{a}}}}{2}}\right)\right\}d\vartheta_{t}\\
&\triangleq\mathsf{W}_{\mathrm{AO};t},
\end{align}
and $\delta_z=0$ in $(\mathsf{j})$ is leveraged to obtain a special case of the maximum search set, and $(\mathsf{k})$ is because $z_\tq$ is a known random variable with PDF $\frac{1}{\mathsf{H}_{\tq}}$. Thus, $\mathrm{ZZB}\left(t_z\right)\geq \mathrm{ZZB}_{\mathrm{AO}}\left(t_z\right)$ is proved. Also, from \eqref{eq:zzbcase3t}, \eqref{eq:mulva}, and \eqref{eq:mulvaclose}, we have $\mathrm{ZZB}\left(t_{z}\right)= \mathrm{ZZB}_{\mathrm{AO}}\left(t_{z}\right)=\frac{1}{12}$ when $\text{SNR}\to 0$, or $D_{\rq}\to 0$, or $\bm{\delta}\to\bm{0}$, which reveals that the lower bound is achievable.

\bibliographystyle{IEEEtran}
\bibliography{reference}

\begin{IEEEbiography}[{\includegraphics[width=1in,height=1.25in,clip,keepaspectratio]{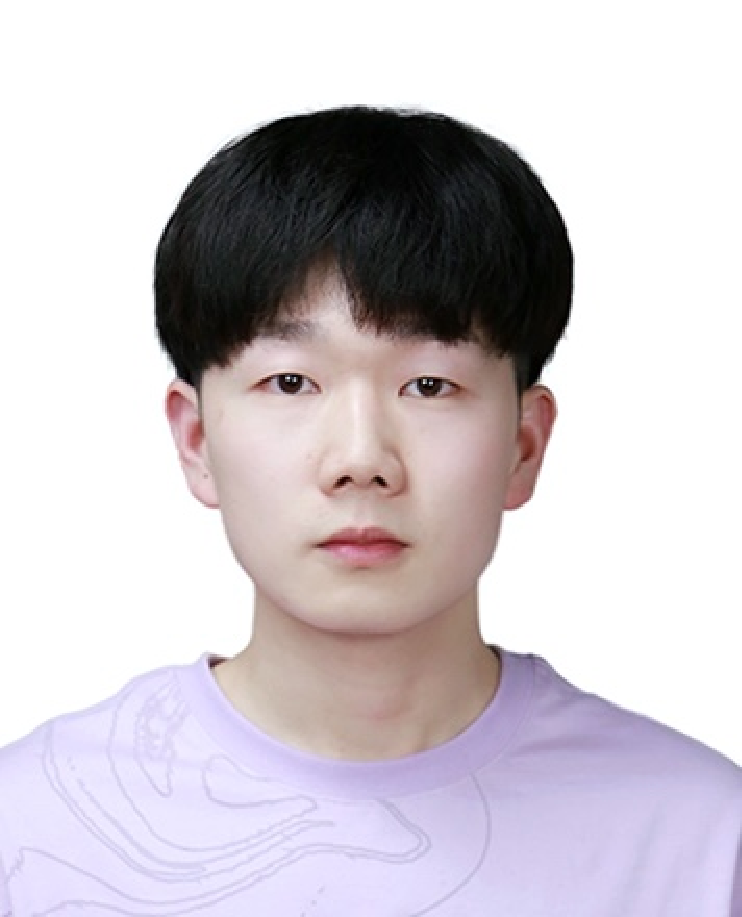}}]{Ang Chen}
    received his bachelor’s degree in communications engineering from the University of Electronic Science and Technology of China (UESTC), Chengdu, China, in 2021. He is pursuing his master’s degree with the Department of Electronic Engineering and Information Science, University of Science and Technology of China (USTC), Hefei, China. His research interests include positioning and sensing, electromagnetic information theory, and near-field communications. He received the National Scholarship in 2023 and the Outstanding Graduate Award of Anhui Province in 2024.
\end{IEEEbiography}

\begin{IEEEbiography}[{\includegraphics[width=1in,height=1.25in,clip,keepaspectratio]{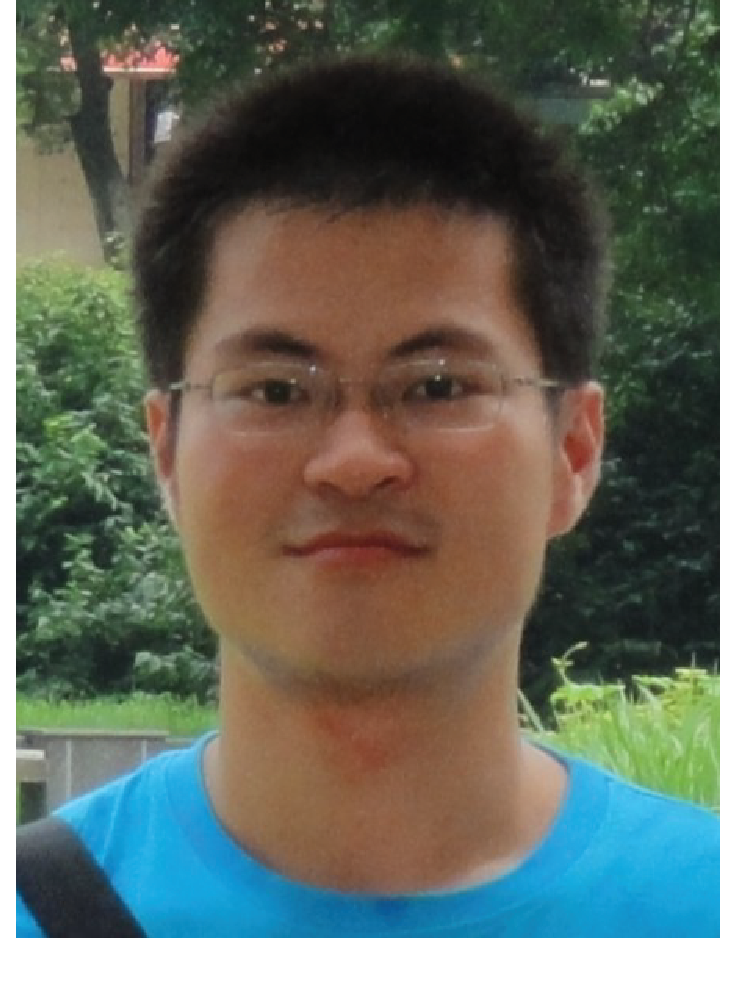}}]{Li Chen}
    (Senior Member, IEEE) received his B.E. degree in electrical and information engineering from the Harbin Institute of Technology (HIT), Harbin, China, in 2009, and his Ph.D. degree in electrical engineering from the University of Science and Technology of China (USTC), Hefei, China, in 2014. He is currently an Associate Professor at the Department of Electronic Engineering and Information Science, University of Science and Technology of China. His research interests include integrated
computation and communication, and integrated sensing and communication.
\end{IEEEbiography}

\begin{IEEEbiography}[{\includegraphics[width=1in,height=1.25in,clip,keepaspectratio]{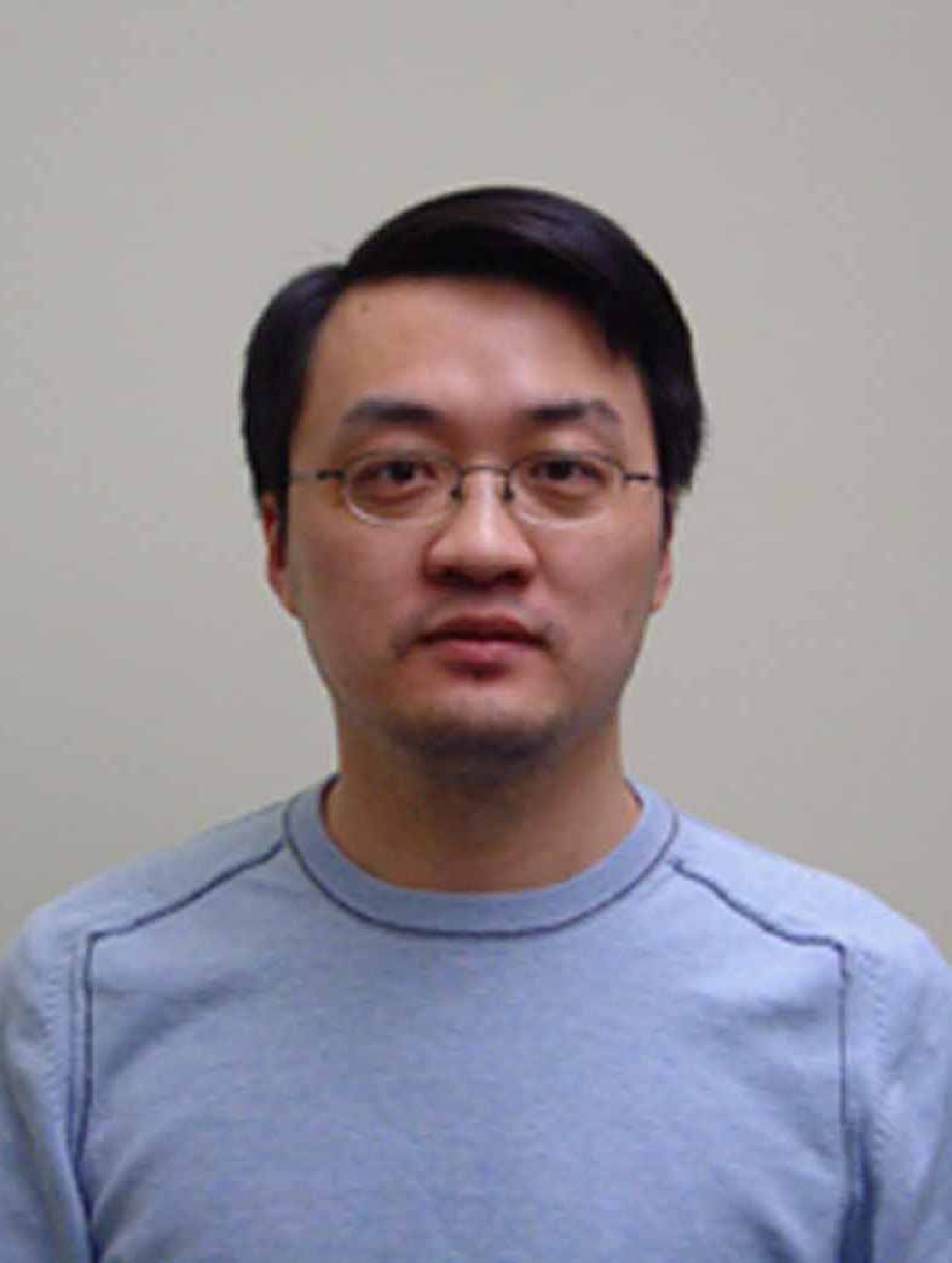}}]
{Yunfei Chen} (Senior Member, IEEE) received his B.E. and M.E. degrees in electronics engineering from Shanghai Jiaotong University (SJTU), Shanghai, P.R.China, in 1998 and 2001, respectively. He received his Ph.D. degree from the University of Alberta in 2006. He is currently working as a Professor in the Department of Engineering at the University of Durham, UK. His research interests include wireless communications, cognitive radios, wireless relaying, and energy harvesting.
\end{IEEEbiography}

\begin{IEEEbiography}[{\includegraphics[width=1in,height=1.25in,clip,keepaspectratio]{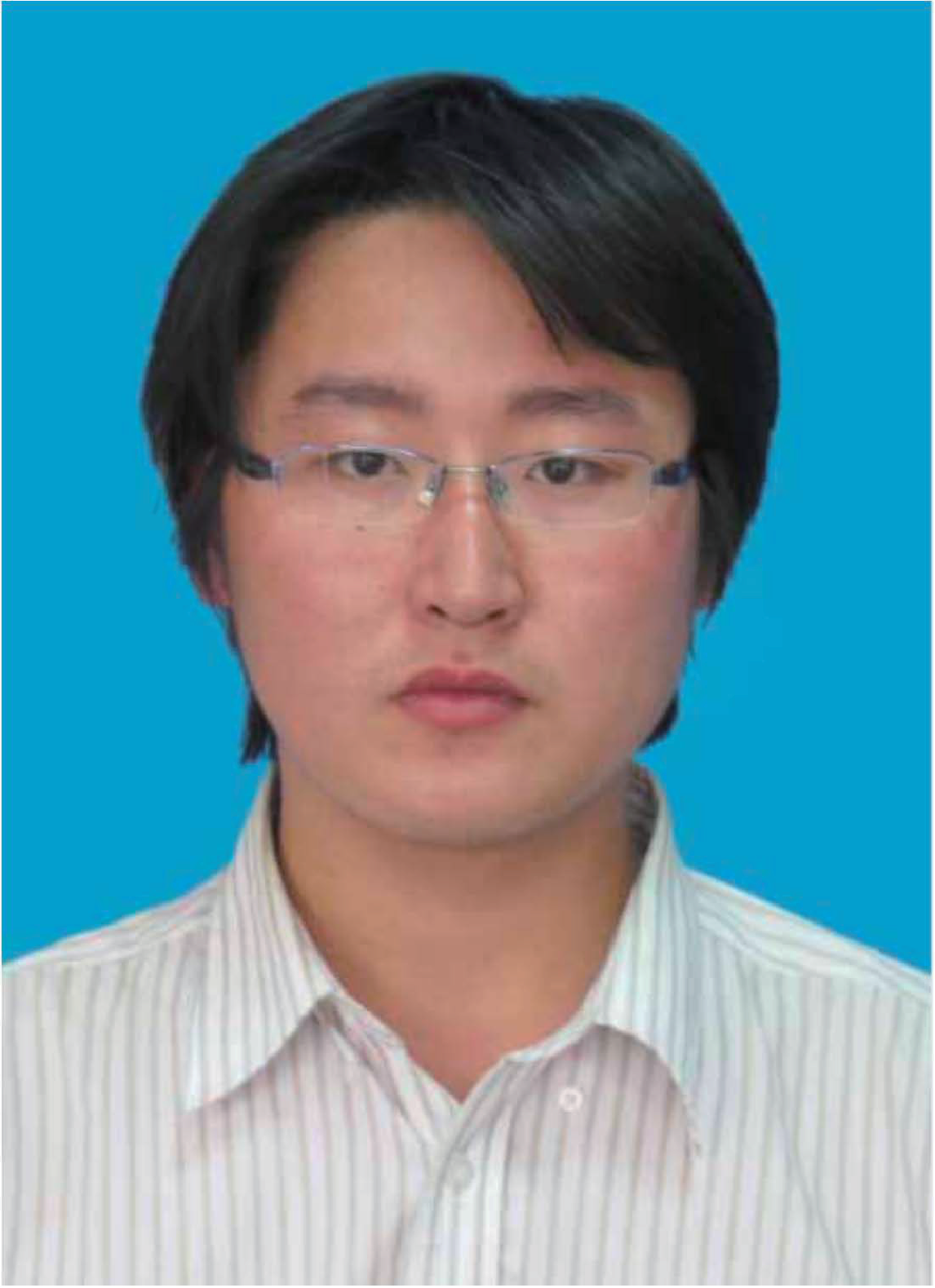}}]
{Nan Zhao} (Senior Member, IEEE) is currently a Professor at Dalian University of Technology (DUT), China. He received his Ph.D. degree in information and communication engineering in 2011 from the Harbin Institute of Technology (HIT), Harbin, China. Dr. Zhao is serving on the editorial boards of IEEE Wireless Communications, IEEE Wireless Communications Letters, and IEEE Transactions on Green Communications and Networking. He won the best paper awards in IEEE VTC in Spring 2017, ICNC in 2018, WCSP in 2018, and WCSP in 2019. He also received the IEEE Communications Society Asia Pacific Board Outstanding Young Researcher Award in 2018.
\end{IEEEbiography}

\begin{IEEEbiography}[{\includegraphics[width=1in,height=1.25in,clip,keepaspectratio]{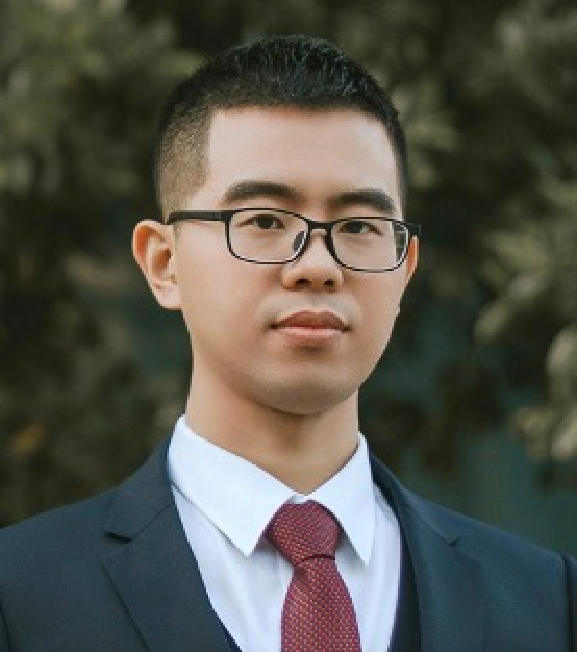}}]
{Changsheng You} (Member, IEEE) received his
B.Eng. degree in 2014 from the University of Science and Technology of China (USTC) and Ph.D. degree in 2018 from the University of Hong Kong (HKU). He is currently an Assistant Professor at the Southern University of Science and Technology (SUSTech) and was a Research Fellow at the National University of Singapore (NUS). His research interests include near-field communications, intelligent reflecting surfaces, UAV communications, and edge computing and learning. Dr. You is a Guest Editor for IEEE Journal on Selected Areas in Communications (JSAC), an editor for IEEE Transactions on Wireless Communications (TWC), IEEE Communications Letters (CL), IEEE Transactions on Green Communications and Networking (TGCN), and 
IEEE Open Journal of the Communications Society (OJ-COMS). He received the IEEE
Communications Society Asia-Pacific Region Outstanding Paper Award in 2019, IEEE ComSoc Best Survey Paper Award in 2021, IEEE ComSoc Best Tutorial Paper Award in 2023. He is listed as the Highly Cited Chinese Researcher, the Exemplary Reviewer of the IEEE Transactions on Communications
(TCOM) and IEEE Transactions on Wireless Communications (TWC).
\end{IEEEbiography}
\end{document}